\documentclass[aps,prb,twocolumn,superscriptaddress,notitlepage]{revtex4-2}
\usepackage{multirow}
\usepackage{xcolor}

\usepackage{overpic}
\usepackage{graphicx}
\usepackage{epsfig}
\usepackage{epstopdf}
\usepackage{float}
\usepackage{subfigure}

\usepackage{dcolumn}
\usepackage{latexsym}
\usepackage{amsmath,amsfonts,amsthm,bm}
\usepackage{wasysym}
\usepackage{bm}

\usepackage[colorlinks,bookmarks=false,citecolor=blue,linkcolor=red,urlcolor=blue]{hyperref}
\usepackage{verbatim}

\usepackage{booktabs}

\usepackage{ulem} 

\usepackage{indentfirst}
\usepackage{listings}
\usepackage{bbold}

\newcommand{\RNum}[1]{\uppercase\expandafter{\romannumeral #1\relax}}

\newcommand{\sgn}{\operatorname{sgn}}

\def\be{\begin{equation}}
\def\ee{\end{equation}}
\def\bea{\begin{eqnarray}}
\def\eea{\end{eqnarray}}

\def\mc{\mathcal}

\usepackage{array}
\newcommand{\PreserveBackslash}[1]{\let\temp=\\#1\let\\=\temp}
\newcolumntype{C}[1]{>{\PreserveBackslash\centering}p{#1}}
\newcolumntype{R}[1]{>{\PreserveBackslash\raggedleft}p{#1}}
\newcolumntype{L}[1]{>{\PreserveBackslash\raggedright}p{#1}}

\usepackage{color}

\definecolor{darkblue}{rgb}{0,0.02,0.45}
\definecolor{darkred}{rgb}{0.45,0.02,0} 

\makeatletter
\newcommand{\thickhline}{%
\noalign {\ifnum 0=`}\fi \hrule height 0.7pt
\futurelet \reserved@a \@xhline
}
\newcolumntype{"}{@{\hskip\tabcolsep\vrrule width 0.7pt\hskip\tabcolsep}}
\makeatother

\usepackage{times}

\begin{document}
\title{Complex orders and chirality in the classical Kitaev-$\bm{\Gamma}$ model}

\author{P. Peter Stavropoulos}
\affiliation{School of Physics and Astronomy, University of Minnesota, Minneapolis, MN 55455, USA}
\author{Yang Yang}
\affiliation{School of Physics and Astronomy, University of Minnesota, Minneapolis, MN 55455, USA}
\author{Ioannis Rousochatzakis}
\affiliation{Department of Physics, Loughborough University, Loughborough, LE11 3TU, UK}
\author{Natalia B. Perkins}
\affiliation{School of Physics and Astronomy, University of Minnesota, Minneapolis, MN 55455, USA}
\affiliation{Technical University of Munich, Germany; Institute for Advanced Study, D-85748 Garching, Germany}
\date{\today}

\begin{abstract}
It is well-recognized that the low-energy physics of many Kitaev materials is governed by two dominant energy scales, the Ising-like Kitaev coupling $K$ and the symmetric off-diagonal $\Gamma$ coupling. An understanding of the interplay between these two scales is therefore the natural starting point toward a quantitative description that includes sub-dominant perturbations that are inevitably present in real materials. This study focuses on the classical $K$-$\Gamma$ model on the honeycomb lattice, with a specific emphasis on the region $K\!<\!0$ and $\Gamma\!>\!0$, which is the most relevant for the available materials and which remains enigmatic in both quantum and classical limits, despite much effort. We employ large-scale Monte Carlo simulations on specially designed finite-size clusters and unravel the presence of a complex multi-sublattice magnetic orders in a wide region of the phase diagram, whose structure is characterized in detail. We show that this order can be quantified in terms of a coarse-grained scalar-chirality order, featuring a counter-rotating modulation on the two spin sublattices. We also provide a comparison to previous studies and discuss the impact of quantum fluctuations on the phase diagram.  
\end{abstract}

\maketitle

\section{Introduction}
Mott insulators with a strong spin-orbit coupling (SOC) have been the subject of a significant research interest in the past decade \cite{Krempa2014ARCMP,Rau2016ARCMP,Winter2016,Winter2017,Takagi2019, Takayama2021JPSJ,Trebst2022,TsirlinPSS2022,Rousochatzakis_2024}. 
In these systems, an interplay of SOC with the crystal fields and strong electron-electron interactions yields anisotropic  bond-dependent exchange interactions between low-energy spin degrees of freedom \cite{Rousochatzakis_2024}.
Most notably, a lot of effort has been devoted to the experimental investigation of the 4$d$ and 5$d$  materials, such as, e.g., $A_2$IrO$_3$ ($A\!=$ Na, Li)~\cite{Singh2010PRB,Singh2012PRL,Choi2012PRL,Ye2012PRB,Chun2015,Williams2016PRB} and $\alpha$-RuCl$_3$~\cite{Plumb2014PRB,Sears2015PRB,Majumder2015}, 
with the goal of finding  candidates  to realize the Kitaev spin liquid \cite{Kitaev2006},  a highly exotic quantum phase of matter. This phase is characterized by fractionalized excitations, non-Abelian anyons, and topological properties that make it of significant interest to physicists. The Kitaev coupling, as originally introduced by Kitaev in 2006 \cite{Kitaev2006}, is indeed the dominant microscopic interaction in all these materials, hence the term `Kitaev materials'~ \cite{Trebst2022}. However, despite the dominance of the Kitaev coupling, most of these materials exhibit magnetic ordering at sufficiently low temperatures \cite{Takagi2019, Takayama2021JPSJ,Trebst2022,TsirlinPSS2022}, indicating the presence of additional interactions in the system.

Studies suggest that a minimal nearest-neighbor (NN) model that effectively describes Kitaev materials is the $J$-$K$-$\Gamma$-$\Gamma'$ model on the honeycomb lattice \cite{Winter2017,MaksimovPRR2020,Rousochatzakis_2024}. 
In most cases, the bond-dependent off-diagonal coupling $\Gamma$ is  of comparable magnitude to the Kitaev interaction $K$, while the  $J$ and $\Gamma'$ interactions are rather small.
The overall predominance of the $K$ and $\Gamma$ interactions can be attributed to two facts:
the indirect superexchange via ligand $p$-orbitals often dominates over
the direct exchange contributions which are responsible for the Heisenberg coupling, and the relatively small trigonal distortion in these systems suppresses the $\Gamma'$.
The $\Gamma$ interaction, stemming from a combination of direct and ligand-mediated hopping, typically exhibits a weaker strength than $K$ but remains larger than other subdominant interactions allowed by symmetry.

The phase diagram of the four dimensional $J$-$K$-$\Gamma$-$\Gamma'$ parameter space is very rich, and the interplay of these additional interactions beyond the Kitaev $K$ coupling plays a crucial role in determining the specific magnetic orderings observed in the Kitaev materials ~\cite{Jackeli2009,Chaloupka2010,Chaloupka2013,
Rau2014,rau2014trigonal,Sizyuk2014,Rousochatzakis2015PRX,Winter2016,
Rousochatzakis2017PRL,Winter2017,MaksimovPRR2020,Rousochatzakis_2024}.
In addition to the compelling question of how close these systems are to the Kitaev quantum spin liquid, probed through a range of dynamical probes with signatures of fractionalization  in low-energy excitations, it is crucial to recognize that Kitaev materials harbor competing magnetic orders characterized by significant complexity and unconventional behavior. This aspect merits in-depth investigation in its own regard, as evident by a substantial body of research in the field \cite{Krempa2014ARCMP,Rau2016ARCMP,Winter2016,Winter2017,Takagi2019, Takayama2021JPSJ,Trebst2022,Rousochatzakis_2024}. For instance, in the lithium allotropes $\alpha,\beta,\gamma$-Li$_2$IrO$_3$, incommensurate phases with intricate internal structures have been observed \cite{TsirlinPSS2022}. Further, some of the predicted nearby magnetic phases exhibit finite chirality, potentially imparting non-trivial topological characteristics to excitations, some others acquire chirality in the presence of the magnetic field. These phases may manifest signatures in thermal transport measurements and give rise to either anomalous  or normal thermal Hall conductivity \cite{Fujimoto2009,LiChern2021,Zhang2021,Zhang2023}.

One of the most interesting regions in the parameter space of the $J$-$K$-$\Gamma$-$\Gamma'$ model and the focus of this work 
is the $K$-$\Gamma$ line, which connects the two types
of strongly correlated regimes, the Kitaev
quantum spin liquid \cite{Kitaev2006} and the $\Gamma$ classical spin liquid \cite{Rousochatzakis2017PRL}.    
Contrary to the Heisenberg interaction on the honeycomb lattice, which leads to simple collinear ordering (ferromagnetic or antiferromagnetic), both Kitaev and $\Gamma$ interactions on generic  tricoordinated lattices are highly frustrated,  with an infinite number of classical ground states, see Refs.~\cite{BaskaranPRB2008,Chandra2010,Samarakoon2017PRB,RousochatzakisNC2018} and \cite{Rousochatzakis2017PRL,Samarakoon2018PRB,Chern2019PRL}, respectively.
Importantly, the local symmetries responsible for the infinite degeneracy remain true symmetries in the quantum regime only for the Kitaev model~\cite{BaskaranPRB2008}, but not for the $\Gamma$ model~\cite{Rousochatzakis2017PRL}. As a result,
the effects of quantum fluctuations in these two models are qualitatively different: while the Kitaev model maintains its spin liquidity in the quantum regime (for any spin $S$~\cite{RousochatzakisNC2018}) due to Elitzur's theorem~\cite{Elitzur1975}, the degeneracy of the $\Gamma$ model is  accidental and is therefore lifted by  quantum fluctuations.  In particular, the temperature scale below which this happens and the type of selected ground state depend on the sign of $\Gamma$~\cite{Rousochatzakis2017PRL}.

When both $K$ and $\Gamma$ are present, the ground states are known only when the two couplings have the same sign~\cite{Rau2014,Rousochatzakis_2024}. The parameter regions of opposite signs remain elusive in both the quantum and the classical regime, despite substantial research efforts~\cite{Rousochatzakis2017PRL,Samarakoon2018PRB,Gohlke2018,Chern2019PRL,Gordon2019NC,Gohlke2020PRR,Wachtel2019PRB,Chern2020PRR,Yamada2020PRB,Liu2021PRR,Buessen2021PRB}. Here we study the opposite sign regime of the $K$-$\Gamma$ model,  with negative $K$ and positive $\Gamma$,  using numerical Monte-Carlo methods and specially designed finite-size cluster minimization procedures. We find an intermediate phase (IP)  that occupies the majority of the phase diagram when $K<0$ and $\Gamma >0$, and is situated between two commensurate phases,  denoted by 18C$_3$ and $6'$ in the literature. The  spatial modulation of the IP is parameter dependent and our finite-size numerics suggests a cascade of incommensurate phases. The latter can be described as long-wavelength modulations of the neighboring 18C$_3$ state,  which essentially interpolate between an 18C$_3^A$ state centered on the $A$ sublattice and a 18C$_3^B$ state centered on the $B$ sublattice. The local inversion symmetry breaking associated with this sublattice center switching is manifested in characteristic counter-rotating structure of the coarse-grained scalar chirality which we analyze in detail.

The rest of the paper is organized as follows: In Sec.~\ref{sec:model} we present the explicit model and its symmetries. A summary of known results for the classical model along the $K\Gamma$ line is summarized in Sec.~\ref{sec:knownres}. Our main results for $K<0$ and $\Gamma >0$ are presented in Sec.~\ref{sec:main_results}, and the IP is analyzed in detail in Sec.~\ref{sec:all_about_the_IS_state}. The scalar and coarse-grained scalar chiralities are defined and analyzed in Sec.~\ref{sec:chir}. In Sec.~\ref{sec:compa} we provide a comparison to previous studies, and in Sec.~\ref{sec:QM} we discuss the impact of quantum-mechanical fluctuations on the phase diagram. Our conclusions are given in Sec.~\ref{sec:conclusions}. Details of the numerical minimization methods are found in Sec.~\ref{sec:numer_methods} and supplementary information is presented in the Appendices.

\section{Model and symmetries}\label{sec:model}
\begin{figure}
\centering
\includegraphics[width=1.0\columnwidth]{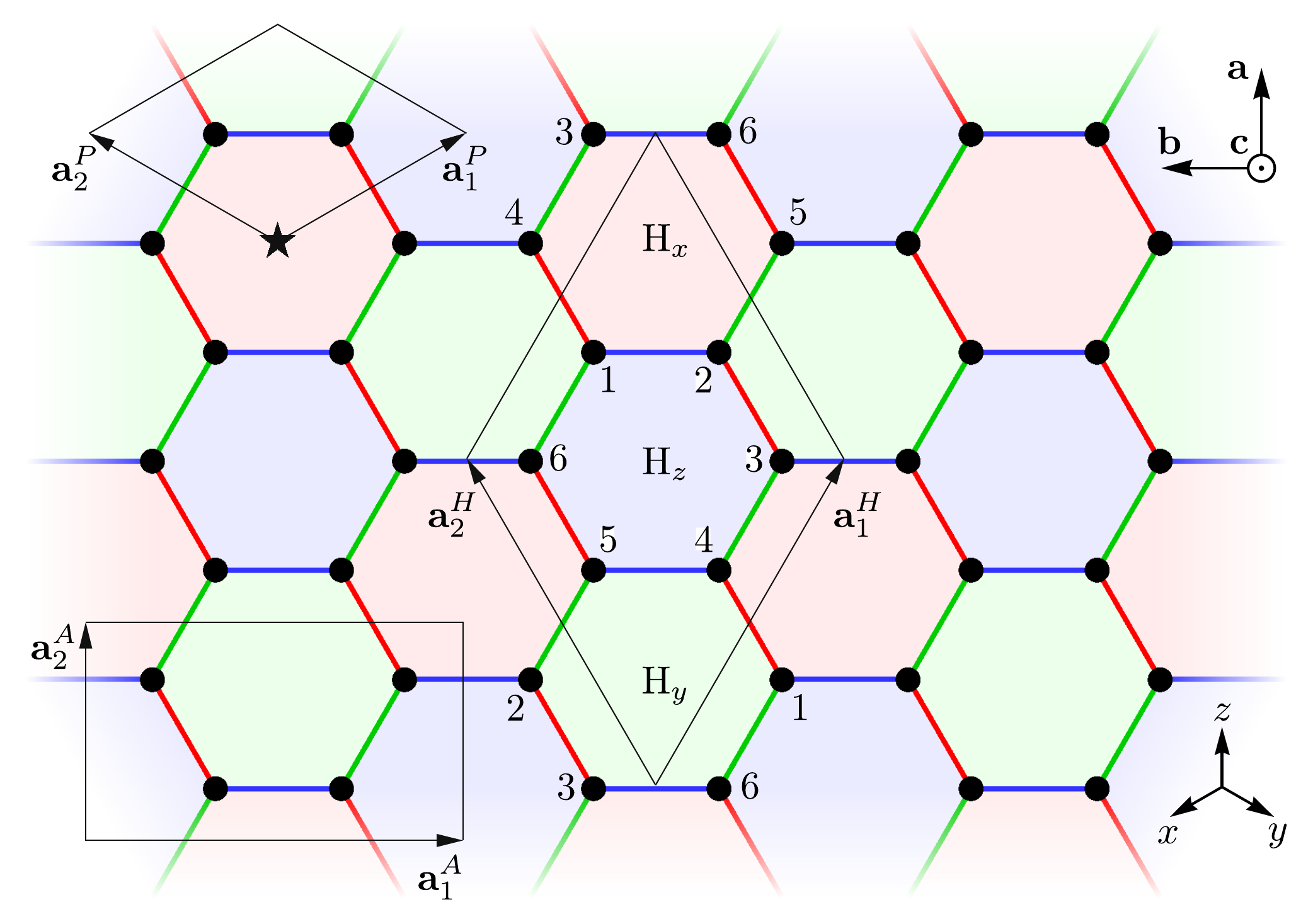}
\caption{The honeycomb structure spans the crystallographic plane $\mathbf{a}\!=\![\overline{1},\overline{1},2]/\sqrt{6}$ and $\mathbf{b}\!=\![1,\overline{1},0]/\sqrt{2}$, perpendicular to $\mathbf{c}\!=\!\mathbf{a}\!\times\!\mathbf{b}\!=\![1,1,1]/\sqrt{3}$, written in cubic coordinates $xyz$. Red, green, and blue indicate NN X-, Y- and Z-bonds respectively, which are parallel to $[0\overline{1}1]$, $[10\overline{1}]$, and $[\overline{1}10]$ directions respectively. A primitive $(\mathbf{a}^P_1,\mathbf{a}^P_2)\!=\!\sqrt{2}([\overline{2},1,1],[1,\overline{2},1])$ unit cell is shown, with the two sublattice in fractional coordinates $(1/3,2/3)$, $(2/3,1/3)$. The crystallographic inversion center is shown by a star. The honeycomb hexagons can be partitioned into three disjoint groups H$_x$, H$_y$, H$_z$, shaded, respectively, by red, green, and blue, up to hexagon conventional unit cell translations $(\mathbf{a}^H_1,\mathbf{a}^H_2)\!=\!(2\mathbf{a}^P_1\!+\!\mathbf{a}^P_2,\mathbf{a}^P_1\!+\!2\mathbf{a}^P_2)$. The armchair unit cell $(\mathbf{a}^A_1,\mathbf{a}^A_2)\!=\!(\mathbf{a}^P_1\!-\!\mathbf{a}^P_2,\mathbf{a}^P_1\!+\!\mathbf{a}^P_2)$ is also shown.}\label{fig:basic_defenitions}
\end{figure}

The $K$-$\Gamma$ model on the honeycomb  lattice reads as
\begin{equation}\label{eq:Model}
\renewcommand{\arraystretch}{1.2}
\begin{array}{c}
\mc{H}=\mc{H}^{(\text{X})}+\mc{H}^{(\text{Y})}+\mc{H}^{(\text{Z})},\\
\mc{H}^{(\text{X})}=\displaystyle\sum\limits_{\langle i j \rangle \in \text{X}}  \!\left[ K S_i^x S_j^x+\Gamma \!\left(\! S_i^y S_j^z \!+\! S_i^z S_j^y \right) \right],\\
\mc{H}^{(\text{Y})}=\displaystyle\sum\limits_{\langle i j \rangle \in \text{Y}}  \!\left[ K S_i^y S_j^y+\Gamma \!\left(\! S_i^z S_j^x \!+\! S_i^x S_j^z \right) \right],\\
\mc{H}^{(\text{Z})}=\displaystyle\sum\limits_{\langle i j \rangle \in \text{Z}}  \!\left[ K S_i^z S_j^z+\Gamma \!\left(\! S_i^x S_j^y \!+\! S_i^y S_j^x \right) \right],
\end{array}
\end{equation}
where $\langle ij\rangle$ denote nearest-neighbor lattice sites  forming an  X-, Y-, or Z-type of bond, shown, respectively, by red, green and blue in Fig.~\ref{fig:basic_defenitions}. This Figure also shows the conventional unit cells and their spanning vectors, as well as the crystallographic ``$abc$'' and cubic ``$xyz$'' frame.

The model  has a $D_{3d}$ point group, consisting of threefold rotations $C_3$ around the out of plane $\mathbf{c}$-axis, twofold rotations $C_{2\text{X}(\text{Y}\text{Z})}$ around the X(YZ)-bond directions,  inversion (through the center of any hexagon plaquette), as well as the mirror symmetries
$m_{\text{X}(\text{Y}\text{Z})}=IC_{2\text{X}(\text{Y}\text{Z})}$ and improper rotations $S_6=IC_3$.

Aside from the space group symmetries, the model has three global, site-dependent, symmetries  $\mc{R}_x$, $\mc{R}_y$, and $\mc{R}_z$, corresponding to six-sublattice operations in the spin-space alone, that map the Hamiltonian to itself (see supplementing material of Ref.~\cite{Rousochatzakis2017PRL} and Ref.~\cite{Rousochatzakis_2024}). To see this one can define a six-sublatice conventional hexagon unit cell ($\mathbf{a}_1^H$, $\mathbf{a}_2^H$) that tiles the entire honeycomb. Based on these translations we can partition the set of all hexagons into three disjoint subsets H$_x$, H$_y$, H$_z$, shown as red, green, and blue subsets, respectively, in Fig.~\ref{fig:basic_defenitions}. Using the site labeling of Fig.~\ref{fig:basic_defenitions}, the $\mc{R}$ symmetries can be written  as the following combinations of two-fold rotations  $\mathsf{C}_{2x(yz)}$ (around cubic $x,\,y,\,z$ axes) in spin space: 
\be\label{eq:Rsymmetries}
\renewcommand{\arraystretch}{1.2}
\begin{array}{l}
\mc{R}_x \!=\! \displaystyle\prod\limits_{i\in \{2, 3\}}\!\mathsf{C}_{2x}(i)\!\prod\limits_{j\in \{1, 6\}}\!\mathsf{C}_{2y}(j)\!\prod\limits_{k\in \{4, 5\}}\!\mathsf{C}_{2z}(k),\\
\mc{R}_y \!=\! \displaystyle\prod\limits_{i\in \{5, 6\}}\!\mathsf{C}_{2x}(i)\!\prod\limits_{j\in \{3, 4\}}\!\mathsf{C}_{2y}(j)\!\prod\limits_{k\in \{1, 2\}}\!\mathsf{C}_{2z}(k),\\
\mc{R}_z \!=\! \displaystyle\prod\limits_{i\in \{1, 4\}}\!\mathsf{C}_{2x}(i)\!\prod\limits_{j\in \{2, 5\}}\!\mathsf{C}_{2y}(j)\!\prod\limits_{k\in \{3, 6\}}\!\mathsf{C}_{2z}(k)\,.
\end{array}
\ee
For instance, the $\mc{R}_z$ transformation  can be visualized as
\be\label{fig:Rsymfig}
\includegraphics[width=0.6\columnwidth]{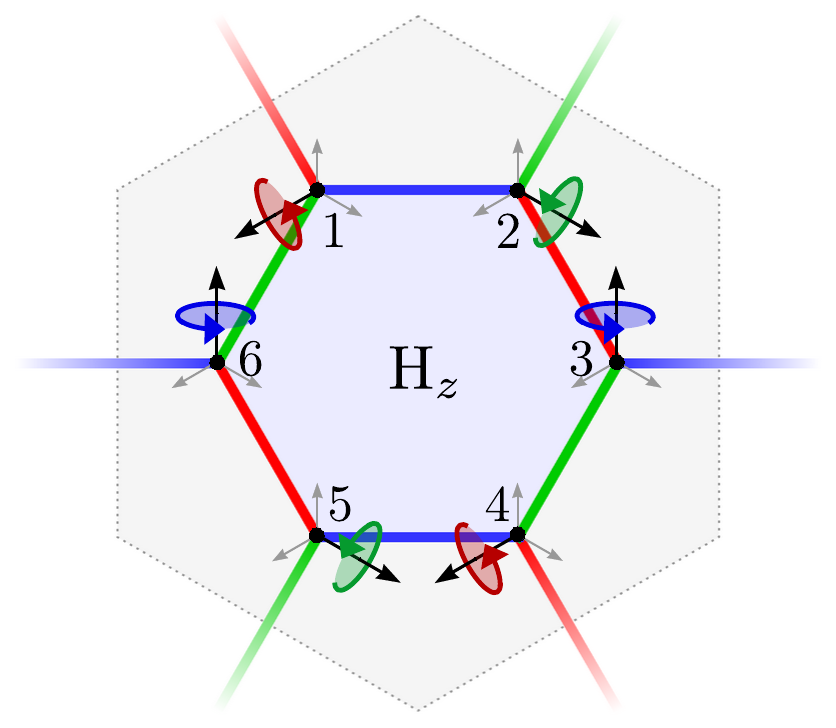}
\ee
Note that the operations $\mc{R}_x$, $\mc{R}_y$ and $\mc{R}_z$ can be thought of as products of the Kitaev's plaquette operators~\cite{Kitaev2006} sitting on the  H$_x$, H$_y$, or H$_z$ subsets.
The three $\mc{R}_\alpha$ operators, together with the identity, form the Klein four-group, i.e., $\mc{R}_\alpha \mc{R}_\beta\!=\!\mc{R}_\gamma$ and $\mc{R}_\alpha \!=\!\mc{R}_\alpha^{-1}$. Due to  their site-dependent nature, these symmetries can mix the local character of any state in a nontrivial way. As we shall see, this leads to degeneracies between very different looking orders.

The $K$-$\Gamma$ model  has a  hidden SU(2) symmetry  when $K\!=\!\Gamma$. The associated transformation is again a six-sublattice operation in spin space alone~\cite{ChaloupkaPRB2015}, which, for the H$_z$ subset of the hexagons, can be written as~\cite{Rousochatzakis_2024}
\be\label{eq:T6ztrans}
\renewcommand{\arraystretch}{1.05}
\begin{array}{lclcrrr}
\multicolumn{7}{c}{\mc{T}_{6z}~\text{transformation:}}\\
\widetilde{\mathbf{S}}_1=\mathbf{S}_1                                          & \Rightarrow & \mathbf{S}_1 &=(&  \widetilde{S}_1^x,&  \widetilde{S}_1^y,&  \widetilde{S}_1^z ),\\
\widetilde{\mathbf{S}}_2=\mathsf{C}_{2\text{Z}}\mathbf{S}_2                    & \Rightarrow & \mathbf{S}_2 &=(& -\widetilde{S}_2^y,& -\widetilde{S}_2^x,& -\widetilde{S}_2^z ),\\
\widetilde{\mathbf{S}}_3=\mathsf{C}_{3}^{-1}\mathbf{S}_3                       & \Rightarrow & \mathbf{S}_3 &=(&  \widetilde{S}_3^y,&  \widetilde{S}_3^z,&  \widetilde{S}_3^x ),\\
\widetilde{\mathbf{S}}_4=\mathsf{C}_{3}\mathsf{C}_{2\text{Z}}\mathbf{S}_4      & \Rightarrow & \mathbf{S}_4 &=(& -\widetilde{S}_4^x,& -\widetilde{S}_4^z,& -\widetilde{S}_4^y ),\\
\widetilde{\mathbf{S}}_5=\mathsf{C}_{3}\mathbf{S}_5                            & \Rightarrow & \mathbf{S}_5 &=(&  \widetilde{S}_5^z,&  \widetilde{S}_5^x,&  \widetilde{S}_5^y ),\\
\widetilde{\mathbf{S}}_6=\mathsf{C}_{3}^{-1}\mathsf{C}_{2\text{Z}}\mathbf{S}_6 & \Rightarrow & \mathbf{S}_6 &=(& -\widetilde{S}_6^z,& -\widetilde{S}_6^y,& -\widetilde{S}_6^x ).\\
\end{array}
\ee
where $(S_i^x, S_i^y, S_i^z)$ are the components in the (global) cubic $xyz$ frame, and $(\widetilde{S}_i^x, \widetilde{S}_i^y, \widetilde{S}_i^z)$ are the components in the rotated,  site-dependent frame.
The transformations $\mc{T}_{6x}$ and $\mc{T}_{6y}$ on the H$_x$ and H$_y$ subsets of hexagons are defined accordingly. To see the hidden SU(2) symmetry, we consider, e.g., the interactions on a Z-bond (1,2). Under $\mc{T}_{6z}$, we have
\begin{equation}
K S_1^z S_2^z + \Gamma (S_1^x S_2^y +S_1^y S_2^x ) \mapsto 
-K \widetilde{S}_1^z \widetilde{S}_2^z - \Gamma (\widetilde{S}_1^x \widetilde{S}_2^x +\widetilde{S}_1^y \widetilde{S}_2^y ),\nonumber
\end{equation}
which, in turn, maps to $-K\ \widetilde{\bf S}_1\!\cdot\!\widetilde{\bf S}_2$ when $K\!=\!\Gamma$. The same holds for the other bonds.  Thus, the point $\!K\!=\!\Gamma\!<\!0$ is a hidden antiferromagnetic (AFM) SU(2) point, and the point $\!K\!=\!\Gamma\!>\!0$ is a hidden ferromagnetic (FM) SU(2) point.

Finally, we note that, in the classical model, there exists a duality transformation which flips the spins of the second sublattice of the honeycomb.  Since this requires acting with time reversal on half of the sites, this transformation is only possible in the classical regime. In this regime, the duality effectively maps $(K,\Gamma)\!\to\!(-K,-\Gamma)$, so the classical phase diagram for a given set of couplings can be obtained from that of the opposite couplings.  This leads to two, qualitatively different regions in the parameter space, the ones where $K$ and $\Gamma$ have the same sign, and  the ones with opposite signs.

\section{Summary of known results about the  Classical phase diagram of the  $K$-$\Gamma$ model}\label{sec:knownres}
Before we present our results, it is instructive to briefly review the main  known features of the phase diagram of $K$-$\Gamma$ model. As usual, we parametrize the two couplings of the model in terms of an angle $\psi$, as 
\be
K = \cos\psi,\ \Gamma = \sin\psi\,,
\ee
and work in energy units of $\sqrt{K^2+\Gamma^2}=1$. In the regions $\psi\!\in\![0,\pi/2]$ and $\psi\!\in\![\pi,3\pi/2]$, where $K$ and $\Gamma$ have the same sign, we can use the following recipe to get the classical ground states~\cite{Rousochatzakis2017PRL,RousochatzakisNC2018}: for every site $i$, write the components of the spin as $\mathbf{S}_i\!=\![x_i,y_i,z_i]$ with $x_i^2\!+\!y_i^2\!+\!z_i^2\!=\!S^2$, and then align its NN spins along $\zeta [x_i,z_i,y_i]$, $\zeta[z_i,y_i,x_i]$, or $\zeta[y_i,x_i,z_i]$, if the two sites share, respectively, an X-, Y-, or Z-bond, and $\zeta\!=\!-\sgn(K)=\!-\sgn(\Gamma)$. Schematically,
\be\label{eqfig:KnegGnegmodelClassGSs}
\includegraphics[width=0.6\columnwidth]{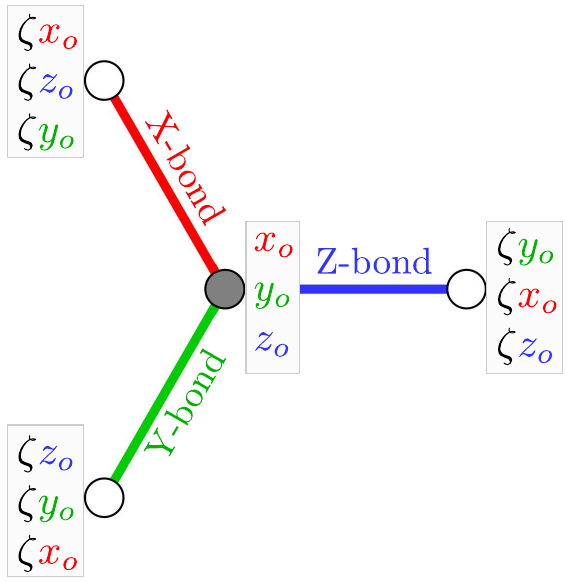}
\ee
The energy contributions from the three bonds emanating from each site $i$ add up to $-(|K|+2|\Gamma|) (x_i^2\!+\!y_i^2\!+\!z_i^2)\!=\!-(|K|+2|\Gamma|)S^2$. Since each bond is shared by two sites, the configurations that are generated by the above recipe saturate the energy lower bound $E_{\text{min}}/N\!=\!-(|K|/2+|\Gamma|) S^2$, and are therefore ground states. 

Denoting by $[x_o,y_o,z_o]$ the spin of a starting, reference site, we can apply the above recipe to its first neighbors, and then to the second neighbors, and so on, until we cover the whole lattice. The resulting configurations take the form
\be\label{eqfig:KGClassGSsSameSign}
\includegraphics[width=0.9\columnwidth]{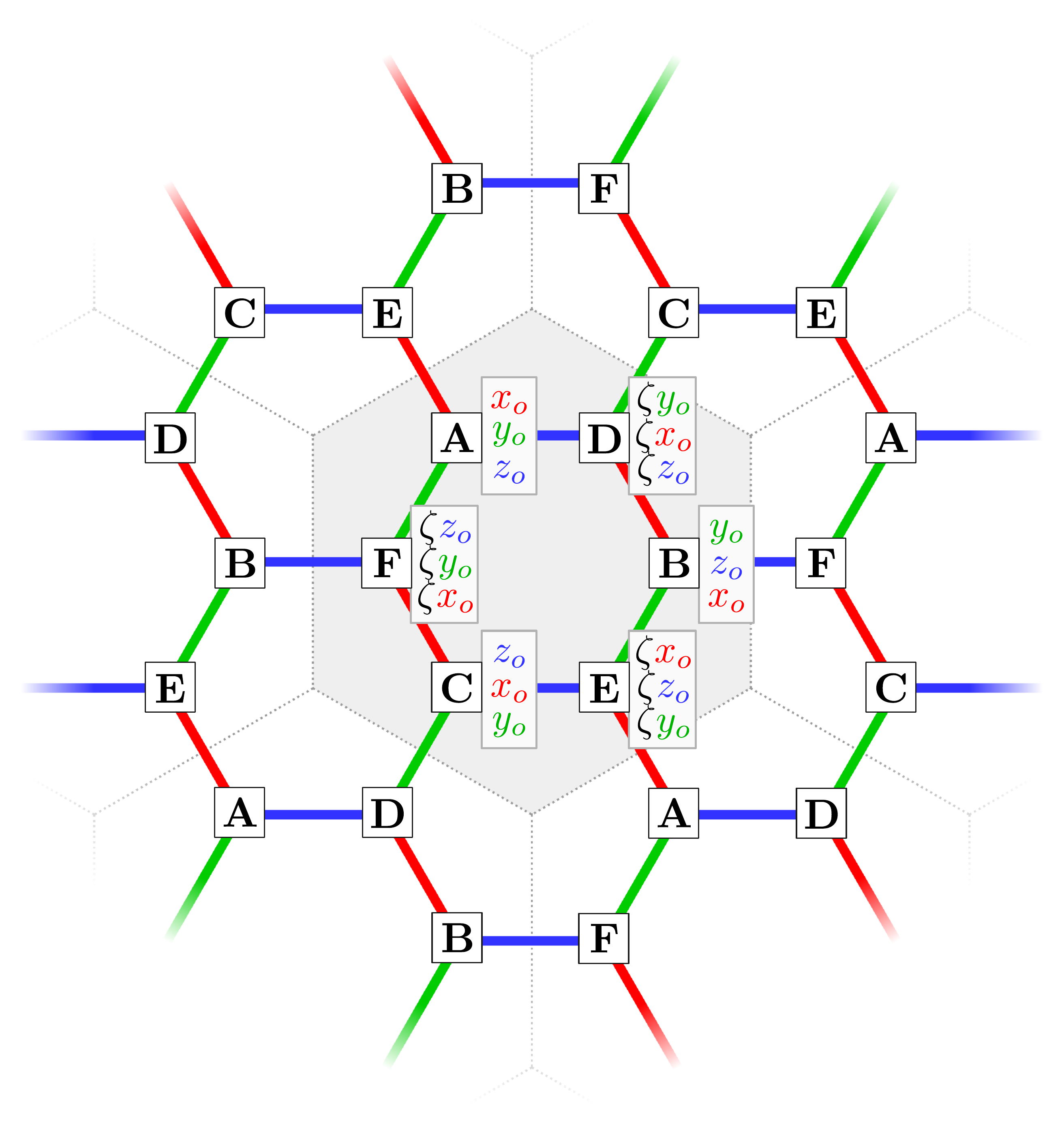}
\ee
with the following features, for general $[x_o,y_o,z_o]$:

i) There are six spin directions, $\mathbf{A}$ through $\mathbf{F}$, with spins pointing along the directions 
\be\label{eq:ABCDEF}
\renewcommand{\arraystretch}{1.2}
\!\!\begin{array}{lll}
{\bf A}\!=\![x_o,y_o,z_o], 
&{\bf B}\!=\![y_o,z_o,x_o], 
&{\bf C}\!=\![z_o,x_o,y_o],
\\
{\bf D}\!=\!\zeta[y_o,x_o,z_o], 
& {\bf E}\!=\!\zeta[x_o,z_o,y_o], 
&{\bf F}\!=\!\zeta[z_o,y_o,x_o]\,.
\end{array}
\ee

ii) The states form a two-parameter manifold, associated with the choice of $[x_o,y_o,z_o]$ of the reference site. 

iii) The states break translational symmetry, with the underlying superlattice shown by dashed lines in Eq.~(\ref{eqfig:KGClassGSsSameSign}) i.e., the Wigner-Seitz cell of the conventional hexagonal unit cell ($\mathbf{a}_1^H$, $\mathbf{a}_2^H$) from Fig.~\ref{fig:basic_defenitions}.

iv)   Going from $\mathbf{A}\!\to\!\mathbf{B}\!\to\!\mathbf{C}$ is achieved  by successive {\it clockwise} 120$^\circ$ spin rotation around the $\mathbf{c}$-axis (even-parity cyclic permutations of $[x_o,y_o,z_o]$), while  for $\mathbf{D}\!\to\!\mathbf{E}\!\to\!\mathbf{F}$ the spin rotation is {\it counter-clockwise} (odd-parity cyclic permutations of $[x_o,y_o,z_o]$).  So, the classical ground states feature a period-3 modulation with two counter-rotating sublattices. This is one of the key ramifications of the interplay between $K$ and $\Gamma$, and here it  arises simply from the requirement to saturate the energy contributions from both couplings along all bonds. 

v) In the local frames defined by the $\mc{T}_6$ transformation of Eq.~(\ref{eq:T6ztrans}), the above states map to the (much simpler) collinear N\'eel  state (with moments along $\widetilde{\mathbf{S}}_i=\pm [x_o,y_o,z_o]$) for $\zeta=-1$, or the FM state (with moments along $\widetilde{\mathbf{S}}_i=[x_o,y_o,z_o]$) for $\zeta=1$. So, for negative $K$ and $\Gamma$, we end up with a ``dual N\'eel'' phase, which includes the hidden SU(2) point $\psi\!=\!5\pi/4$, and, similarly, for positive $K$ and $\Gamma$, we get a ``dual FM'' phase, which includes the hidden SU(2) point $\psi\!=\!\pi/4$.

It should be noted that at the isolated Kitaev points or the isolated $\Gamma$ points, there are infinitely more ground states, which can be generated by the two recipes described in Refs.~\cite{Rousochatzakis2017PRL} and \cite{RousochatzakisNC2018}. The construction of (\ref{eqfig:KnegGnegmodelClassGSs}) is essentially a combination of those two recipes, which works when the two couplings have the same sign~\cite{Rousochatzakis_2024}.

When $K$ and $\Gamma$ have opposite signs, the two recipes cannot be combined due to frustration, and the phase diagram is more complex. Various classical energy  minimization approaches, such as  iterative simulated annealing \cite{Chern2020PRR}, classical Monte Carlo simulations \cite {RayyanPRB2021,Chen2023NJP}, and machine learning approach \cite{Liu2021PRR,Rau2021PRR} have been employed to study the phase diagram in this region. These studies coherently show that a big part of the phase diagram is occupied by magnetic orders with large unit cells and/or truly incommensurate phases, but overall there is no consensus on the overall  structure  of the classical phase diagram in this region. This part of the phase diagram is the main focus of  our study.

\begin{figure*}
\centering
    \begin{overpic}[width=1.0\textwidth,percent]{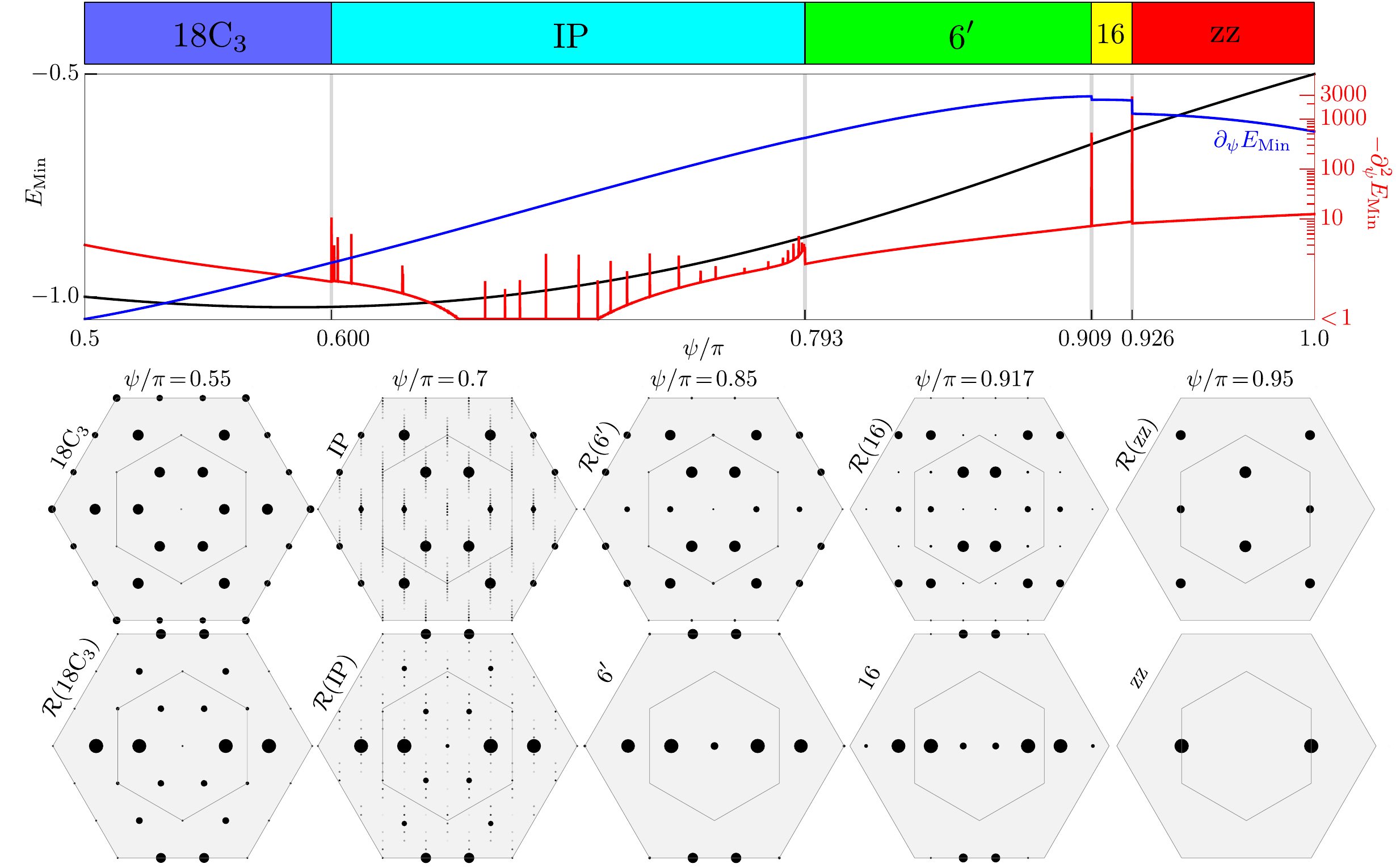}
       \put(0.,61){(a)}
       \put(0.,56.5){(b)}
       \put(0.,34.5){(c)}
    \end{overpic}  
\caption{(a) Phase diagram in the region $\psi\!/\!\pi\!\in\![0.5,1.0]$. 
(b) Minimum classical energy, $E_{\text{Min}}$ as a function of $\psi/\pi$ (black curve),  as obtained by stitching together results from Monte Carlo runs and trial ansatz minimizations. Blue and red curves show the first and second derivatives of $E_{\text{Min}}$,  respectively. The peaks in the second derivative (red curve) indicate 
{
across all the  clusters considered}.
Note  multiple weak peaks in the IP, subleading by two order of magnitude from the strong transitions $6'$ $\rightarrow$ 16.
   (c)  Static structure factors (SSF) of the  minimum energy state at representative $\psi$,  one from each of the five regions. (SSFs decomposed in the cubic coordinates for the same values of the representative values $\psi$ are also shown Fig.~\ref{fig:SSF_decomposition}.)
   The magnitudes of the SSF peaks are shown in the logarithmic scale from largest black dot to smallest gray dot, with the over all values normalized individually for every $\psi$ to the largest peak of that case.}\label{fig:all_the_energies}
\end{figure*}

\section{Classical phase  diagram for $\bm{K\!<\!0}$ and  $\bm{\Gamma\!>\!0}$ }\label{sec:results}

\begin{figure}
	\centering
    \begin{overpic}[width=1.0\columnwidth,percent]{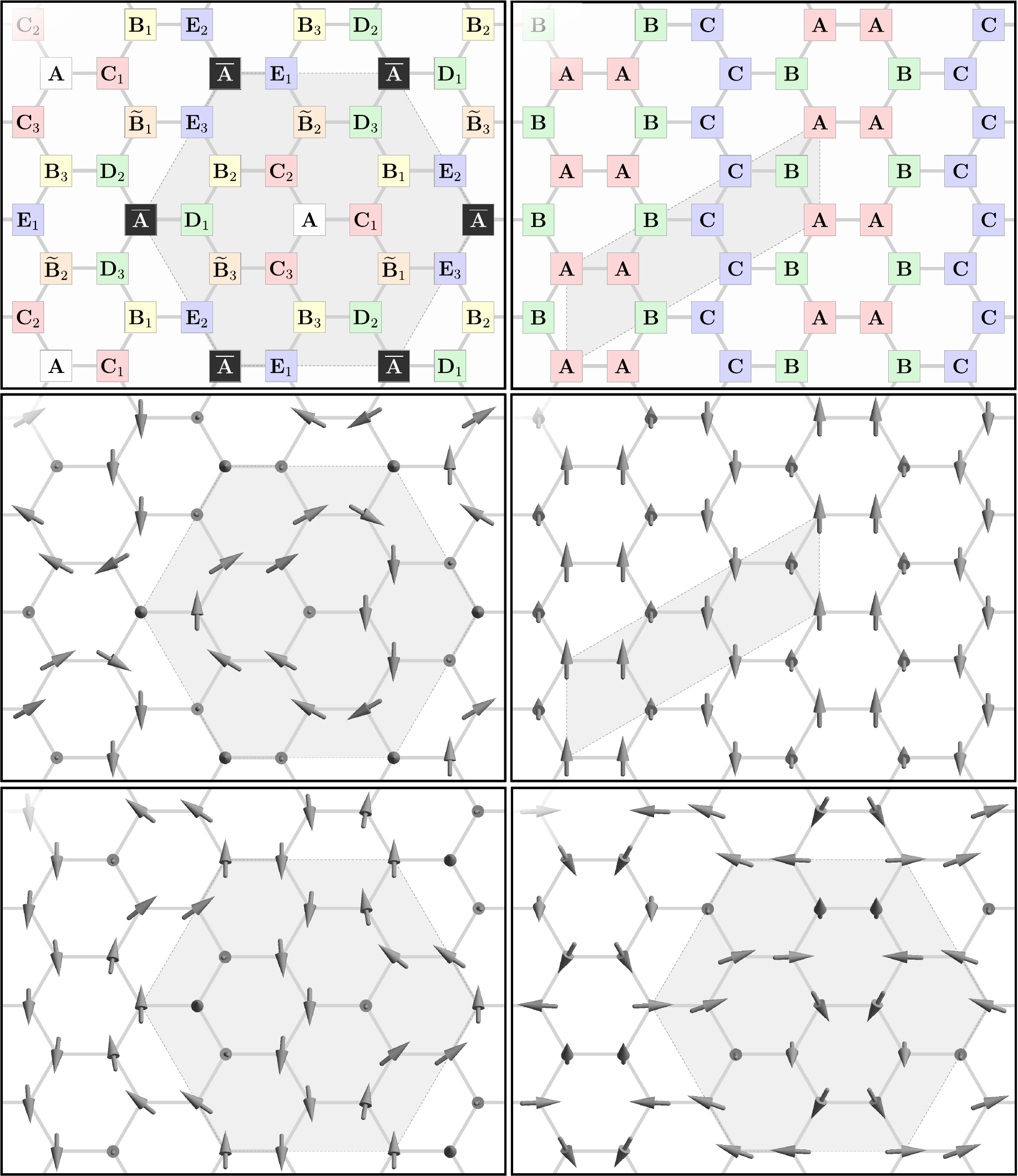}
       \put(0.5,30){(c)}
       \put(0.5,63.5){(b)}
       \put(0.5,97){(a)}
       \put(43.8,30){(f)}
       \put(43.8,63.5){(e)}
       \put(43.8,97){(d)}
    \end{overpic}  
 \caption{The ansatz 18C$_3$ state (a)-(c) and $6'$ state (d)-(f). For each of the ansatz we show: a color coded sketch of the ansatz (a) and (d), that corresponds to Eq.~(\ref{eq:18c3anz}) and (\ref{eq:Kanz}), the real space configuration of the spin state (b) at $\psi\!/\!\pi=0.55$ and (e) at $\psi\!/\!\pi=0.85$, and their $\mc{R}$ dual state (c) and (f). Magnetic unit cells are indicated by a faint gray area. The $\mc{R}$ symmetries reveal dual states, with the 18C$_3$ state being degenerate with a dual 18 site order, and the $6'$ state degenerate with another different dual 18 site order. }\label{fig:three_classical_ansatz_A}
\end{figure}

\subsection{Computational methods}\label{sec:numer_methods}

\subsubsection{Parallel Tempering}

We performed classical Monte Carlo simulation of the $K$-$\Gamma$ model employing a combination of simulated annealing and parallel tempering approaches. In the simulated annealing we used an exponential cooling scheduler, performing $100-200$ cooling steps, with $10^5-10^6$ sweeps per step, cooling from $T=3$ down to $T=10^{-7}$  (all energies and temperatures are given in units of $\sqrt{K^2+\Gamma^2}$). In parallel tempering, we simulate $256$ replicas, with temperature logarithmically spread between $T=3$ and $T=10^{-5}$, and performing $10^5$ temperature updates for smaller clusters and up to $0.5 \times10^6$ for the largest clusters, with 100 sweeps between updates. One sweep consists of system-size number of Metropolis–Hastings trial updates.

\subsubsection{Subsequent refinement with numerical optimization}
The states delivered by Monte Carlo simulations were subsequently refined by two independent numerical minimization methods.
The first involves using the non-linear optimization Ceres Library~\cite{Agarwal_Ceres_Solver_2022} to solve the extrema equations
\be
(\partial_{S_{i}^{\alpha}} \widetilde{\mathcal{H}}, \partial_{\lambda_i}\widetilde{\mathcal{H}})=0\,,
\ee
where $\widetilde{H}$ is related to the Hamiltonian $\mc{H}$ by 
\be
\widetilde{\mc{H}} = \mc{H} + \sum_i \lambda_{i}\left(\mathbf{S}_i\!\cdot\!\mathbf{S}_i-1\right),
\ee 
with $\lambda_i$ being the Lagrange multipliers that enforce the spin length constraints at each site.   

The second method, called  ``torque updates'' in the following, is based on an iterative numerical minimization scheme, whereby  we select spin sites at random and rotate them in the direction of the corresponding local mean field ${\bf h}_i=-\partial\mc{H}/\partial{\bf S}_i$. The updates are performed repeatedly, until the convergence criterion $\mathrm{Max}\left\{\left\Vert \mathbf{S}'_i-\mathbf{S}_i \right\Vert\right\}<10^{-15}$ is satisfied for all sites $i$. 

Within numerical precision, the results from the non-linear optimization and the torque update methods agree with each other for all parameters points that we have studied.

\subsubsection{Trial ansatz identification}
The trial ansatz, to be discussed in the next sections, were verified by Monte Carlo simulations plus refinement on symmetric $(n \mathbf{a}^P_1,n \mathbf{a}^P_2)$ clusters, which we denote as $P_{n\times n}$, with $n\!=\!6$, $12$, and $18$. A few selected points were also verified in $n\!=\!36$. The energies resulting from these clusters where compared against the trial ansatz energy, and were found to match to machine precision whenever the ansatz was the minimum energy state. 


The armchair $A_{3\times n}$ clusters with $n$ up to 60 were used for Monte Carlo simulations at five  points: $ \psi=$0.65, 0.7, 0.725, 0.75, 0.775. These results were then used  as seeds for the non-linear optimization at the intermediate $\psi$ points. We perform this procedure  for every  considered $n$. 
Using all generated data, the minimum energy in the IP state was determined by consistently  selecting the lowest energy for each $\psi$.
We also conducted checks with the ansatz energy of nearby commensurate phases 
(18C3, $6'$, 16, and zz), which were consistently  higher in energy in the IP region.
The energy $E_{\rm{Min}}$, used in the energy and second-derivative plots, is the minimum energy  at a given $ \psi$ selected from all available data.

\subsection{General aspects of the phase diagram}\label{sec:main_results}
Combining all our numerical results, the resulting phase diagram for $K\!<0\!$ and $\Gamma\!>\!0$ is shown in Fig.~\ref{fig:all_the_energies}\,(a) and contains 5 extended regions, labeled as 18C$_3$, IP, $6'$, 16 and zz. It is determined by the minimum classical energy $E_{\text{Min}}$ containing the combined information from all Monte Carlo runs, as well as ansatz minimization, across 10000 points in $\psi\!\in\![\pi/2,\pi]$, by always selecting the smallest energy state.
In Fig.~\ref{fig:all_the_energies}\,(b) we show the evolution of the classical energy $E_{\text{Min}}$ (black curve), along with the first and second derivatives with respect to $\psi$ (blue and red curves, respectively), which signify the phase boundaries. 
For a selection of representative values of $\psi$, in Fig.~\ref{fig:all_the_energies}(c) we show the static structure factors (SSF) 
\be
S(\mathbf{q}) = \dfrac{1}{N}\sum\limits_{i,j}\mathrm{e}^{i \mathbf{q}\cdot(\mathbf{r}_i-\mathbf{r}_j)}\mathbf{S}_i\!\cdot\!\mathbf{S}_j
\ee
up to the second Brillouin zone (BZ), for the corresponding ground states and their counterparts (called `$\mc{R}$ dual' in the following) resulting from the symmetry operations $\mc{R}_{x(yz)}$ of Eq.~(\ref{eq:Rsymmetries}). The decomposition of the SSF in various polarization channels is shown in  Appendix~\ref{appx:SSF_decomposition}.

Among the five extended regions, the 18C$_3$, $6'$, 16, and zz phases (whose structure will be discussed in more detail below) have been identified previously in the literature~\cite{Chern2020PRR,RayyanPRB2021,Chen2023NJP}, and the special points $\psi\!=\!\pi/2$ and $\pi$ have infinite classical ground states~\cite{BaskaranPRB2008,Rousochatzakis2017PRL}. The wide intermediate region occupies almost 40\% of the phase diagram and comprises a cascade of transitions between states with varying periodicity. Evidence for this is revealed by the series of peaks in the second derivative of the ground state energy, $\partial^2_\psi E_{\text{Min}}$ shown in Fig.~\ref{fig:all_the_energies}\,(b). Similar peaks in the second derivative mark the boundary of the IP region to the 18C$_3$ phase on the left and to the $6'$ phase on the right, as well as the transitions between the $6'$, 16 and zz phases. The peaks for the latter transitions are three orders of magnitude higher that those in the IP. While other global aspects of the various states are discussed in Appendix~\ref{appx:additionla_numb_info}, in the following we analyze the main characteristics of each phase separately.

\subsection{ The 18C$_\mathbf{3}$ phase}
Let us now examine the  phase diagram more closely. In the region $\psi/\pi\!\in\!(0.5 , 0.6]$, we obtain  the threefold-symmetric order with 18 spin sublattices called 18C$_3$ order in Ref.~[\citenum{Chern2020PRR}], 18$^{\eta}$ order in Ref.~[\citenum{RayyanPRB2021}] and triple-meron crystal in  Ref.~[\citenum{Chen2023NJP}]. This state was also noted earlier in Ref.~[\citenum{JanssenPRB2017}] and denoted as a multi-$Q$ state. Its region of stability is much smaller compared to that reported in the literature~\cite{Chern2020PRR,RayyanPRB2021,Chen2023NJP}.  Its magnetic unit cell is the $(3\mathbf{a}^P_1,3\mathbf{a}^P_2)$ cluster,  which we denote as $P_{3\times3}$ in the following. Figure~\ref{fig:three_classical_ansatz_A}\,(a) shows the sublattice decomposition of this state, while   Figs.~\ref{fig:three_classical_ansatz_A}(b) and \ref{fig:three_classical_ansatz_A}(c) show the actual spin directions of this state and its dual counterpart for the representative point $\psi=0.55\pi$. The Cartesian components of the 18 sublattices have the general form
{
\small
\begin{equation}\label{eq:18c3anz}
\begin{array}{lll}
\multicolumn{3}{c}{\mathbf{A} \!=\! -[1,1,1]/\sqrt{3}\,,~~~\overline{\mathbf{A}} \!=\! [1,1,1]/\sqrt{3},} \\
\mathbf{B}_1 \!=\! [x,y,z], & \mathbf{B}_2 \!=\! [z,x,y], & \mathbf{B}_3 \!=\! [y,z,x], \\
\widetilde{\mathbf{B}}_1 \!=\! [y,x,z], & \widetilde{\mathbf{B}}_2 \!=\! [z,y,x], & \widetilde{\mathbf{B}}_3 \!=\! [x,z,y], \\
\mathbf{C}_1 \!=\! [x_{\mathrm{r}},x_{\mathrm{r}},z_{\mathrm{r}}], & \mathbf{C}_2 \!=\! [z_{\mathrm{r}},x_{\mathrm{r}},x_{\mathrm{r}}], & \mathbf{C}_3 \!=\! [x_{\mathrm{r}},z_{\mathrm{r}},x_{\mathrm{r}}], \\
\mathbf{D}_1 \!=\! [x_{\mathrm{g}},x_{\mathrm{g}},z_{\mathrm{g}}], & \mathbf{D}_2 \!=\! [z_{\mathrm{g}},x_{\mathrm{g}},x_{\mathrm{g}}], & \mathbf{D}_3 \!=\! [x_{\mathrm{g}},z_{\mathrm{g}},x_{\mathrm{g}}], \\
\mathbf{E}_1 \!=\! [x_{\mathrm{b}},x_{\mathrm{b}},z_{\mathrm{b}}], & \mathbf{E}_2 \!=\! [z_{\mathrm{b}},x_{\mathrm{b}},x_{\mathrm{b}}], & \mathbf{E}_3 \!=\! [x_{\mathrm{b}},z_{\mathrm{b}},x_{\mathrm{b}}]. \\
\end{array}
\end{equation}
}

and, apart from ${\mathbf{A}}$ and $\overline{\mathbf{A}}$, all other  components vary with $\psi$. The sublattices ${\mathbf{A}}$ and $\overline{\mathbf{A}}$ reside at the center of the magnetic unit cell shown by the shaded hexagons. The remaining sublattices can be grouped into several flavors ($\mathbf{B}$ through $\mathbf{E}$), and  are depicted by different colors in  Fig.~\ref{fig:three_classical_ansatz_A}\,(a). Within each group, there are three sites that are $\mathsf{C}_3$ related, indicated by subscripts 1,2,3, and in this sense the sites swirl around the $\mathbf{A}$ center [see Fig.~\ref{fig:three_classical_ansatz_A}(b)]. The  sites of $\mathbf{B}$ and $\widetilde{\mathbf{B}}$ groups  have three non-zero spin components and are related to each other by $\mathsf{C}_2$ rotations, i.e., $\widetilde{\mathbf{B}}_1\!=\!-\mathsf{C}_{2\text{Z}}\mathbf{B}_1$, so we denote them by the similar letters.  The sites on both magnetic $\mathbf{B}$  and $\widetilde{\mathbf{B}}$ groups, as well as  the ${\mathbf{A}}$ and $\overline{\mathbf{A}}$ groups,  reside  on the same sublattice of the honeycomb lattice. The other  sublattice of the honeycomb lattice hosts spins belonging to $\mathbf{C}$, $\mathbf{D}$, and $\mathbf{E}$ groups [depicted by red, green, blue colors in Fig.~\ref{fig:three_classical_ansatz_A}(a)]. These sites have $[x,x,z]$ character, i.e., they live on the $\mathbf{ac}$ plain. From the ansatz structure, we clearly see that honeycomb sublattice symmetry is broken in the 18C$_3$ state. We also note that 18C3 breaks the inversion symmetry. The inversion symmetry related state of 18C3 state with the center on the A sublattice (dubbed  18C$_3^A$ in the following) will  have its center on the B sublattice (dubbed 18C$_3^B$). 
 
The dual counterpart of the 18C$_3$ state, denoted by $\mc{R}(\text{18C$_3$})$, also involves 18 sites, however, its internal structure is different and breaks the $\mathsf{C}_3$ symmetry [see Fig.~\ref{fig:three_classical_ansatz_A}(c)]. This is due to the nontrivial, site-dependent nature of the $\mc{R}$ transformations.

 \begin{figure}
	\centering
     \begin{overpic}[width=1.0\columnwidth,percent]{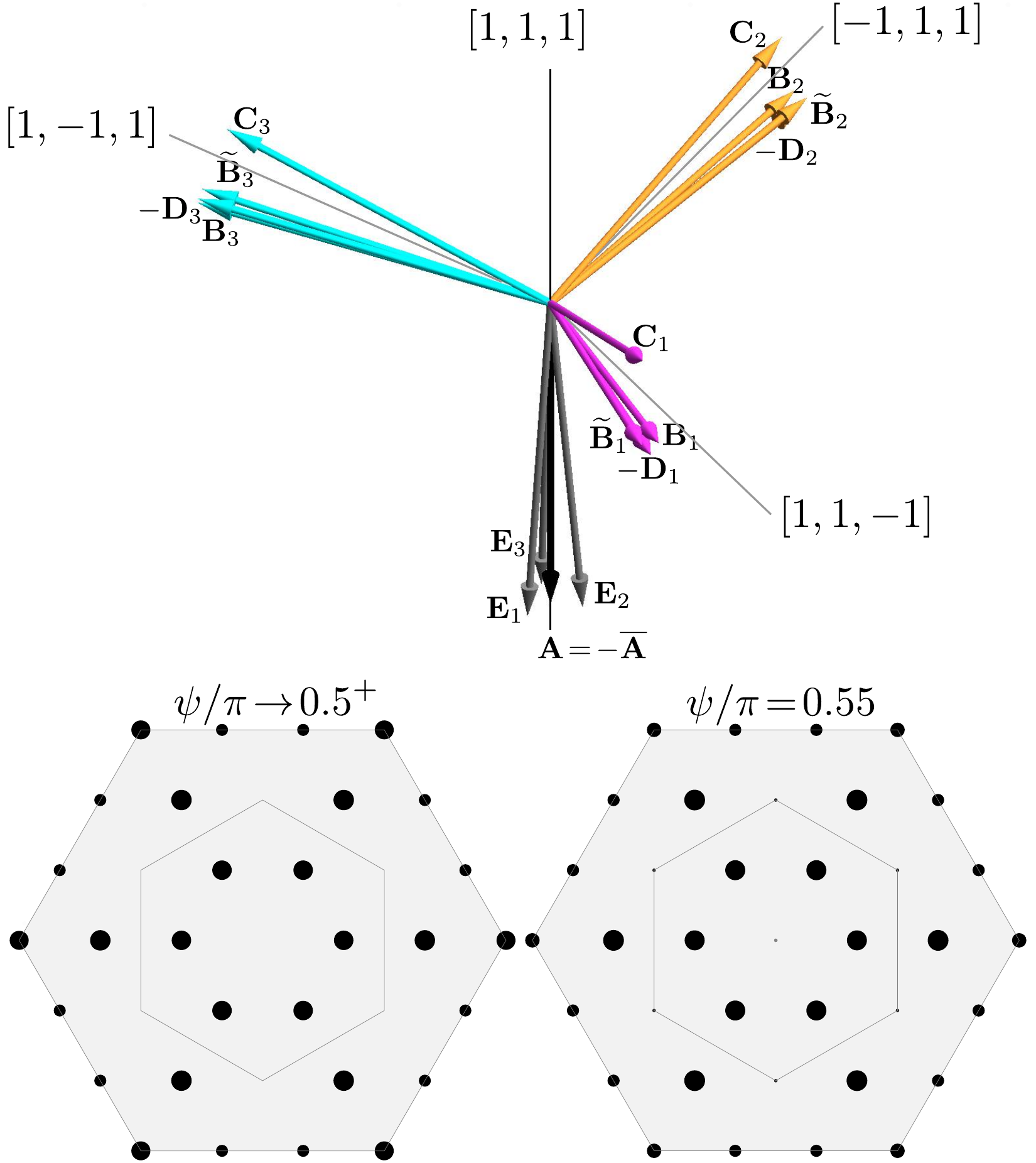}
       \put(0.5,97.5){(a)}
       \put(0.5,39){(b)}
    \end{overpic}  
\caption{
(a) Common-origin plot of the spin directions in the 18C$_3$ state at $\psi\!=\!0.55\pi$ (arrows) and in its `parent' state at $\psi\!\to\!(\pi/2)^{+}$, given by Eq.~(\ref{eq:18c3limit}) $\psi\!\to\!(\pi/2)^{+}$ (thin lines). b) Comparison between the corresponding SSFs of the two states. }\label{fig:18C3_common_and_ssf}
\end{figure}

The SSFs of the 18C$_3$ state and its dual are shown in the first column of Fig.~\ref{fig:all_the_energies}\,(c). The $C_3$ character of the 18C$_3$ state is characterized by a 3$Q$-pattern inside both the first and second BZ, where peaks reside at the points $\pm2{\bf M}/3$ and $\pm4{\bf M}/3$, along with subdominant peaks at the boundary of the second BZ, and with all $C_3$ related momenta having the same weight (as expected by symmetry). By contrast, the dual $\mc{R}(\text{18C$_3$})$ loses this behavior. Its SSF is dominated by the $2{\bf M}/3$ momentum points, with subdominant contribution from the $\mathbf{K}$ points of the BZ and a residual $\mathbf{q=0}$  contribution. Note that if we were to disregard the influence of the residual points, the system would exhibit a state characterized by only 6 magnetic sublattices, as in the  $6'$ state, which will be discussed in Sec. \ref{sub6'}. However, due to the presence of these residual points, we observe a truly intricate 18-magnetic-sublattice order within this region of the phase diagram,  both in the original and the dual spin space. As we discuss in Sec.~\ref{sec:chir} below, due to this complex spin structure,
the 18$C_3$ state 
possesses a non-zero total scalar chirality, the largest among the various states of the phase diagram.

{\it The `parent 18C$_3$' state.} 
We will now show that the 18C$_3$ state can be thought of as a slightly distorted version of its limiting structure at $\psi\!\to\!(\pi/2)^{+}$. In this limit, the 18$C_3$ state becomes a member of the infinitely degenerate ground state manifold of the pure $\Gamma$ point~\cite{Rousochatzakis2017PRL}. Indeed, in this limit, the Cartesian components of Eq.~(\ref{eq:18c3anz}) tend to 
\begin{equation}\label{eq:18c3limit}
\begin{array}{c}
\mathbf{A} = -\overline{\mathbf{A}} = \mathbf{E}_1 = \mathbf{E}_2 =  \mathbf{E}_3 = \!=\! -[1,1,1]/\sqrt{3}, \\
\mathbf{B}_1 = \widetilde{\mathbf{B}}_1 = \mathbf{C}_1 = -\mathbf{D}_1 = [1,1,-1]/\sqrt{3}, \\
\mathbf{B}_2 = \widetilde{\mathbf{B}}_2 = \mathbf{C}_2 = -\mathbf{D}_2 = [-1,1,1]/\sqrt{3}, \\
\mathbf{B}_3 = \widetilde{\mathbf{B}}_3 = \mathbf{C}_3 = -\mathbf{D}_3 = [1,-1,1]/\sqrt{3}. 
\end{array}
\end{equation}
These directions satisfy the general recipe discussed in Ref.~\cite{Rousochatzakis2017PRL} and the state is therefore one of the ground states.  Now, as shown in Fig.~\ref{fig:18C3_common_and_ssf}(a), the 18C$_3$ state at a representative point inside the phase ($\psi\!=\!0.55\pi$) is quite similar to that of Eq.~(\ref{eq:18c3limit}). This is further reflected in the almost identical SSFs shown in Fig.~\ref{fig:18C3_common_and_ssf}(b). Therefore, the characteristic 3$Q$ SSF profile of the 18C$_3$ state originates in the special, four-sublattice structure of the ``parent 18C$_3$'' state, with spins pointing along the different [111] axes. The weak deviations of the spins away from these four primary directions of the parent state, which are caused by a negative $K$, amount to weak signals at ${\bf q}\!=\!0$ (evidencing a small nonzero magnetization) and the two corners of the 1\textsuperscript{st} BZ (${\bf K}$-points). 

\subsection{The $\mathbf{6'}$ phase}\label{sub6'}
The $6'$ state is stable on the right side of the IP phase, in the region $\psi/\pi\!\in\! [0.793, 0.909]$. These boundaries  align with those reported in the literature [\citenum{RayyanPRB2021,Chen2023NJP}], although they appear slightly shifted on the left side. The magnetic unit cell comprises six sublattices with only three different spin directions, all in the same plane. There are three versions of the $6'$ state, related to each other by threefold rotations.  Figure~\ref{fig:three_classical_ansatz_A}\,(d) shows the sublattice decomposition of one of them, with spins  living on the $\mathbf{ac}$ plane, and sublattices 
\be\label{eq:Kanz}
\begin{array}{lll}
\mathbf{A}\!=\![x_{\mathrm{r}},x_{\mathrm{r}},z_{\mathrm{r}}], &
\mathbf{B}\!=\![x_{\mathrm{g}},x_{\mathrm{g}},z_{\mathrm{g}}], &
\mathbf{C}\!=\![x_{\mathrm{b}},x_{\mathrm{b}},z_{\mathrm{b}}]. 
\end{array}
\ee
The spatial profile of these sublattices shows a counter-rotating modulation, $\mathbf{ABC}\cdots$ vs $\mathbf{ACB}\cdots$, of the two honeycomb sublattices, similar to the so-called ``$K$-state'' discussed in the 3D material $\beta$-Li$_2$IrO$_3$~\cite{Ducatman2018}.

The actual magnetization profiles of the $6'$ state and its dual are shown in Figs.~\ref{fig:three_classical_ansatz_A}(e) and \ref{fig:three_classical_ansatz_A}(f) for the representative point $\psi\!=\!0.85\pi$. The corresponding SSFs are shown in the third column of Fig.~\ref{fig:all_the_energies}\,(c). The magnetic unit cell of the dual state $\mc{R}(6')$ has 18 sites, three times larger than that of the $6'$  state. The SSF of the $6'$ state is dominated by four peaks at $\pm2{\bf M}/3$ and $\pm4{\bf M}/3$, and has residual contributions at the boundary of the second BZ and at $\mathbf{q=0}$. The latter shows that this state features a nonzero total moment.  The SSF of the dual state is similar to the SSF of the $\mc{R}(\text{18C}_3)$ state if we ignore subdominant peaks, 
suggesting a deeper connection between the two phases. Indeed, we have found that the characteristic ABC-structure in the $6'$ configuration, with spins in the ${\bf ab}$-plane, can be reproduced (modulo a spin-length normalization for the vectors ${\bf A}$, ${\bf B}$ and ${\bf C}$) by the following linear combination of different replica of the 18C$_3$ state:
\be\label{eq:18C3vs6p}
6'\!\propto\! \Big(1+\mc{T}_{{\bf a}^A_2}+\mc{T}_{2{\bf a}^A_2}\Big)\cdot
\Big(
\text{18C}_3^A\!+\!\text{18C}_3^B\Big)\,,
\ee
where the translation operator $\mc{T}_{\bf a}$ moves the spins by lattice vector distances $\mc{T}_{\bf a}\cdot{\bf S}_{A(B),{\bf R}}\!=\!{\bf S}_{A(B),{\bf R}+{\bf a}}$, and $\mathbf{a}_2^A\!=\!\mathbf{a}_1^P\!+\!\mathbf{a}_2^P$ one of the armchair lattice vectors, (see Fig.~\ref{fig:basic_defenitions}). This construction,   elaborated in Appendix~\ref{appx:6pfrom18C3}, shows that contrary to  the  18C3 state, $6'$ state  restores the inversion symmetry.

\subsection{The 16 phase}
\begin{figure}[t]
	\centering
    \begin{overpic}[width=1.0\columnwidth,percent]{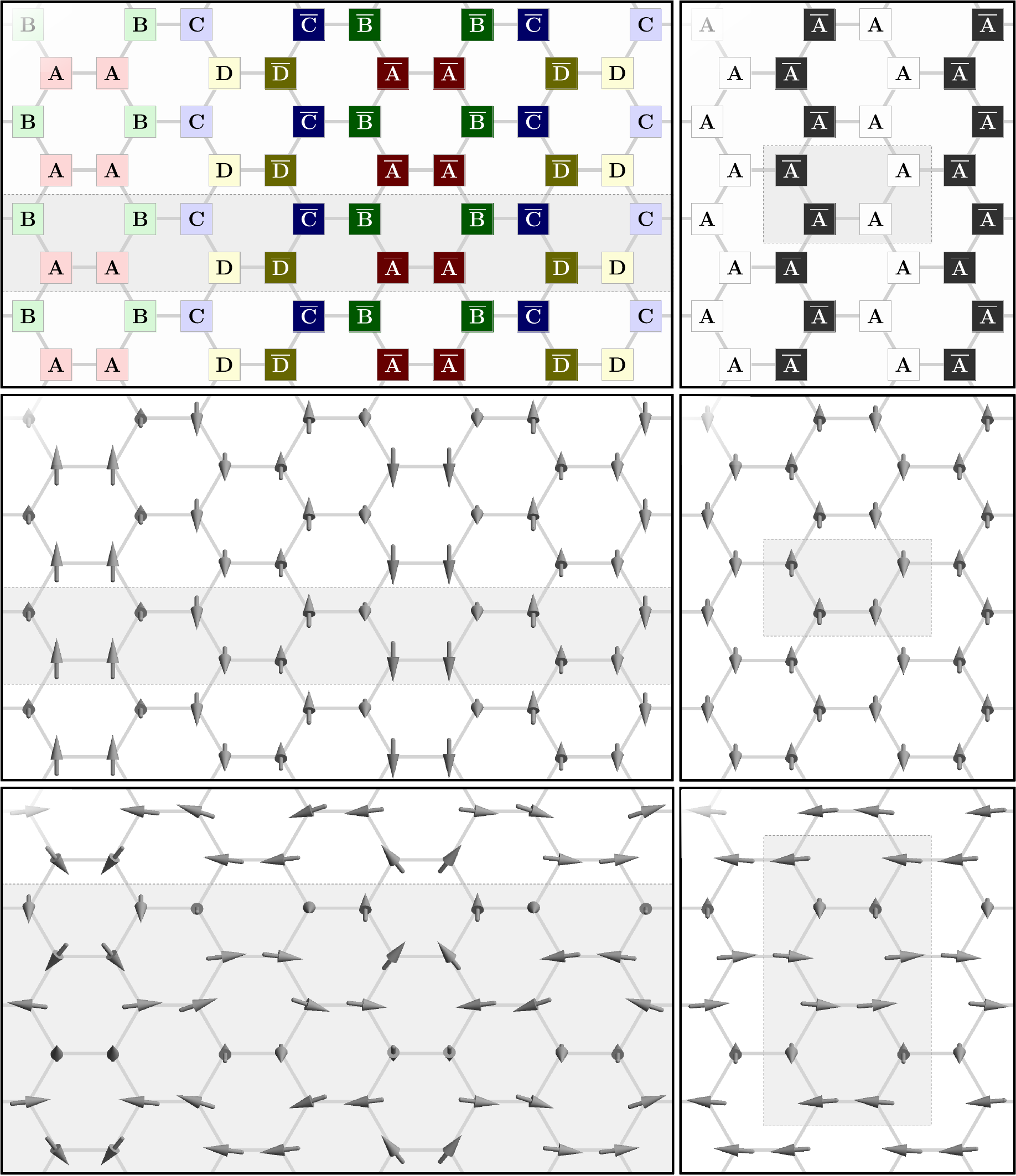}
       \put(0.5,30){(c)}
       \put(0.5,63.5){(b)}
       \put(0.5,97){(a)}
       \put(58.5,30){(f)}
       \put(58.5,63.5){(e)}
       \put(58.5,97){(d)}
    \end{overpic}  
  \caption{The ansatz 16 state (a)-(c) and zz state (d)-(f). For each of the ansatz we show: a color coded sketch of the ansatz (a) and (d), that corresponds to Eqs.~(\ref{eq:16anz}) and (\ref{eq:ZZanz}), the real space configuration of the spin state (b) and (e), and the $\mathbf{R}$ dual state (c) and (f). Magnetic unit cells are indicated by a faint gray area. The $\mc{R}$ symmetries reveal dual state, with the that 16 state being a 48 site order, and the zz state degenerate with a 12 site order. }\label{fig:three_classical_ansatz_B}
\end{figure}

The phase labeled as ``16'' in Fig.~\ref{fig:all_the_energies}\,(a) is stable in the narrow region  $\psi/\pi\in [0.909, 0.926]$ and has a magnetic unit cell with 16 sites, which in turn belong to 8 different spin directions. 
As in the $6'$ state, the spins are again coplanar and there are three versions of the 16 order, related to each other by threefold rotations. Figure~\ref{fig:three_classical_ansatz_B}\,(a) shows the sublattice decomposition of one of them, with spins living on the ${\bf ac}$ plane, and sublattices given by
\be\label{eq:16anz}
\begin{array}{ll}
\mathbf{A}\!=\![x_{\mathrm{r}},x_{\mathrm{r}},z_{\mathrm{r}}], & \overline{\mathbf{A}}\!=\!-{\bf A}, \\
\mathbf{B}\!=\![x_{\mathrm{g}},x_{\mathrm{g}},z_{\mathrm{g}}], & \overline{\mathbf{B}}\!=\!-{\bf B}, \\
\mathbf{C}\!=\![x_{\mathrm{b}},x_{\mathrm{b}},z_{\mathrm{b}}], & \overline{\mathbf{C}}\!=\!-{\bf C}, \\
\mathbf{D}\!=\![x_{\mathrm{y}},x_{\mathrm{y}},z_{\mathrm{y}}], & \overline{\mathbf{D}}\!=\!-{\bf D},
\end{array}
\ee
The magnetization profiles of this state and its dual counterpart $\mc{R}(16)$  (whose magnetic unit cell contains 48 sites) are shown in Figs.~\ref{fig:three_classical_ansatz_B}\,(b-c), for the representative point $\psi\!=\!0.917\pi$. Their corresponding SSFs are shown in the fourth column of Fig.~\ref{fig:all_the_energies}\,(c). The SSF of the 16 state shows dominant peaks at $\pm 3{\bf M}/4$ and $\pm 5{\bf M}/4$, and a bit smaller peaks at $\pm {\bf M}/4$ and $\pm 7{\bf M}/4$.
 Comparing to the neighboring phases, it appears as if the $6'$ state with dominant  peaks at $\pm 2{\bf M}/3$ is attempting to move towards
the zz state with peaks at
the ${\bf M}$ points, however  it happens through the intermediate small window of the 16 state with  dominant $\pm 3{\bf M}/4$ peaks [with visual representation in Fig.\ref{fig:all_the_energies} (c)].

\begin{figure*}[ht!]
	\centering
     \begin{overpic}[width=1.0\textwidth,percent]{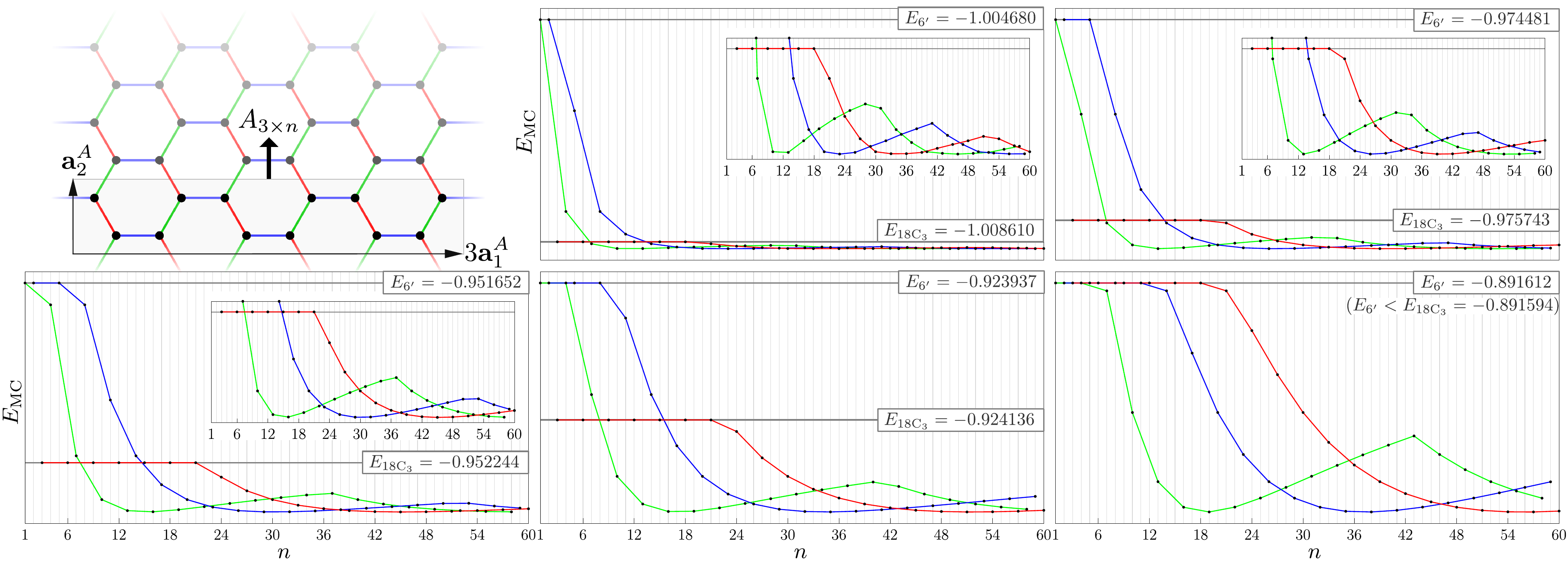}
       \put(2,33.5){(a)}
       \put(35,33.5){(b)}
       \put(67.8,33.5){(c)}
       \put(2,16.5){(d)}
       \put(35,16.5){(e)}
       \put(67.8,16.5){(f)}
    \end{overpic}  
\caption{ Energy  per site plots as a function of armchair cluster length $n$, as found from Monte Carlo simulations. The simulated clusters are $A_{3\times n}$ clusters, as illustrated in (a). Total energies obtained by the Monte Carlo simulations for $\psi/\pi=$ $0.65$ (b), $0.7$ (c), $0.725$ (d), $0.75$ (e), and $0.775$ (f) show deviations from the energy of the 18C$_3$ and $6'$ states indicated by horizontal gray lines. Note that in panel (f) the energy of the 18C$_3$ has become higher than the $6'$. We see a universal behavior: the results partition into three curves $3|n$ (red), $3|(n-1)$ (green), and $3|(n-2)$ (blue). The oscillatory behavior of the curves as a function of $n$ is interpreted as a signature of incommensuration.}\label{fig:mc_results}
\end{figure*}
\subsection{The zz phase}
The zigzag (zz) order is stable for $\psi/\pi\!\in\![0.926,1)$. The appearance of this order may be attributed to the nearby hidden symmetry point $\mathcal{T}_1\mathcal{T}_4$ in the enlarged $J$-$K$-$\Gamma$-$\Gamma'$ model \cite{ChaloupkaPRB2015}, and has been seen to stabilize a sizable zz region even for very small $\Gamma'\!<\!0$ interactions compared to the $K$ and $\Gamma$ parameters~\cite{rau2014trigonal}.

The magnetic unit cell comprises four sites which belong to two different spin directions, aligned opposite to each other (i.e., the state is collinear). 
As in the phases $6'$ and 16, there are three different versions of the zz phase, related to each other by threefold rotations. Figure~\ref{fig:three_classical_ansatz_B}(d) shows one of them, with spins living on the $\mathbf{ac}$ plane and sublattices given by
\be\label{eq:ZZanz}
\mathbf{A}\!=\![x, x, z]\,,~~ 
\overline{\mathbf{A}}\!=\!-{\bf A}
\ee
The magnetization profiles of the zz state and its dual counterpart $\mc{R}(\text{zz})$ (whose magnetic unit cell contains 12 sites) are shown in Figs.~\ref{fig:three_classical_ansatz_B}(e) and \ref{fig:three_classical_ansatz_B}(f) for the representative point $\psi\!=\!0.95\pi$. The corresponding SSFs are shown in the last column of Fig.~\ref{fig:all_the_energies}\,(c), with the SSF of the zz state having the standard Bragg peaks at one of the three ${\bf M}$ points of the BZ.

Finally, we note that in the limit of $\psi\!\to\!\pi^{-}$, the zz state becomes a member of the infinitely degenerate ground state manifold of the pure Kitaev point~\cite{BaskaranPRB2008,Chandra2010,RousochatzakisNC2018}. Indeed, in this limit, the Cartesian components of Eq.~(\ref{eq:ZZanz}) tend to 
\be
\overline{\mathbf{A}}\!=\!-{\bf A}=[1,1,0]/\sqrt{2}.
\ee
These directions satisfy the general recipe discussed in Ref.~\cite{RousochatzakisNC2018} and the state is therefore one of the ground states.

\subsection{The intermediate phase}\label{sec:all_about_the_IS_state}

\subsubsection{ Energetics of finite-size clusters}
The IP occupies a considerable region of the phase diagram $\psi/\pi\!\in\! [0.6, 0.793]$. Previous studies ~\cite{Chern2020PRR,Rousochatzakis2020KITP,RayyanPRB2021,Liu2021PRR,JhengArxiv2023,Chen2023NJP} suggested that this region  consists of phases with large magnetic unit cells or long-wavelength, incommensurate modulations.  While such orders with large (or infinite)  unit cells are challenging to study in finite-size simulations, the associated multi-peaked SSFs offer distinctive fingerprints for their experimental detection. This is why here we perform a detailed  analysis of this phase  by employing numerical simulations on the most suitable elongated armchair clusters.
 
The armchair building block is a conventional rectangle unit cell, spanned by $(\mathbf{a}_1^A, \mathbf{a}_2^A)$ lattice vectors, parallel to $\mathbf{a}$ and $\mathbf{b}$ directions (see Fig.~\ref{fig:basic_defenitions}). The elongated clusters that host the IP state  have  a structure of  $(3\mathbf{a}_1^A, n\mathbf{a}_2^A)\!\equiv\!A_{3\times n}$ [see Fig.~\ref{fig:mc_results}(a)]. By construction the zz state fits on any armchair. The 16-site state needs at least the $A_{4\times 1}$ cluster and is not compatible with any $A_{3\times n}$ clusters. The $6'$ state commensurates with the $A_{3\times 1}$ cluster, exactly two copies of it. The 18C$_3$ state lives on the $P_{3\times3}$ cluster, which commensurates with $A_{3\times n}$ clusters only if  $n$ is divisible by 3, or in math notation $3|n$. The $\mc{R}$ operations need at least the hexagonal $(\mathbf{a}_1^H, \mathbf{a}_2^H)$ unit cell, which commensurates with the primitive $P_{3\times3}$, and consequently commensurates with $A_{3\times n}$ when $3|n$. Further details on commensurability of the $P_{m\times m}$ clusters with $A_{3\times n}$ clusters can be found in Appendix~\ref{appx:commensuration_of_clu}.

We performed Monte Carlo simulations on the $A_{3\times n}$ clusters with $1\!\leq\!n\!\leq\!60$ and for $\psi/\pi\!=$0.65, 0.7, 0.725, 0.75, 0.775. Results for the energy per site as a function of $n$ are shown in  Fig.~\ref{fig:mc_results}(b) and \ref{fig:mc_results}(f).  For comparison, we also indicate, by labeled horizontal gray lines, the (variational) energies per site of the 18C$_3$ and $6'$ states for each given parameter point. Some general trends, that are particularly well seen in the insets of Fig.~\ref{fig:mc_results}(b) and \ref{fig:mc_results}(f),  emerge from our simulations. The energy curves naturally partition into three families: $3|n$ (red curve), $3|(n-1)$ (green curve), and $3|(n-2)$ (blue curve). The $3|n$ family fits the 18C$_3$ state and stabilizes it for a large window of $n$,  until eventually the energy gets lower than that of 18C$_3$, suggesting states with even larger unit cells, or, even incommensurate states. Similarly, the energies of the $3|(n-1)$ and $3|(n-2)$ families, after a short window of $n$, get below the energy of the $6'$ state and eventually below that of the 18C$_3$ for $n$ sightly larger than 6. The first curve to drop, for any $\psi$, is $3|(n-1)$, then $3|(n-2)$, and finally $3|n$. For each curve, the energy appears to oscillate. Moreover, the minimum switches between families. These results are suggestive of an incommensurate state being accommodated on finite-size clusters.

\subsubsection{Static structure factors}

We computed the SSFs for several $\psi$ points in the intermediate phase,  and they  appear to be similar in all cases. For all points, magnetic orders obtained by the Monte Carlo simulations on the finite  $A_{3\times n}$ clusters appear to utilize the entire cluster length with no sub-periods. To illustrate their structure,  we plot in the second column of Fig.~\ref{fig:all_the_energies}\,(c) the SSF of the IP and its $\mc{R}$-dual (diagonal components of SSH  in cubic coordinates are shown Fig.~\ref{fig:SSF_decomposition}), as  obtained for $\psi/\pi\!=\!0.7$ on the cluster with $n\!=\!39$ belonging to the $3|n$ family.  The SSFs  exhibit a sequence of well-defined Bragg peaks, accompanied by additional trailing points  in the vicinity of these peaks. Notably, the positions and relative intensities of the sharp Bragg peaks resemble closely those of the neighboring 18C$_3$ and $\mc{R}(6')$ states, while all trailing points exhibit vertical displacement. These observations indicate a one-dimensional incommensurate modulation (here in the vertical direction) of the neighboring commensurate states.

\subsubsection{ Local 18C$_3$ character}
\begin{figure}
    \centering
    \begin{overpic}[width=1.0\columnwidth,percent]{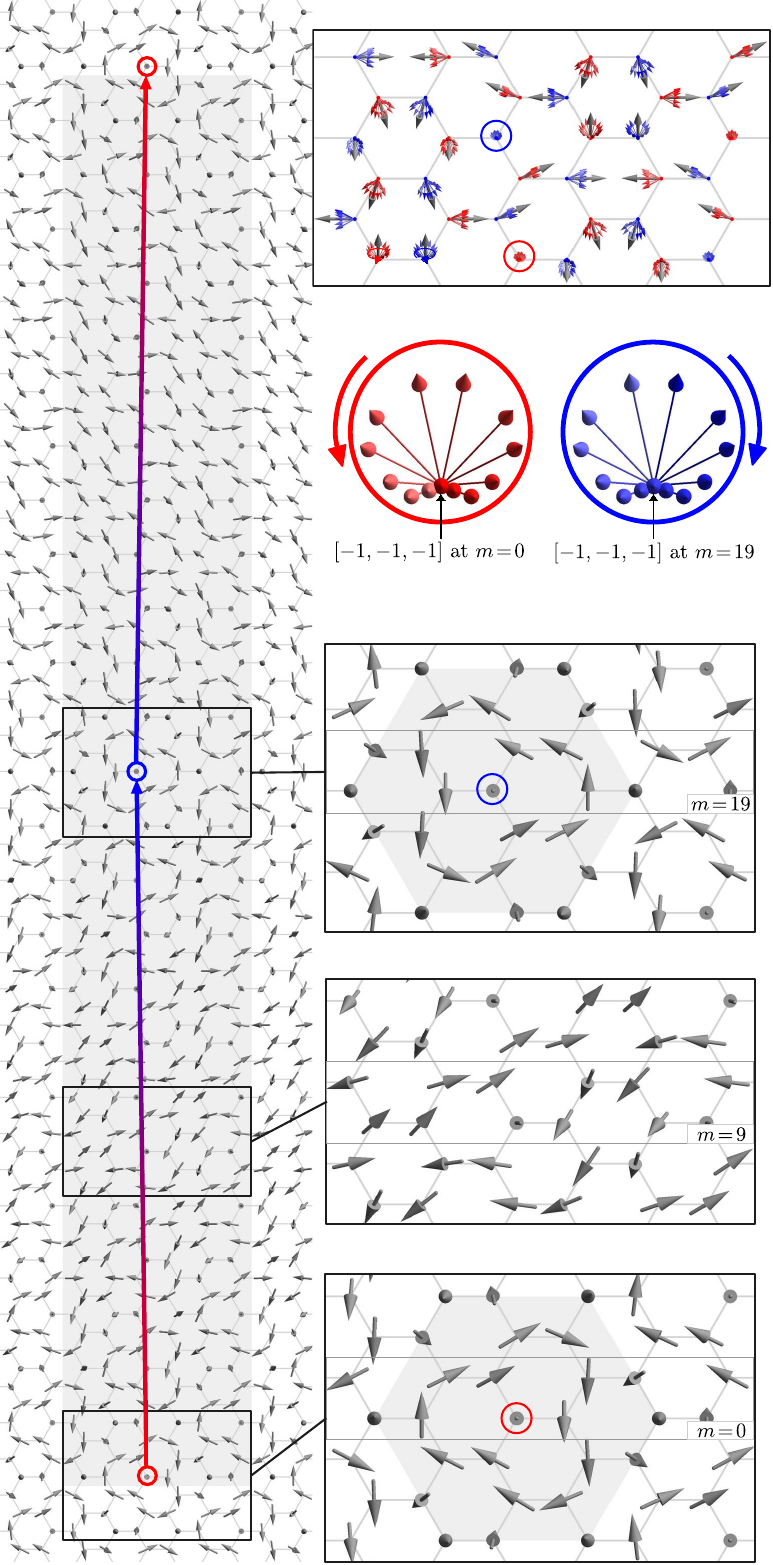}
        \put(0.5,98.5){(a)}
        \put(20,98.5){(b)}
        \put(20,78){(c)}
        \put(20,60){(d)}
    \end{overpic}  
\caption{(a) The IP state, for $\psi/\pi\!=\!0.7$ hosted on the $A_{3\times39}$ cluster ($3|n$ familiarly). (b) Common origin plot of spins, resolved on the $A_{3\times3}$ subcluster ($13A_{3\times3}=A_{3\times39}$), where sublattice $A(B)$ is colored in red(blue). Gray arrows indicate the average direction of the ``near-cone'' formed by the precessing spins [note that these gray arrows follow the directions of the $\mc{R}(6')$ state, compare to Fig.~\ref{fig:three_classical_ansatz_A} (f)]. The red ($A$ sublattice) and blue ($B$ sublattice) near-cones counter rotate to each other, shown more clearly in (c) for two example common origin points. (d) Zoom in windows of the elongated state, highlighting the switch of 18C$_3$ like local windows, from $A$ (red circle) to  $B$ (blue circle) sublattice, approximately half way through the cluster.}\label{fig:chir2}
\end{figure}

The corresponding magnetic state needs a  large  $3|n$ unit  cell, so the real space visualization is rather difficult.   Fig.~\ref{fig:chir2}(a) shows a zoomed out overview of the intermediate state for $\psi/\pi\!=\!0.7$, along with some zoomed in windows in Fig.~\ref{fig:chir2}(c). We observe a distinctive modulation of the magnetic order which can be best described by the switching between windows of local 18C$_3$-like character.  Going sliver by sliver, at the bottom of the cluster we have a fragment of the 18C$_3$  state centered  on the  $A$ sublattice of the honeycomb lattice (referred to as 18$C_3^A$), then by moving vertically up along the cluster, all sublattices of 18C$_3$  state  slowly undergo a gradual precession such that at about the middle  of the cluster we see that the center of   the  18C$_3$  switches to the honeycomb $B$ sublattice (referred to as 18$C_3^B$). These two regions are related by inversion symmetry i.e., 18$C_3^B$=$I($18$C_3^A)$. This multi-sublattice magnetic order mimics the intermediate state and  lowers  the energy of the  18C$_3$ like centers via restoring the sublattice  and inversion symmetry of the honeycomb lattice. 

The switching of sublattices in the 18C$_3$-like centers is achieved through a non-trivial precession of the spins. To simplify the representation, we can collapse all the sites of the $A_{3\times 39}$ lattice down to a single $A_{3\times 3}$ window.  Then each  site serves as a common origin for plotting the spins.  Those corresponding to the centers of the 18C$_3$ order shown in Fig.~\ref{fig:chir2}\,(b) are  depicted in Fig.~\ref{fig:chir2}\,(c), with spins colored in red for the $A$ sublattice and blue for the $B$ sublattice. The spin configurations almost form cones but exhibit some deviation, and we shall refer to them as `near-cones'.

\begin{figure}
	\centering
\includegraphics[width=1.0\columnwidth]{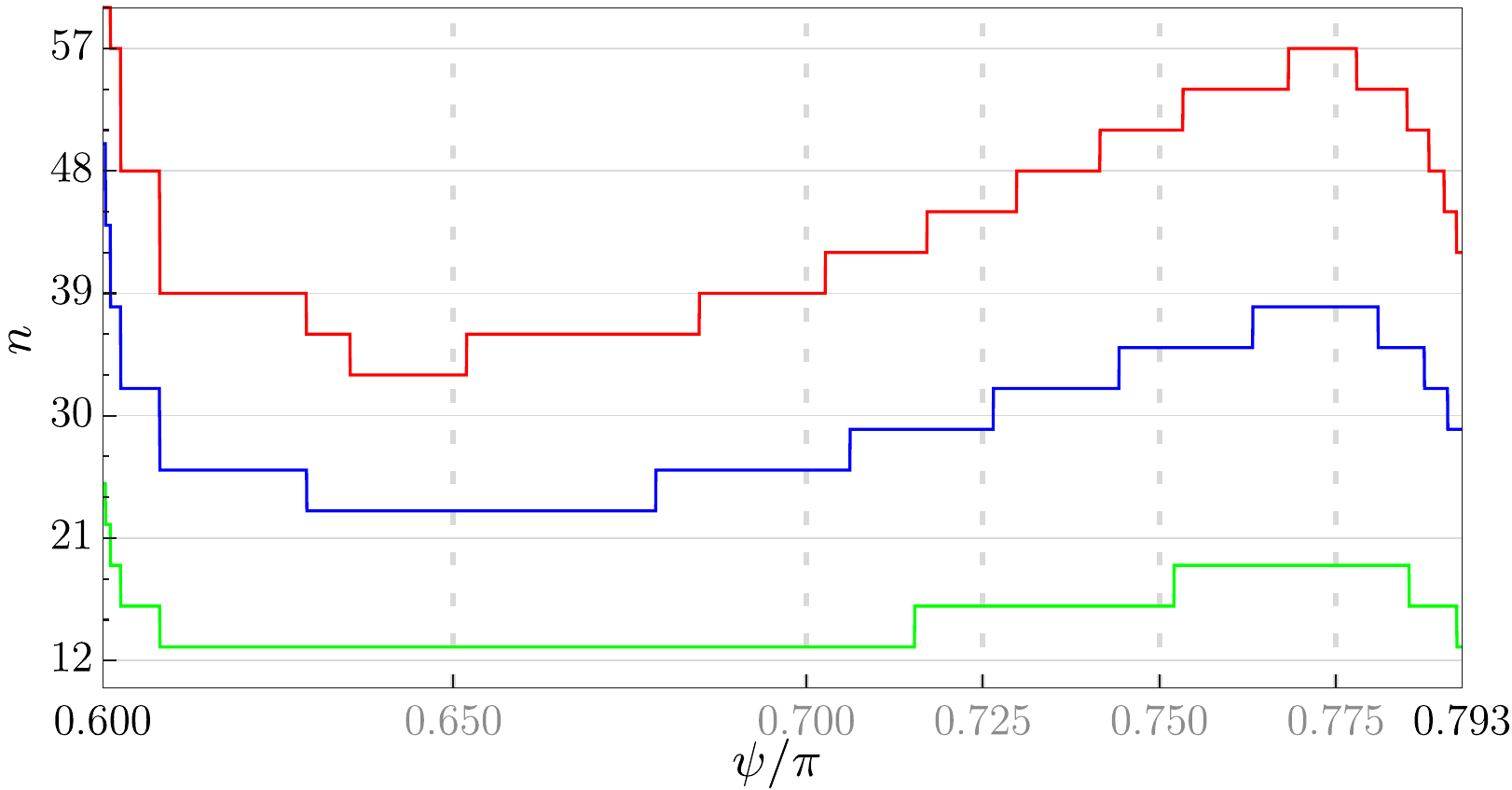}
   \caption{Tracking the cluster's length $n$ of the  minimum energy state  within the three families: $3|n$ (red), $3|(n-1)$ (green), and $3|(n-2)$ (blue). Dashed gray lines indicate the $\psi$ points for which $A_{3\times n}$ clusters were simulated in Monte Carlo for $n$ up to 60. The rest of the points were calculated by Lagrange multiplier minimization using Monte Carlo result as an input, also supplemented with torque updates.  }\label{fig:torq_and_cers_results}
\end{figure}

Let us first focus on the centers within the cluster that underwent a mid-cluster switch. Specifically, one center originated from the $A$ sublattice, while the other originated from the $B$ sublattice, in  Fig.~\ref{fig:chir2}\,(a) marked with red and blue circles, respectively. Interestingly, the near-cone associated with the $A$ sublattice and the one linked to the $B$ sublattice align perfectly in their orientations. However, a noteworthy observation is that these two near-cones exhibit counter-rotational precession relative to each other.
 
To gain further insight, we can calculate the average directions of the near-cones, which reveal the axis of precession. These average directions are indicated by the gray arrows in  Fig.~\ref{fig:chir2}\,(b). Upon closer examination, it becomes evident that this counter-rotational behavior is not limited to a specific region; instead, it persists throughout the system. Specifically, the near-cones associated with the $A$ sublattice exhibit a counter-clockwise traversal around their respective axes of precession, while those related to the $B$ sublattice undergo clockwise precession. Notably, the directions of precession within the system  aligns with the directions  of the spins observed in the $\mc{R}(6')$ state. This observation implies that the dominant peaks in  the SSF of the intermediate state stem from the magnetic structure described by the averaged directions of the spins,  and  the trailing points that extend away from these dominant peaks are due  to the gradual precession of spins occurring on the near-cones.

Next, using the cluster length as an indication of the incommensurate repetition period and in order to examine the dependence of optimal cluster length on the  model parameters, we use the Monte Carlo numerical states as the input  for a  more refined Lagrange multiplier minimization, in combination  with torque updates.  Our results are summarized in Fig.~\ref{fig:torq_and_cers_results} where we track the energy minimum in every family and plot the optimal value of  $n$  as a function of $\psi$. All three families  in  Fig.~\ref{fig:torq_and_cers_results} (b) show a similar behavior, indicative that all families are attempting to fit the same type of state as the ground state. Within each family, the cluster length  $n$  for which the energy is minimized  generally increases as $\psi$ increases. 
  Note that the steps corresponding to each family of armchair clusters shown by different colors in Fig.~\ref{fig:torq_and_cers_results} do not exactly match the positions of the peaks in $\partial^2_\psi E_{\mathrm{min}}$ seen in  Fig.~\ref{fig:all_the_energies}, because each such peak corresponds to the minimum energy across all cluster families.

Along with the smooth behavior of cluster length in the majority of the IP, Fig.~\ref{fig:torq_and_cers_results}\,(b) shows an increase in the optimal cluster's size towards the left boundary to  the 18C$_3$ state and a decrease towards the right boundary to the $6'$ state. The sharp increase in cluster length as we approach the left boundary of the IP can be attributed to the 18C$_3^A$ centres pushing away from the 18C$_3^B$ centers in order to leave a single clean 18C$_3$ state. The transition between the 18C$_3$ and the IP can therefore be thought of as a continuous commensurate-incommensurate (C-I) transition, whereby the spin structure `rotates' between different `domains' (here 18C$_3^A$ and 18C$_3^B$) of a parent commensurate state (here the 18C$_3$ state), with the distance between the domains going to infinity at the transition~\cite{DesGennes1975,Bak1982,McMillan1976,Schaub1985}. 

By contrast, the decrease of the cluster length $n$ at around $\psi/\pi\!=\!0.793$  suggests that the transition to the $6'$ state is not a continuous C-I transition.  In particular, the deeper connection between the $6'$ and the 18C$_3$ states, as follows from  Eq.~(\ref{eq:18C3vs6p}), suggests that the sharp decrease of cluster length towards the $6'$ boundary can be interpreted as merging of several 18$C_3$ windows to form the  $6'$ state. 

\subsection{Evolution of scalar spin chirality profiles}\label{sec:chir}

 \begin{figure}[t]
	\centering
     \begin{overpic}[width=1.0\columnwidth,percent]{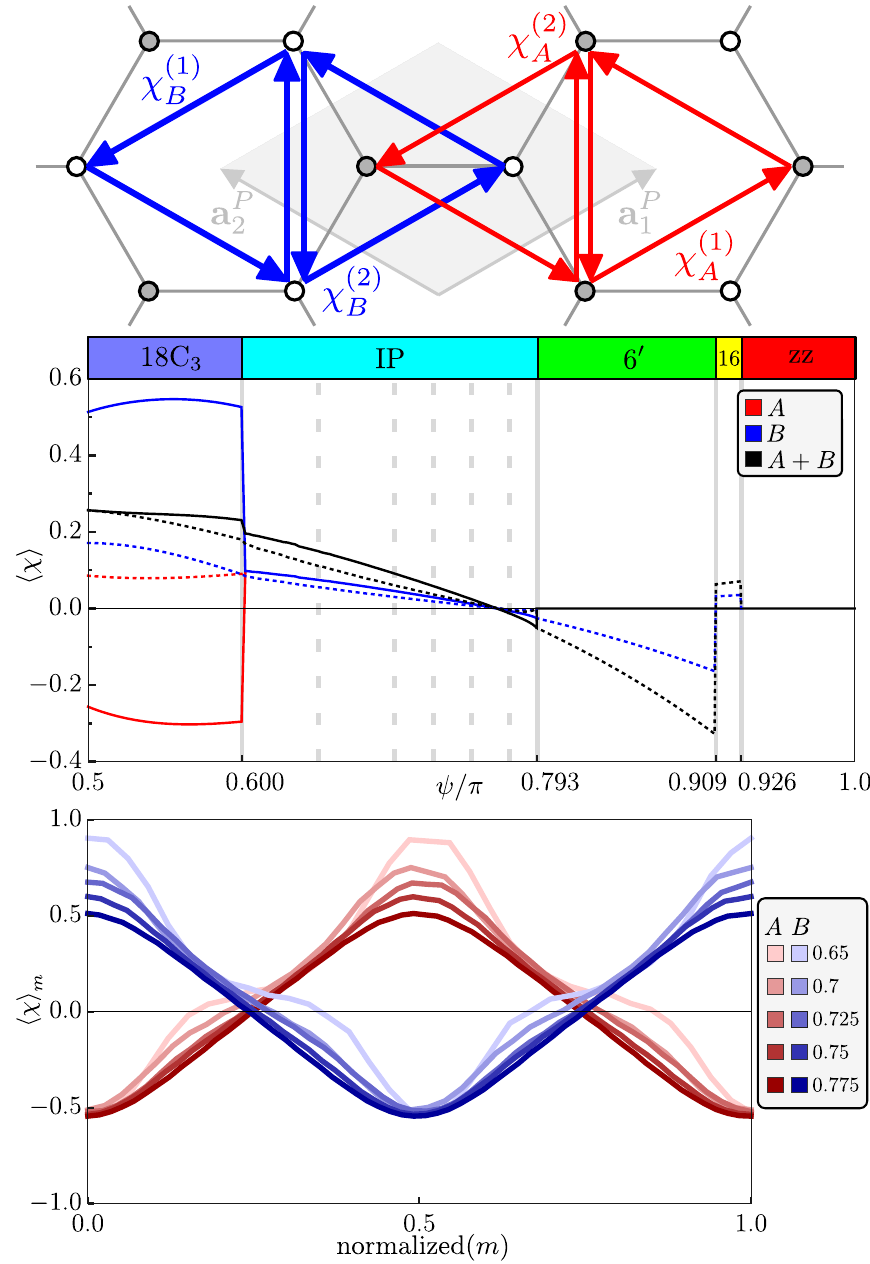}
        \put(0.5,98.5){(a)}
        \put(0.5,71){(b)}
        \put(0.5,33.5){(c)}
    \end{overpic}  
   \caption{(a) Definition of chiralities $ \chi^{(1)/(2)}_{A/B,\mathbf{R}}$ for sublattice $A$ (red) and $B$ (blue).(b) Plot of $\left\langle \chi \right\rangle$ for the original (solid lines) and $\mc{R}$ dual (doted lines) states, resolved on sublattices (red and blue) as well as the total (black). Dashed gray lines indicate the $\psi$ points for which $A_{3\times n}$ clusters where simulated in Monte Carlo, while other points in the IP are evaluated from Lagrange multiplier minimization. (c) Coarse-grained chirality $\left\langle \chi_{A(B)} \right\rangle_{m}$ 
   with $0\!\leq\! m \!\leq\! n$. Each $\psi/\pi\!=\!0.65, 0.7, 0.725, 0.75, 0.775$ realizes the ground state on a cluster of different length,$n_{\psi}=33, 39, 45, 51, 57$ respectively, and the  coarse-grained chirality is plotted as a function of the $\text{normalized}(m)=m/n_{\psi}$.}\label{fig:chir1}
\end{figure}

The visual representation of the intermediate state in Fig.~\ref{fig:chir2} provides some hints that the nature of the IP can be understood as an incommensuration between 18C$_3^{A(B)}$ like centers. Aiming to quantify this behavior, in this section we  will discuss how inequivalence of the scalar chiralities for $A$ and $B$ triangular sublattices  can be used to describe  this incommensuration.

Scalar spin chirality is defined for each triangular plaquette formed by second-neighbor sites $i$, $j$, and $k$ as $\mathbf{S}_i\!\cdot\left(\mathbf{S}_j\!\times\!\mathbf{S}_k\right)$. If the moments on the plaquette are non-coplanar, the
resulting scalar spin chirality is non-zero, 
$\langle\mathbf{S}_i\!\cdot\left(\mathbf{S}_j\!\times\!\mathbf{S}_k\right)\rangle\neq 0$ and breaks the time-reversal and the inversion symmetries.
On the honeycomb lattice, with two sublattices $A$ and $B$ in the primitive unit cell located at position $\mathbf{R}$, four chiralities can be defined:
\be
\renewcommand{\arraystretch}{1.3}
\begin{array}{l}
         \chi^{(1)}_{A,\mathbf{R}} \!\equiv\! \mathbf{S}_{A,\mathbf{R}} \!\cdot\! \left(\mathbf{S}_{A,\mathbf{R}+\mathbf{a}^P_1} \!\times\! \mathbf{S}_{A,\mathbf{R}+\mathbf{a}^P_1+\mathbf{a}^P_2}\right), \\
         \chi^{(2)}_{A,\mathbf{R}} \!\equiv\! \mathbf{S}_{A,\mathbf{R}} \!\cdot\! \left(\mathbf{S}_{A,\mathbf{R}+\mathbf{a}^P_1+\mathbf{a}^P_2} \!\times\! \mathbf{S}_{A,\mathbf{R}+\mathbf{a}^P_2}\right), \\
         \chi^{(1)}_{B,\mathbf{R}} \!\equiv\! \mathbf{S}_{B,\mathbf{R}} \!\cdot\! \left(\mathbf{S}_{B,\mathbf{R}+\mathbf{a}^P_1+\mathbf{a}^P_2} \!\times\! \mathbf{S}_{B,\mathbf{R}+\mathbf{a}^P_2}\right), \\
         \chi^{(2)}_{B,\mathbf{R}} \!\equiv\! \mathbf{S}_{B,\mathbf{R}} \!\cdot\! \left(\mathbf{S}_{B,\mathbf{R}+\mathbf{a}^P_1} \!\times\! \mathbf{S}_{B,\mathbf{R}+\mathbf{a}^P_1+\mathbf{a}^P_2}\right), 
\end{array}
\ee
where superscript $(1)$ indicates triangles of $A(B)$ sublattice sites centered over an empty hexagon, and superscript $(2)$ triangles of $A(B)$ sublattice sites centered over a $B(A)$ sublattice site. The site ordering is chosen such that the sites are traversed anti-clockwise. In Fig.~\ref{fig:chir1}\,(a) we show these quantities as red triangles for sublattice $A$, and blue triangles for sublattice $B$. Together, $\chi^{(1)}_{A(B),\mathbf{R}}$ and $\chi^{(2)}_{A(B),\mathbf{R}}$ tessellate the entire sublattice $A(B)$. The average chirality on each sublattice is
\be
\left\langle \chi_{A(B)} \right\rangle \equiv \dfrac{1}{2 N_c}\displaystyle\sum\limits_{\mathbf{R}}\left[  \chi^{(1)}_{A(B),\mathbf{R}} + \chi^{(2)}_{A(B),\mathbf{R}} \right],
\label{eqn:avg_chirality}
\ee
with $N_c$ the number of unit cells, and the factor of $\frac{1}{2}$ to normalize the average to unity, since $\chi^{(1)}$ and $\chi^{(2)}$ each have a maximum value of 1. We further define the coarse-grained chirality $\left\langle \chi_{A(B)} \right\rangle_{m}$ as the partial averaging of the scalar chirality over a window of $A_{3\times3}$ centered at vertical location $m$,
\be
\left\langle \chi_{A(B)} \right\rangle_{m}\!=\! \dfrac{1}{36}\displaystyle\sum\limits_{\mathbf{R}\in A_{3\times3}}\left[  \chi^{(1)}_{A(B),\mathbf{R} +m\mathbf{a}_2^A} \!+\! \chi^{(2)}_{A(B),\mathbf{R} +m\mathbf{a}_2^A}\right]. 
\ee 
Similarly, we can define the coarse-grained chirality $\left\langle \chi_{A(B)} \right\rangle_{m}$ for a $A_{3\times n}$ cluster with $0 \!\leq\! m \!\leq\! n-1$ running vertically throughout the entire cluster. We can compute, alternatively to Eq.~(\ref{eqn:avg_chirality}, the average chirality of the entire cluster as
\be
  \left\langle \chi_{A(B)} \right\rangle  = \dfrac{1}{n}\displaystyle\sum\limits_{m = 0}^{n-1}
  \left\langle \chi_{A(B)} \right\rangle_{m}.
\ee

In Fig.~\ref{fig:chir1}(b) we show the average chirality $\langle\chi_{A(B)}\rangle$ by a red(blue) solid curve for the original state and dotted lines for the $\mc{R}$ dual states. In the 18C$_3$ state, due to the sublattice imbalance indicated earlier, we see that the average chirality is also imbalanced with $\left\langle \chi_{A} \right\rangle \neq \pm\left\langle \chi_{B} \right\rangle$, however, all other states including the IP show a sublattice equitable behavior of $\left\langle \chi_{A} \right\rangle = \left\langle \chi_{B} \right\rangle$. The chirality for states $6'$, 16, and zz are all zero, and can be attributed to the surviving $m_{\text{Z}}$ mirror symmetry in all three states. A jump is seen going from the 18C$_3$ to the IP, however, going from the IP to the dual $\mc{R}(6')$  18-sublattice state looks continuous. This is consistent with the picture of $\mc{R}(6')$ state being a commensurate limit of the IP.

For the minimum energy  states of the $3|n$ family obtained   for a  set  of points $\psi/\pi\!=$0.65, 0.7, 0.725, 0.75, 0.775, we examine the coarse-grained chirality $\left\langle \chi_{A(B)} \right\rangle_{m}$ while traversing the cluster vertically with $m$. Since for every $\psi$ the minima is on a different cluster length $n_{\psi}$, we normalize the length and plot $\left\langle \chi_{A(B)} \right\rangle_{m}$  as a function of  normalized $(m)$=$m/n_{\psi}$, with $0 \!\leq\! m \!\leq\! n_{\psi}$. In Fig.~\ref{fig:chir1}\,(c) we see the coarse-grained chirality for the IP, plotted against $m$. Running through the cluster, we see that $\left\langle \chi_{A} \right\rangle_{m}\!=\!-\left\langle \chi_{B} \right\rangle_{m}$ but with their average centered around the same value consistent with the average chirality behavior. 
We also see that approximately $\left\langle \chi_{A(B)} \right\rangle_{m}\!\propto\!\cos\left(\frac{2\pi}{n}m\right)$, with the smaller $\psi$ having additional harmonic contributions, while the larger $\psi$ being essentially a clean sinusoidal behavior. This harmonic behavior is capturing a clear switching between sublattice $A$ to $B$ halfway through the cluster.

\section{Discussion}

\subsection{Comparison with known results}\label{sec:compa}

In a recent study by Liu et al. \cite{Liu2021PRR}, 
{
the region $\psi/\pi\!\in[0.5,0.92]$}
was classified as a single state denoted as $S_3\times Z_3$. This classification was based on the identification of an internal 18-site symmetry that is parameter-dependent and was discovered using machine learning techniques. It is important to note that the machine learning algorithm employed in their study was trained on temperatures, T, of approximately $10^{-3}\sqrt{K^2+\Gamma^2}$.
 However, it is worth highlighting that the states 18C$_3$, IP, and $6'$ may differ in energy only at the third decimal place, meaning the $S_3\times Z_3$ characterization was carried out at
 finite temperature,
 where the energy resolution was at its limit. Consequently, this characterization may not be entirely conclusive.

In another work, Li {\it et al.}~\cite{JhengArxiv2023} used Monte-Carlo methods and identified an incommensurate helical order, which shares some qualitative features with our IP phase, most notably, the description in terms of long-distance modulations of 18 sublattices.
{
Additionally, the Monte Carlo calculations of Ref.~\cite{JhengArxiv2023} on the P$_{72\times 72}$ cluster at $\psi\!=\!0.75\pi$ yielded an energy per site equal to -0.92424917 (8 digits reported), which is 
identical to the energy -0.924249168580$\simeq$-0.92424917 (rounding to 8 digits) that we obtain on the A$_{3\times 36}$ armchair cluster, which is commensurate with the P$_{72\times 72}$ cluster. }
Finally, the helical state is modulated along the primitive directions $\bm{a}_1^P$ and $\bm{a}_2^P$, which is 
{
compatible}
with the one-dimensional modulation of the IP state along ${\bf a}_2^A$. 
So, it is possible that the IP state and the helical order tend to the same state in the thermodynamic limit.


{
However, the helical state of Ref.~\cite{JhengArxiv2023} is claimed to span the entire region $\psi/\pi\!\in[0.5,0.92]$, whereas our IP phase is stable in a narrower region.}
Unfortunately, the only energy reported in Ref.~\cite{JhengArxiv2023} is the one at $\psi\!=\!0.75\pi$, so a direct comparison beyond the IP region is not possible. To enable such direct comparisons to future studies, we provide our energies per site on a number of representative points throughout the phase diagram in Table~\ref{tab:raw_num_mc} of App.~\ref{sec:appendixC3}.



The phase diagram of various other extended extended $K$-$\Gamma$ models has also been studied. The anisotropic $K$-$\Gamma$ model, where one bond is allowed to be stronger than the other two, was studied by Rayyan et al.~\cite{RayyanPRB2021}. Additionally, Chen et al.~\cite{Chen2023NJP} explored the effects of introducing a single-ion anisotropy term. In these studies the $K$-$\Gamma$ line is a critical line owing to the breaking of the $\mc{R}$ symmetries from these additional interactions. Although we are unable to provide detailed insights into the extended regions, we can draw comparisons with their findings along the $K$-$\Gamma$ line. In the region $\psi/\pi\in[0.5,0.909]$ two phases where identified;  Phase I,  referred to as $18^{\eta}V$/$18^{\eta}{x(yz)}$ in Ref.~[\citenum{RayyanPRB2021}], TmX/18$_{\mathrm{B}}$ in Ref.[\citenum{Chen2023NJP}], and 18C$_3$/$\mc{R}(\text{18C}_3)$ in this work; Phase  II, denoted as  6/18 in Ref.~[\citenum{RayyanPRB2021}], 6$_{\mathrm{A}}$/18$_{\mathrm{A}}$ in Ref.~[\citenum{Chen2023NJP}], $6'$/$\mc{R}(6')$ in this work. Notably, our findings deviate from previous studies. In Refs.~[\citenum{RayyanPRB2021}] and [\citenum{Chen2023NJP}], Phase I is found to extend to the whole region of  $\psi\!/\!\pi\!\in\![0.5,0.77]$, subsequently transitioning into Phase II occupying $\psi\!/\!\pi\!\in\![0.77,0.909]$.  In our study,  we have observed a notable reduction in the size of these phases, opening a large  region  $\psi/\pi\in[0.6,0.793]$ for the  intermediate phase. Lastly we note that in all these  works,  including the current one, the region $\psi/\pi\in[0.909,1]$ is consistently characterized by the presence of the zigzag phase. {
However, we note that the stability of the zig-zag phase is claimed to go away from thermal fluctuations in Refs.~\cite{Liu2021PRR,JhengArxiv2023}}.

\subsection{Role of quantum fluctuations}\label{sec:QM}
Let us now discuss the fate of the above phase diagram when we include quantum fluctuations. We start by focusing on the semiclassical, large-$S$ limit.
On general grounds, in this limit the phase diagram is expected to remain qualitatively the same almost everywhere,  except in the vicinity of the classical spin liquid regions, namely, the pure Kitaev and the pure $\Gamma$ points.
In some of these regions, there is a competition between states that are selected at the mean field level (at order $S^2$) and states that are stabilized by the leading (order $S$) quantum fluctuations.

\begin{figure}[!t]
	\centering
     \begin{overpic}[width=1.0\columnwidth,percent]{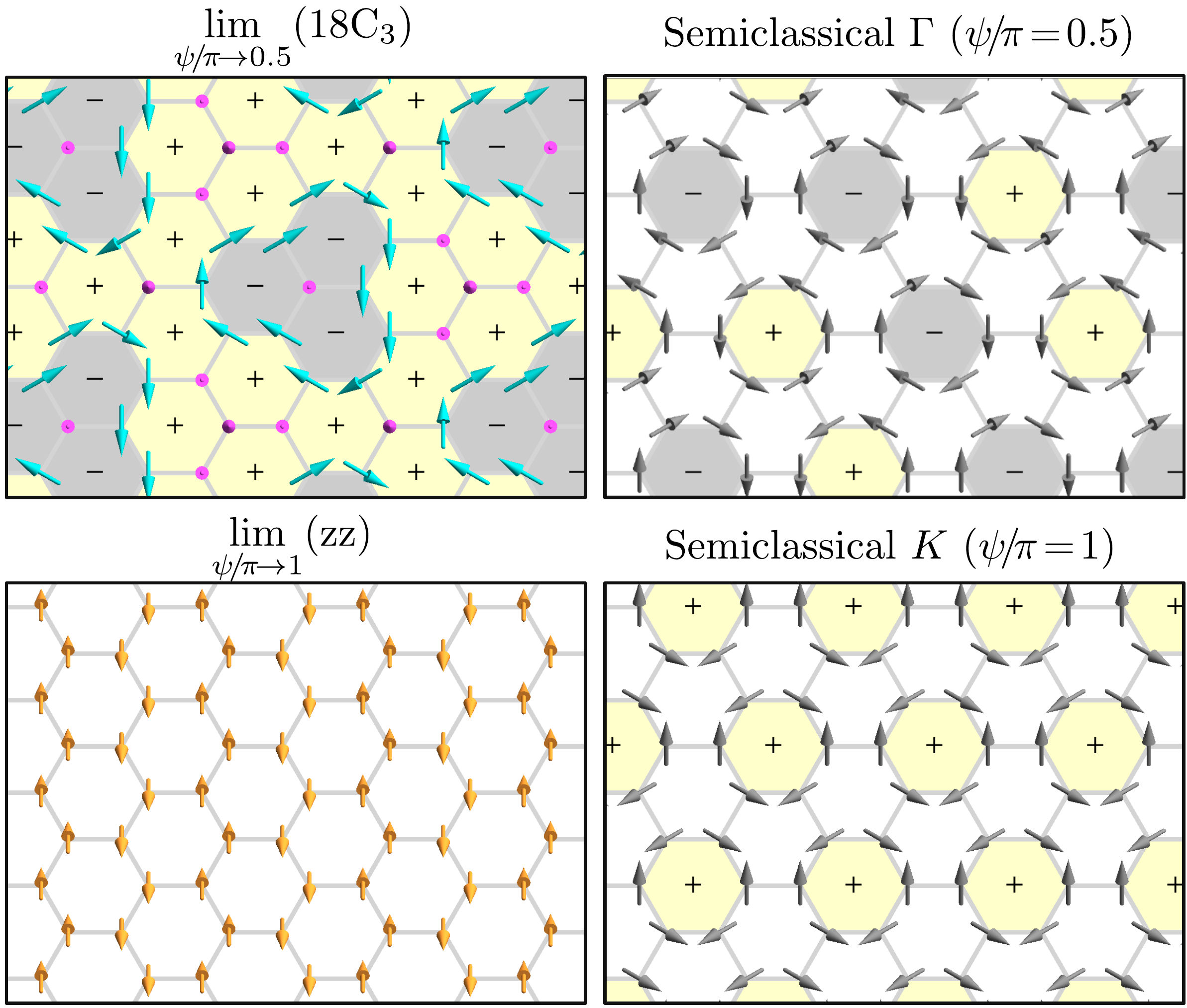}
        \put(0.3,81.2){(a)}
        \put(50.55,81.2){(b)}
        \put(0.3,38.2){(c)}
        \put(50.55,38.2){(d)}
    \end{overpic} 
 \caption{
 Comparison between the spin configurations selected at the mean field level (a), (c) and one of the states selected by the leading (order $S$) quantum fluctuations (b), (d) in the vicinity of the pure  $\Gamma\!>\!0$ point ($\psi/\pi\!\to\!0.5^+$) and the pure FM Kitaev point ($\psi/\pi\!\to\!1^-$). In (b) and (d), the spin directions (shown by gray arrows) are pinned along the cubic directions $[\pm1,0,0]$, $[0,\pm1,0]$, $[0,0,\pm1]$, whereas in (a) and (c) the spins point along the various $[111]$ and $[110]$ axes.}\label{fig:limits_of_anz}
\end{figure}
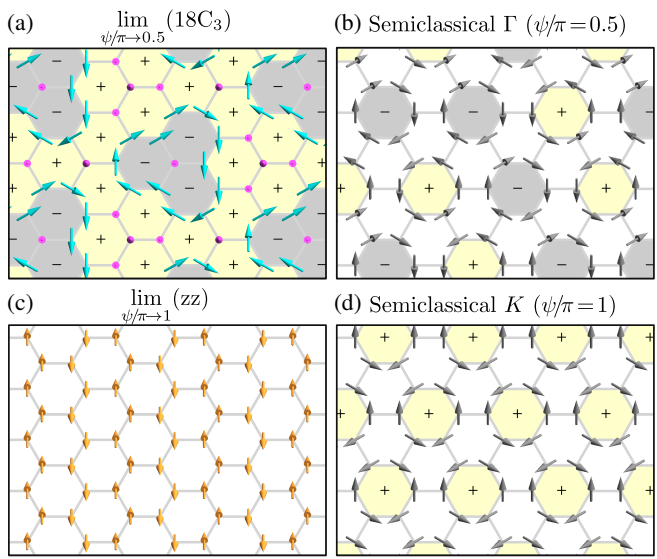

Take, for example, the vicinity of the pure  positive $\Gamma$ model, with a small negative $K$. The $\Gamma$ point hosts an infinite number of ground states~\cite{Rousochatzakis2017PRL} and, as mentioned above, a weak negative $K$ selects, at the mean field level, the `parent 18C$_3$' state of Eq.~(\ref{eq:18c3limit}), which is shown in the top right panel of Fig.~\ref{fig:limits_of_anz}. The energy scale associated with the selection of this state is $\propto KS^2$. This state is qualitatively different from the family of states that are generated by the leading quantum fluctuations at the $\Gamma$ point itself. Indeed, according to Ref.~\cite{Rousochatzakis2017PRL}, the leading order-by-disorder mechanism gives rise to emergent Ising degrees of freedom $\tau$, residing on the triangular superlattice formed by 1/3 of the hexagonal plaquettes of the lattice. For positive $\Gamma$, the interactions between the $\tau$ variables are described in terms of an AF Ising model, one of the prototype models of frustration. This model is  characterized  by  an infinitely large sub-family of ground states of $\tau$'s, featuring two majority and one minority $\tau$ in each triangle. One of these states is shown in the top left panel of Fig.~\ref{fig:limits_of_anz}, where $\pm$ signs inside the shaded plaquettes indicate the direction (up or down) of the corresponding $\tau$ variables. Given that the order-by-disorder energy scale leading to $\tau$ states is $\propto\Gamma S$, we anticipate that these states will be energetically favored compared to the parent 18C$_3$ state for sufficiently weak $K$. The width of this region, which is controlled by quantum mechanical corrections, will increase with decreasing $S$. The precise nature of the actual ground state within this window or whether the $\tau$ picture survives down to $S\!=\!1/2$ is still under debate. 
We note finally that in the opposite side of the  positive $\Gamma$ point (i.e., the one with a small positive $K$) the $\tau$ objects interact with each other ferromagnetically, leading to a fully polarized state of $\tau$ variables, which coincides with the state obtained at the mean field level~\cite{Rousochatzakis2017PRL}. So the semiclassical corrections do not alter the classical picture in this region.

Let us now turn to the vicinity of the FM Kitaev point, with a small positive $\Gamma$. The Kitaev point hosts an infinite number of classical ground states~\cite{BaskaranPRB2008,Chandra2010,RousochatzakisNC2018}, and, as mentioned above, a small positive $\Gamma$ selects, at the mean field level, the zigzag state, which is shown  in the bottom right panel of Fig.~\ref{fig:limits_of_anz}. The energy scale associated with the selection of this state is $\propto\Gamma S^2$.
This state is qualitatively different from the family of states that are generated by the leading spin-wave corrections at the Kitaev point. Indeed, according to Refs.~\cite{BaskaranPRB2008,RousochatzakisNC2018}, the leading spin-wave corrections give rise to emergent Ising degrees of freedom $\eta$, which now leave on the midpoints of nearest-neighbor dimers, which in turn form a ``star pattern''. Including  quantum tunneling corrections on top of the spin-wave, potential-like, terms, gives rise to an effective toric-code  Hamiltonian on the honeycomb superlattice of $\eta$'s~\cite{RousochatzakisNC2018}. The latter has a quantum spin liquid (QSL) ground state with topological degeneracy and fractionalized excitations. 
This QSL state is a linear superposition of an infinite number of special states (the ones selected by the potential terms alone~\cite{RousochatzakisNC2018}), a member of which is shown in the bottom left panel of Fig.~\ref{fig:limits_of_anz}. Now, the energy scale of the potential terms of the toric-code is $\propto KS$, while the tunneling terms are exponentially small in $S$. So, we anticipate that the QSL state of the $\eta$ variables will be energetically favored compared to the zigzag state for sufficiently weak $\Gamma$. The width of this region will increase with decreasing $S$. As discussed in Ref.~\cite{RousochatzakisNC2018}, the QSL state of $\eta$ variables survives down to $S\sim3/2$, below which other types of spin liquids are expected (including the exactly solvable QSL at $S=1/2$~\cite{Kitaev2006}).

\section{Conclusion}\label{sec:conclusions}

We provide a comprehensive theoretical description of the classical phase diagram of the Kitaev-$\Gamma$ model, with particular emphasis on the region with negative $K$ and positive $\Gamma$. This region not only holds crucial relevance to the existing materials, but also poses a formidable challenge for theoretical and numerical investigations. Within this region, the frustration between $K$ and $\Gamma$ gives rise to remarkably 
rich classical phase diagram, characterized by a plethora of magnetic orders with large unit cells. In addition to previously established 18C$_3$, $6'$, $16$ and zz phases~\cite{Chern2020PRR,RayyanPRB2021,Chen2023NJP}, for which we provide further insights, we also identify an intermediate phase (IP) which occupies a substantial portion  (about 40\%) of the parameter region $\psi/\pi\in[0.5,1]$.  To characterize the complex structure of this phase, we employ large-scale numerical minimizations on specially designed finite-size clusters (with periodic boundary conditions), which are significantly elongated along the crystallographic $\mathbf{a}$ direction.

The IP can be qualitatively understood as a long-distance twisting of the neighboring 18C$_3$ state, which essentially interpolates from a 18C$_3^{A}$ state centered on sublattice $A$ to the state 18C$_3^{B}\!=\!I$(18C$_3^{A}$), centered on sublattice $B$.

To reduce the energy, the IP introduces an incommensurate separation between 18C$_3^{A}$ and 18C$_3^{B}$, causing sublattices $A$ and $B$ to undergo counter-precession relative to each other. This precession is further qualified and quantified by a detailed study of the scalar chirality profiles. Within this phase, we have observed a cosine-like behavior in the coarse-grained chirality of sublattice $A$, accompanied by a corresponding negative cosine-like pattern in sublattice $B$. This behavior is attributed to the switching of 18C$_3$-like centers.  We anticipate that the presence of a finite average chirality, along with the intrinsic internal structure of coarse-grained chirality, can give rise to non-trivial topological features in spin-wave excitations. This, in turn, can lead to the emergence of a thermal Hall effect even in the absence of an external magnetic field.

The boundaries of the IP reflect fundamental changes in the spatial relationship between the two 18C$_3$-like centers. Also, the distinctiveness of the boundaries that define the IP serves as a clear indicator of its intricate relationship with neighboring phases. The transition from the IP to the 18C$_3$, occurring at approximately $\psi\!/\!\pi\!=\!0.6$, can be conceptually understood   as a continuous commensurate-incommensurate transition, whereby the distance between the centers of 18C$_3^{A}$ and 18C$_3^{B}$ tends to to infinity at the transition.  This separation of the 18C$_3^{A}$ and 18C$_3^{B}$ centers aligns with an increase in the total chirality within the system, ultimately reaching its maximum at the 18C$_3$ phase. Conversely, the boundary between the IP state and the $6'$ state, found at around $\psi\!/\!\pi=0.793$, corresponds to a situation in which 18C$_3^{A}$ and 18C$_3^{B}$ regions overlap perfectly, effectively reducing their separation distance to zero,  thereby restoring the inversion symmetry. As such, the transition to the $6'$ state is also accompanied by the vanishing of the total chirality.

We also address the role of quantum fluctuations at a qualitative level, and identify distinct regions, the vicinities of the pure $K$ and pure $\Gamma$ models, where quantum fluctuations play a non-trivial role already at the large-$S$ limit. The extent to which the resulting ``semiclassical'' predictions for these regions survives down to $S=1/2$ is a significant open problem, which warrants further theoretical investigation.

\section{Acknowledgements}
The authors thank  Cristian Batista and  Hae-Young Kee for valuable discussions. The work by  P.P.S., Y.Y. and N.B.P.  was supported by the U.S. Department of Energy, Office of Science, Basic Energy Sciences under Award No. DE-SC0018056.  IR acknowledges the support by the Engineering and Physical Sciences Research Council, Grant No. EP/V038281/1. Y.Y., I.R. and N.B.P.  acknowledge  the hospitality of KITP during the  Qmagnets23 program and partial support by the National Science Foundation under Grants No. NSF PHY-1748958 and PHY-2309135. N.B.P. also acknowledges the hospitality and partial support  of the Technical University of Munich – Institute for Advanced Study and  the Alexander von Humboldt Foundation.

\bibliography{jour_name_abbreviation.bib,thy_refs.bib}

\begin{thebibliography}{61}%
\makeatletter
\providecommand \@ifxundefined [1]{%
 \@ifx{#1\undefined}
}%
\providecommand \@ifnum [1]{%
 \ifnum #1\expandafter \@firstoftwo
 \else \expandafter \@secondoftwo
 \fi
}%
\providecommand \@ifx [1]{%
 \ifx #1\expandafter \@firstoftwo
 \else \expandafter \@secondoftwo
 \fi
}%
\providecommand \natexlab [1]{#1}%
\providecommand \enquote  [1]{``#1''}%
\providecommand \bibnamefont  [1]{#1}%
\providecommand \bibfnamefont [1]{#1}%
\providecommand \citenamefont [1]{#1}%
\providecommand \href@noop [0]{\@secondoftwo}%
\providecommand \href [0]{\begingroup \@sanitize@url \@href}%
\providecommand \@href[1]{\@@startlink{#1}\@@href}%
\providecommand \@@href[1]{\endgroup#1\@@endlink}%
\providecommand \@sanitize@url [0]{\catcode `\\12\catcode `\$12\catcode
  `\&12\catcode `\#12\catcode `\^12\catcode `\_12\catcode `\%12\relax}%
\providecommand \@@startlink[1]{}%
\providecommand \@@endlink[0]{}%
\providecommand \url  [0]{\begingroup\@sanitize@url \@url }%
\providecommand \@url [1]{\endgroup\@href {#1}{\urlprefix }}%
\providecommand \urlprefix  [0]{URL }%
\providecommand \Eprint [0]{\href }%
\providecommand \doibase [0]{https://doi.org/}%
\providecommand \selectlanguage [0]{\@gobble}%
\providecommand \bibinfo  [0]{\@secondoftwo}%
\providecommand \bibfield  [0]{\@secondoftwo}%
\providecommand \translation [1]{[#1]}%
\providecommand \BibitemOpen [0]{}%
\providecommand \bibitemStop [0]{}%
\providecommand \bibitemNoStop [0]{.\EOS\space}%
\providecommand \EOS [0]{\spacefactor3000\relax}%
\providecommand \BibitemShut  [1]{\csname bibitem#1\endcsname}%
\let\auto@bib@innerbib\@empty
\bibitem [{\citenamefont {Witczak-Krempa}\ \emph {et~al.}(2014)\citenamefont
  {Witczak-Krempa}, \citenamefont {Chen}, \citenamefont {Kim},\ and\
  \citenamefont {Balents}}]{Krempa2014ARCMP}%
  \BibitemOpen
  \bibfield  {author} {\bibinfo {author} {\bibfnamefont {W.}~\bibnamefont
  {Witczak-Krempa}}, \bibinfo {author} {\bibfnamefont {G.}~\bibnamefont
  {Chen}}, \bibinfo {author} {\bibfnamefont {Y.~B.}\ \bibnamefont {Kim}},\ and\
  \bibinfo {author} {\bibfnamefont {L.}~\bibnamefont {Balents}},\ }\bibfield
  {title} {\bibinfo {title} {{Correlated Quantum Phenomena in the Strong
  Spin-Orbit Regime}},\ }\href
  {https://doi.org/10.1146/annurev-conmatphys-020911-125138} {\bibfield
  {journal} {\bibinfo  {journal} {Annu. Rev. Condens. Matter Phys.}\ }\textbf
  {\bibinfo {volume} {5}},\ \bibinfo {pages} {57} (\bibinfo {year}
  {2014})}\BibitemShut {NoStop}%
\bibitem [{\citenamefont {Rau}\ \emph {et~al.}(2016)\citenamefont {Rau},
  \citenamefont {Lee},\ and\ \citenamefont {Kee}}]{Rau2016ARCMP}%
  \BibitemOpen
  \bibfield  {author} {\bibinfo {author} {\bibfnamefont {J.~G.}\ \bibnamefont
  {Rau}}, \bibinfo {author} {\bibfnamefont {E.~K.-H.}\ \bibnamefont {Lee}},\
  and\ \bibinfo {author} {\bibfnamefont {H.-Y.}\ \bibnamefont {Kee}},\
  }\bibfield  {title} {\bibinfo {title} {Spin-{Orbit} {Physics} {Giving} {Rise}
  to {Novel} {Phases} in {Correlated} {Systems}: {Iridates} and {Related}
  {Materials}},\ }\href
  {https://doi.org/10.1146/annurev-conmatphys-031115-011319} {\bibfield
  {journal} {\bibinfo  {journal} {Annu. Rev. Condens. Matter Phys.}\ }\textbf
  {\bibinfo {volume} {7}},\ \bibinfo {pages} {195} (\bibinfo {year}
  {2016})}\BibitemShut {NoStop}%
\bibitem [{\citenamefont {Winter}\ \emph {et~al.}(2016)\citenamefont {Winter},
  \citenamefont {Li}, \citenamefont {Jeschke},\ and\ \citenamefont
  {Valent\'{\i}}}]{Winter2016}%
  \BibitemOpen
  \bibfield  {author} {\bibinfo {author} {\bibfnamefont {S.~M.}\ \bibnamefont
  {Winter}}, \bibinfo {author} {\bibfnamefont {Y.}~\bibnamefont {Li}}, \bibinfo
  {author} {\bibfnamefont {H.~O.}\ \bibnamefont {Jeschke}},\ and\ \bibinfo
  {author} {\bibfnamefont {R.}~\bibnamefont {Valent\'{\i}}},\ }\bibfield
  {title} {\bibinfo {title} {Challenges in design of kitaev materials: Magnetic
  interactions from competing energy scales},\ }\href
  {https://doi.org/10.1103/PhysRevB.93.214431} {\bibfield  {journal} {\bibinfo
  {journal} {Phys. Rev. B}\ }\textbf {\bibinfo {volume} {93}},\ \bibinfo
  {pages} {214431} (\bibinfo {year} {2016})}\BibitemShut {NoStop}%
\bibitem [{\citenamefont {Winter}\ \emph {et~al.}(2017)\citenamefont {Winter},
  \citenamefont {Tsirlin}, \citenamefont {Daghofer}, \citenamefont {van~den
  Brink}, \citenamefont {Singh}, \citenamefont {Gegenwart},\ and\ \citenamefont
  {Valent\'{\i}}}]{Winter2017}%
  \BibitemOpen
  \bibfield  {author} {\bibinfo {author} {\bibfnamefont {S.~M.}\ \bibnamefont
  {Winter}}, \bibinfo {author} {\bibfnamefont {A.~A.}\ \bibnamefont {Tsirlin}},
  \bibinfo {author} {\bibfnamefont {M.}~\bibnamefont {Daghofer}}, \bibinfo
  {author} {\bibfnamefont {J.}~\bibnamefont {van~den Brink}}, \bibinfo {author}
  {\bibfnamefont {Y.}~\bibnamefont {Singh}}, \bibinfo {author} {\bibfnamefont
  {P.}~\bibnamefont {Gegenwart}},\ and\ \bibinfo {author} {\bibfnamefont
  {R.}~\bibnamefont {Valent\'{\i}}},\ }\bibfield  {title} {\bibinfo {title}
  {{Models and materials for generalized Kitaev magnetism}},\ }\href
  {http://stacks.iop.org/0953-8984/29/i=49/a=493002} {\bibfield  {journal}
  {\bibinfo  {journal} {J. Phys.: Condens. Matter}\ }\textbf {\bibinfo {volume}
  {29}},\ \bibinfo {pages} {493002} (\bibinfo {year} {2017})}\BibitemShut
  {NoStop}%
\bibitem [{\citenamefont {Takagi}\ \emph {et~al.}(2019)\citenamefont {Takagi},
  \citenamefont {Takayama}, \citenamefont {Jackeli}, \citenamefont
  {Khaliullin},\ and\ \citenamefont {Nagler}}]{Takagi2019}%
  \BibitemOpen
  \bibfield  {author} {\bibinfo {author} {\bibfnamefont {H.}~\bibnamefont
  {Takagi}}, \bibinfo {author} {\bibfnamefont {T.}~\bibnamefont {Takayama}},
  \bibinfo {author} {\bibfnamefont {G.}~\bibnamefont {Jackeli}}, \bibinfo
  {author} {\bibfnamefont {G.}~\bibnamefont {Khaliullin}},\ and\ \bibinfo
  {author} {\bibfnamefont {S.~E.}\ \bibnamefont {Nagler}},\ }\bibfield  {title}
  {\bibinfo {title} {{Concept and realization of Kitaev quantum spin
  liquids}},\ }\href {https://doi.org/10.1038/s42254-019-0038-2} {\bibfield
  {journal} {\bibinfo  {journal} {Nat. Rev. Phys.}\ }\textbf {\bibinfo {volume}
  {1}},\ \bibinfo {pages} {264} (\bibinfo {year} {2019})}\BibitemShut {NoStop}%
\bibitem [{\citenamefont {Takayama}\ \emph {et~al.}(2021)\citenamefont
  {Takayama}, \citenamefont {Chaloupka}, \citenamefont {Smerald}, \citenamefont
  {Khaliullin},\ and\ \citenamefont {Takagi}}]{Takayama2021JPSJ}%
  \BibitemOpen
  \bibfield  {author} {\bibinfo {author} {\bibfnamefont {T.}~\bibnamefont
  {Takayama}}, \bibinfo {author} {\bibfnamefont {J.}~\bibnamefont {Chaloupka}},
  \bibinfo {author} {\bibfnamefont {A.}~\bibnamefont {Smerald}}, \bibinfo
  {author} {\bibfnamefont {G.}~\bibnamefont {Khaliullin}},\ and\ \bibinfo
  {author} {\bibfnamefont {H.}~\bibnamefont {Takagi}},\ }\bibfield  {title}
  {\bibinfo {title} {Spin-orbit-entangled electronic phases in {4\textit{d}}
  and {5\textit{d}} transition-metal compounds},\ }\href
  {https://doi.org/10.7566/JPSJ.90.062001} {\bibfield  {journal} {\bibinfo
  {journal} {J. Phys. Soc. Jpn.}\ }\textbf {\bibinfo {volume} {90}},\ \bibinfo
  {pages} {062001} (\bibinfo {year} {2021})}\BibitemShut {NoStop}%
\bibitem [{\citenamefont {Trebst}\ and\ \citenamefont
  {Hickey}(2022)}]{Trebst2022}%
  \BibitemOpen
  \bibfield  {author} {\bibinfo {author} {\bibfnamefont {S.}~\bibnamefont
  {Trebst}}\ and\ \bibinfo {author} {\bibfnamefont {C.}~\bibnamefont
  {Hickey}},\ }\bibfield  {title} {\bibinfo {title} {Kitaev materials},\ }\href
  {https://doi.org/https://doi.org/10.1016/j.physrep.2021.11.003} {\bibfield
  {journal} {\bibinfo  {journal} {Phys. Rep.}\ }\textbf {\bibinfo {volume}
  {950}},\ \bibinfo {pages} {1} (\bibinfo {year} {2022})}\BibitemShut {NoStop}%
\bibitem [{\citenamefont {Tsirlin}\ and\ \citenamefont
  {Gegenwart}(2022)}]{TsirlinPSS2022}%
  \BibitemOpen
  \bibfield  {author} {\bibinfo {author} {\bibfnamefont {A.~A.}\ \bibnamefont
  {Tsirlin}}\ and\ \bibinfo {author} {\bibfnamefont {P.}~\bibnamefont
  {Gegenwart}},\ }\bibfield  {title} {\bibinfo {title} {Kitaev magnetism
  through the prism of lithium iridate},\ }\href
  {https://doi.org/https://doi.org/10.1002/pssb.202100146} {\bibfield
  {journal} {\bibinfo  {journal} {Phys. Status Solidi B}\ }\textbf {\bibinfo
  {volume} {259}},\ \bibinfo {pages} {2100146} (\bibinfo {year}
  {2022})}\BibitemShut {NoStop}%
\bibitem [{\citenamefont {Rousochatzakis}\ \emph {et~al.}(2024)\citenamefont
  {Rousochatzakis}, \citenamefont {Perkins}, \citenamefont {Luo},\ and\
  \citenamefont {Kee}}]{Rousochatzakis_2024}%
  \BibitemOpen
  \bibfield  {author} {\bibinfo {author} {\bibfnamefont {I.}~\bibnamefont
  {Rousochatzakis}}, \bibinfo {author} {\bibfnamefont {N.~B.}\ \bibnamefont
  {Perkins}}, \bibinfo {author} {\bibfnamefont {Q.}~\bibnamefont {Luo}},\ and\
  \bibinfo {author} {\bibfnamefont {H.-Y.}\ \bibnamefont {Kee}},\ }\bibfield
  {title} {\bibinfo {title} {{Beyond Kitaev physics in strong spin-orbit
  coupled magnets}},\ }\href {https://doi.org/10.1088/1361-6633/ad208d}
  {\bibfield  {journal} {\bibinfo  {journal} {Reports on Progress in Physics}\
  }\textbf {\bibinfo {volume} {87}},\ \bibinfo {pages} {026502} (\bibinfo
  {year} {2024})}\BibitemShut {NoStop}%
\bibitem [{\citenamefont {Singh}\ and\ \citenamefont
  {Gegenwart}(2010)}]{Singh2010PRB}%
  \BibitemOpen
  \bibfield  {author} {\bibinfo {author} {\bibfnamefont {Y.}~\bibnamefont
  {Singh}}\ and\ \bibinfo {author} {\bibfnamefont {P.}~\bibnamefont
  {Gegenwart}},\ }\bibfield  {title} {\bibinfo {title} {{Antiferromagnetic Mott
  insulating state in single crystals of the honeycomb lattice material
  ${\text{Na}}_{2}{\text{IrO}}_{3}$}},\ }\href
  {https://doi.org/10.1103/PhysRevB.82.064412} {\bibfield  {journal} {\bibinfo
  {journal} {Phys. Rev. B}\ }\textbf {\bibinfo {volume} {82}},\ \bibinfo
  {pages} {064412} (\bibinfo {year} {2010})}\BibitemShut {NoStop}%
\bibitem [{\citenamefont {Singh}\ \emph {et~al.}(2012)\citenamefont {Singh},
  \citenamefont {Manni}, \citenamefont {Reuther}, \citenamefont {Berlijn},
  \citenamefont {Thomale}, \citenamefont {Ku}, \citenamefont {Trebst},\ and\
  \citenamefont {Gegenwart}}]{Singh2012PRL}%
  \BibitemOpen
  \bibfield  {author} {\bibinfo {author} {\bibfnamefont {Y.}~\bibnamefont
  {Singh}}, \bibinfo {author} {\bibfnamefont {S.}~\bibnamefont {Manni}},
  \bibinfo {author} {\bibfnamefont {J.}~\bibnamefont {Reuther}}, \bibinfo
  {author} {\bibfnamefont {T.}~\bibnamefont {Berlijn}}, \bibinfo {author}
  {\bibfnamefont {R.}~\bibnamefont {Thomale}}, \bibinfo {author} {\bibfnamefont
  {W.}~\bibnamefont {Ku}}, \bibinfo {author} {\bibfnamefont {S.}~\bibnamefont
  {Trebst}},\ and\ \bibinfo {author} {\bibfnamefont {P.}~\bibnamefont
  {Gegenwart}},\ }\bibfield  {title} {\bibinfo {title} {{Relevance of the
  Heisenberg-Kitaev Model for the Honeycomb Lattice Iridates
  ${A}_{2}{\mathrm{IrO}}_{3}$}},\ }\href
  {https://doi.org/10.1103/PhysRevLett.108.127203} {\bibfield  {journal}
  {\bibinfo  {journal} {Phys. Rev. Lett.}\ }\textbf {\bibinfo {volume} {108}},\
  \bibinfo {pages} {127203} (\bibinfo {year} {2012})}\BibitemShut {NoStop}%
\bibitem [{\citenamefont {Choi}\ \emph {et~al.}(2012)\citenamefont {Choi},
  \citenamefont {Coldea}, \citenamefont {Kolmogorov}, \citenamefont
  {Lancaster}, \citenamefont {Mazin}, \citenamefont {Blundell}, \citenamefont
  {Radaelli}, \citenamefont {Singh}, \citenamefont {Gegenwart}, \citenamefont
  {Choi}, \citenamefont {Cheong}, \citenamefont {Baker}, \citenamefont
  {Stock},\ and\ \citenamefont {Taylor}}]{Choi2012PRL}%
  \BibitemOpen
  \bibfield  {author} {\bibinfo {author} {\bibfnamefont {S.~K.}\ \bibnamefont
  {Choi}}, \bibinfo {author} {\bibfnamefont {R.}~\bibnamefont {Coldea}},
  \bibinfo {author} {\bibfnamefont {A.~N.}\ \bibnamefont {Kolmogorov}},
  \bibinfo {author} {\bibfnamefont {T.}~\bibnamefont {Lancaster}}, \bibinfo
  {author} {\bibfnamefont {I.~I.}\ \bibnamefont {Mazin}}, \bibinfo {author}
  {\bibfnamefont {S.~J.}\ \bibnamefont {Blundell}}, \bibinfo {author}
  {\bibfnamefont {P.~G.}\ \bibnamefont {Radaelli}}, \bibinfo {author}
  {\bibfnamefont {Y.}~\bibnamefont {Singh}}, \bibinfo {author} {\bibfnamefont
  {P.}~\bibnamefont {Gegenwart}}, \bibinfo {author} {\bibfnamefont {K.~R.}\
  \bibnamefont {Choi}}, \bibinfo {author} {\bibfnamefont {S.-W.}\ \bibnamefont
  {Cheong}}, \bibinfo {author} {\bibfnamefont {P.~J.}\ \bibnamefont {Baker}},
  \bibinfo {author} {\bibfnamefont {C.}~\bibnamefont {Stock}},\ and\ \bibinfo
  {author} {\bibfnamefont {J.}~\bibnamefont {Taylor}},\ }\bibfield  {title}
  {\bibinfo {title} {{Spin Waves and Revised Crystal Structure of Honeycomb
  Iridate ${\mathrm{Na}}_{2}{\mathrm{IrO}}_{3}$}},\ }\href
  {https://doi.org/10.1103/PhysRevLett.108.127204} {\bibfield  {journal}
  {\bibinfo  {journal} {Phys. Rev. Lett.}\ }\textbf {\bibinfo {volume} {108}},\
  \bibinfo {pages} {127204} (\bibinfo {year} {2012})}\BibitemShut {NoStop}%
\bibitem [{\citenamefont {Ye}\ \emph {et~al.}(2012)\citenamefont {Ye},
  \citenamefont {Chi}, \citenamefont {Cao}, \citenamefont {Chakoumakos},
  \citenamefont {Fernandez-Baca}, \citenamefont {Custelcean}, \citenamefont
  {Qi}, \citenamefont {Korneta},\ and\ \citenamefont {Cao}}]{Ye2012PRB}%
  \BibitemOpen
  \bibfield  {author} {\bibinfo {author} {\bibfnamefont {F.}~\bibnamefont
  {Ye}}, \bibinfo {author} {\bibfnamefont {S.}~\bibnamefont {Chi}}, \bibinfo
  {author} {\bibfnamefont {H.}~\bibnamefont {Cao}}, \bibinfo {author}
  {\bibfnamefont {B.~C.}\ \bibnamefont {Chakoumakos}}, \bibinfo {author}
  {\bibfnamefont {J.~A.}\ \bibnamefont {Fernandez-Baca}}, \bibinfo {author}
  {\bibfnamefont {R.}~\bibnamefont {Custelcean}}, \bibinfo {author}
  {\bibfnamefont {T.~F.}\ \bibnamefont {Qi}}, \bibinfo {author} {\bibfnamefont
  {O.~B.}\ \bibnamefont {Korneta}},\ and\ \bibinfo {author} {\bibfnamefont
  {G.}~\bibnamefont {Cao}},\ }\bibfield  {title} {\bibinfo {title} {{Direct
  evidence of a zigzag spin-chain structure in the honeycomb lattice: A neutron
  and x-ray diffraction investigation of single-crystal
  Na${}_{2}$IrO${}_{3}$}},\ }\href
  {http://link.aps.org/doi/10.1103/PhysRevB.85.180403} {\bibfield  {journal}
  {\bibinfo  {journal} {Phys. Rev. B}\ }\textbf {\bibinfo {volume} {85}},\
  \bibinfo {pages} {180403} (\bibinfo {year} {2012})}\BibitemShut {NoStop}%
\bibitem [{\citenamefont {Hwan~Chun}\ \emph {et~al.}(2015)\citenamefont
  {Hwan~Chun}, \citenamefont {Kim}, \citenamefont {Kim}, \citenamefont {Zheng},
  \citenamefont {Stoumpos}, \citenamefont {Malliakas}, \citenamefont
  {Mitchell}, \citenamefont {Mehlawat}, \citenamefont {Singh}, \citenamefont
  {Choi}, \citenamefont {Gog}, \citenamefont {Al-Zein}, \citenamefont {Sala},
  \citenamefont {Krisch}, \citenamefont {Chaloupka}, \citenamefont {Jackeli},
  \citenamefont {Khaliullin},\ and\ \citenamefont {Kim}}]{Chun2015}%
  \BibitemOpen
  \bibfield  {author} {\bibinfo {author} {\bibfnamefont {S.}~\bibnamefont
  {Hwan~Chun}}, \bibinfo {author} {\bibfnamefont {J.-W.}\ \bibnamefont {Kim}},
  \bibinfo {author} {\bibfnamefont {J.}~\bibnamefont {Kim}}, \bibinfo {author}
  {\bibfnamefont {H.}~\bibnamefont {Zheng}}, \bibinfo {author} {\bibfnamefont
  {C.~C.}\ \bibnamefont {Stoumpos}}, \bibinfo {author} {\bibfnamefont {C.~D.}\
  \bibnamefont {Malliakas}}, \bibinfo {author} {\bibfnamefont {J.~F.}\
  \bibnamefont {Mitchell}}, \bibinfo {author} {\bibfnamefont {K.}~\bibnamefont
  {Mehlawat}}, \bibinfo {author} {\bibfnamefont {Y.}~\bibnamefont {Singh}},
  \bibinfo {author} {\bibfnamefont {Y.}~\bibnamefont {Choi}}, \bibinfo {author}
  {\bibfnamefont {T.}~\bibnamefont {Gog}}, \bibinfo {author} {\bibfnamefont
  {A.}~\bibnamefont {Al-Zein}}, \bibinfo {author} {\bibfnamefont {M.~M.}\
  \bibnamefont {Sala}}, \bibinfo {author} {\bibfnamefont {M.}~\bibnamefont
  {Krisch}}, \bibinfo {author} {\bibfnamefont {J.}~\bibnamefont {Chaloupka}},
  \bibinfo {author} {\bibfnamefont {G.}~\bibnamefont {Jackeli}}, \bibinfo
  {author} {\bibfnamefont {G.}~\bibnamefont {Khaliullin}},\ and\ \bibinfo
  {author} {\bibfnamefont {B.~J.}\ \bibnamefont {Kim}},\ }\bibfield  {title}
  {\bibinfo {title} {{Direct evidence for dominant bond-directional
  interactions in a honeycomb lattice iridate Na$_2$IrO$_3$}},\ }\href
  {http://dx.doi.org/10.1038/nphys3322} {\bibfield  {journal} {\bibinfo
  {journal} {Nat. Phys.}\ }\textbf {\bibinfo {volume} {11}},\ \bibinfo {pages}
  {462 } (\bibinfo {year} {2015})}\BibitemShut {NoStop}%
\bibitem [{\citenamefont {Williams}\ \emph {et~al.}(2016)\citenamefont
  {Williams}, \citenamefont {Johnson}, \citenamefont {Freund}, \citenamefont
  {Choi}, \citenamefont {Jesche}, \citenamefont {Kimchi}, \citenamefont
  {Manni}, \citenamefont {Bombardi}, \citenamefont {Manuel}, \citenamefont
  {Gegenwart},\ and\ \citenamefont {Coldea}}]{Williams2016PRB}%
  \BibitemOpen
  \bibfield  {author} {\bibinfo {author} {\bibfnamefont {S.~C.}\ \bibnamefont
  {Williams}}, \bibinfo {author} {\bibfnamefont {R.~D.}\ \bibnamefont
  {Johnson}}, \bibinfo {author} {\bibfnamefont {F.}~\bibnamefont {Freund}},
  \bibinfo {author} {\bibfnamefont {S.}~\bibnamefont {Choi}}, \bibinfo {author}
  {\bibfnamefont {A.}~\bibnamefont {Jesche}}, \bibinfo {author} {\bibfnamefont
  {I.}~\bibnamefont {Kimchi}}, \bibinfo {author} {\bibfnamefont
  {S.}~\bibnamefont {Manni}}, \bibinfo {author} {\bibfnamefont
  {A.}~\bibnamefont {Bombardi}}, \bibinfo {author} {\bibfnamefont
  {P.}~\bibnamefont {Manuel}}, \bibinfo {author} {\bibfnamefont
  {P.}~\bibnamefont {Gegenwart}},\ and\ \bibinfo {author} {\bibfnamefont
  {R.}~\bibnamefont {Coldea}},\ }\bibfield  {title} {\bibinfo {title}
  {{Incommensurate counterrotating magnetic order stabilized by Kitaev
  interactions in the layered honeycomb
  $\ensuremath{\alpha}\text{-}{\mathrm{Li}}_{2}{\mathrm{IrO}}_{3}$}},\ }\href
  {https://doi.org/10.1103/PhysRevB.93.195158} {\bibfield  {journal} {\bibinfo
  {journal} {Phys. Rev. B}\ }\textbf {\bibinfo {volume} {93}},\ \bibinfo
  {pages} {195158} (\bibinfo {year} {2016})}\BibitemShut {NoStop}%
\bibitem [{\citenamefont {Plumb}\ \emph {et~al.}(2014)\citenamefont {Plumb},
  \citenamefont {Clancy}, \citenamefont {Sandilands}, \citenamefont {Shankar},
  \citenamefont {Hu}, \citenamefont {Burch}, \citenamefont {Kee},\ and\
  \citenamefont {Kim}}]{Plumb2014PRB}%
  \BibitemOpen
  \bibfield  {author} {\bibinfo {author} {\bibfnamefont {K.~W.}\ \bibnamefont
  {Plumb}}, \bibinfo {author} {\bibfnamefont {J.~P.}\ \bibnamefont {Clancy}},
  \bibinfo {author} {\bibfnamefont {L.~J.}\ \bibnamefont {Sandilands}},
  \bibinfo {author} {\bibfnamefont {V.~V.}\ \bibnamefont {Shankar}}, \bibinfo
  {author} {\bibfnamefont {Y.~F.}\ \bibnamefont {Hu}}, \bibinfo {author}
  {\bibfnamefont {K.~S.}\ \bibnamefont {Burch}}, \bibinfo {author}
  {\bibfnamefont {H.-Y.}\ \bibnamefont {Kee}},\ and\ \bibinfo {author}
  {\bibfnamefont {Y.-J.}\ \bibnamefont {Kim}},\ }\bibfield  {title} {\bibinfo
  {title} {{$\ensuremath{\alpha}-{\mathrm{RuCl}}_{3}$: A spin-orbit assisted
  Mott insulator on a honeycomb lattice}},\ }\href
  {https://doi.org/10.1103/PhysRevB.90.041112} {\bibfield  {journal} {\bibinfo
  {journal} {Phys. Rev. B}\ }\textbf {\bibinfo {volume} {90}},\ \bibinfo
  {pages} {041112} (\bibinfo {year} {2014})}\BibitemShut {NoStop}%
\bibitem [{\citenamefont {Sears}\ \emph {et~al.}(2015)\citenamefont {Sears},
  \citenamefont {Songvilay}, \citenamefont {Plumb}, \citenamefont {Clancy},
  \citenamefont {Qiu}, \citenamefont {Zhao}, \citenamefont {Parshall},\ and\
  \citenamefont {Kim}}]{Sears2015PRB}%
  \BibitemOpen
  \bibfield  {author} {\bibinfo {author} {\bibfnamefont {J.~A.}\ \bibnamefont
  {Sears}}, \bibinfo {author} {\bibfnamefont {M.}~\bibnamefont {Songvilay}},
  \bibinfo {author} {\bibfnamefont {K.~W.}\ \bibnamefont {Plumb}}, \bibinfo
  {author} {\bibfnamefont {J.~P.}\ \bibnamefont {Clancy}}, \bibinfo {author}
  {\bibfnamefont {Y.}~\bibnamefont {Qiu}}, \bibinfo {author} {\bibfnamefont
  {Y.}~\bibnamefont {Zhao}}, \bibinfo {author} {\bibfnamefont {D.}~\bibnamefont
  {Parshall}},\ and\ \bibinfo {author} {\bibfnamefont {Y.-J.}\ \bibnamefont
  {Kim}},\ }\bibfield  {title} {\bibinfo {title} {{Magnetic order in
  $\ensuremath{\alpha}-{\text{RuCl}}_{3}$: A honeycomb-lattice quantum magnet
  with strong spin-orbit coupling}},\ }\href
  {https://doi.org/10.1103/PhysRevB.91.144420} {\bibfield  {journal} {\bibinfo
  {journal} {Phys. Rev. B}\ }\textbf {\bibinfo {volume} {91}},\ \bibinfo
  {pages} {144420} (\bibinfo {year} {2015})}\BibitemShut {NoStop}%
\bibitem [{\citenamefont {Majumder}\ \emph {et~al.}(2015)\citenamefont
  {Majumder}, \citenamefont {Schmidt}, \citenamefont {Rosner}, \citenamefont
  {Tsirlin}, \citenamefont {Yasuoka},\ and\ \citenamefont
  {Baenitz}}]{Majumder2015}%
  \BibitemOpen
  \bibfield  {author} {\bibinfo {author} {\bibfnamefont {M.}~\bibnamefont
  {Majumder}}, \bibinfo {author} {\bibfnamefont {M.}~\bibnamefont {Schmidt}},
  \bibinfo {author} {\bibfnamefont {H.}~\bibnamefont {Rosner}}, \bibinfo
  {author} {\bibfnamefont {A.~A.}\ \bibnamefont {Tsirlin}}, \bibinfo {author}
  {\bibfnamefont {H.}~\bibnamefont {Yasuoka}},\ and\ \bibinfo {author}
  {\bibfnamefont {M.}~\bibnamefont {Baenitz}},\ }\bibfield  {title} {\bibinfo
  {title} {Anisotropic ${\mathrm{ru}}^{3+} 4{d}^{5}$ magnetism in the
  $\ensuremath{\alpha}\ensuremath{-}{\mathrm{rucl}}_{3}$ honeycomb system:
  Susceptibility, specific heat, and zero-field nmr},\ }\href
  {https://doi.org/10.1103/PhysRevB.91.180401} {\bibfield  {journal} {\bibinfo
  {journal} {Phys. Rev. B}\ }\textbf {\bibinfo {volume} {91}},\ \bibinfo
  {pages} {180401} (\bibinfo {year} {2015})}\BibitemShut {NoStop}%
\bibitem [{\citenamefont {Kitaev}(2006)}]{Kitaev2006}%
  \BibitemOpen
  \bibfield  {author} {\bibinfo {author} {\bibfnamefont {A.}~\bibnamefont
  {Kitaev}},\ }\bibfield  {title} {\bibinfo {title} {Anyons in an exactly
  solved model and beyond},\ }\href {https://doi.org/10.1016/j.aop.2005.10.005}
  {\bibfield  {journal} {\bibinfo  {journal} {Annals of Physics}\ }\textbf
  {\bibinfo {volume} {321}},\ \bibinfo {pages} {2} (\bibinfo {year}
  {2006})}\BibitemShut {NoStop}%
\bibitem [{\citenamefont {Maksimov}\ and\ \citenamefont
  {Chernyshev}(2020)}]{MaksimovPRR2020}%
  \BibitemOpen
  \bibfield  {author} {\bibinfo {author} {\bibfnamefont {P.~A.}\ \bibnamefont
  {Maksimov}}\ and\ \bibinfo {author} {\bibfnamefont {A.~L.}\ \bibnamefont
  {Chernyshev}},\ }\bibfield  {title} {\bibinfo {title} {Rethinking
  $\ensuremath{\alpha}\text{\ensuremath{-}}{\mathrm{rucl}}_{3}$},\ }\href
  {https://doi.org/10.1103/PhysRevResearch.2.033011} {\bibfield  {journal}
  {\bibinfo  {journal} {Phys. Rev. Res.}\ }\textbf {\bibinfo {volume} {2}},\
  \bibinfo {pages} {033011} (\bibinfo {year} {2020})}\BibitemShut {NoStop}%
\bibitem [{\citenamefont {Jackeli}\ and\ \citenamefont
  {Khaliullin}(2009)}]{Jackeli2009}%
  \BibitemOpen
  \bibfield  {author} {\bibinfo {author} {\bibfnamefont {G.}~\bibnamefont
  {Jackeli}}\ and\ \bibinfo {author} {\bibfnamefont {G.}~\bibnamefont
  {Khaliullin}},\ }\bibfield  {title} {\bibinfo {title} {{Mott Insulators in
  the Strong Spin-Orbit Coupling Limit: From Heisenberg to a Quantum Compass
  and Kitaev Models}},\ }\href {https://doi.org/10.1103/PhysRevLett.102.017205}
  {\bibfield  {journal} {\bibinfo  {journal} {Phys. Rev. Lett.}\ }\textbf
  {\bibinfo {volume} {102}},\ \bibinfo {pages} {017205} (\bibinfo {year}
  {2009})}\BibitemShut {NoStop}%
\bibitem [{\citenamefont {Chaloupka}\ \emph {et~al.}(2010)\citenamefont
  {Chaloupka}, \citenamefont {Jackeli},\ and\ \citenamefont
  {Khaliullin}}]{Chaloupka2010}%
  \BibitemOpen
  \bibfield  {author} {\bibinfo {author} {\bibfnamefont {J.}~\bibnamefont
  {Chaloupka}}, \bibinfo {author} {\bibfnamefont {G.}~\bibnamefont {Jackeli}},\
  and\ \bibinfo {author} {\bibfnamefont {G.}~\bibnamefont {Khaliullin}},\
  }\bibfield  {title} {\bibinfo {title} {Kitaev-heisenberg model on a honeycomb
  lattice: Possible exotic phases in iridium oxides
  ${A}_{2}{\mathrm{iro}}_{3}$},\ }\href
  {https://doi.org/10.1103/PhysRevLett.105.027204} {\bibfield  {journal}
  {\bibinfo  {journal} {Phys. Rev. Lett.}\ }\textbf {\bibinfo {volume} {105}},\
  \bibinfo {pages} {027204} (\bibinfo {year} {2010})}\BibitemShut {NoStop}%
\bibitem [{\citenamefont {Chaloupka}\ \emph {et~al.}(2013)\citenamefont
  {Chaloupka}, \citenamefont {Jackeli},\ and\ \citenamefont
  {Khaliullin}}]{Chaloupka2013}%
  \BibitemOpen
  \bibfield  {author} {\bibinfo {author} {\bibfnamefont {J.}~\bibnamefont
  {Chaloupka}}, \bibinfo {author} {\bibfnamefont {G.}~\bibnamefont {Jackeli}},\
  and\ \bibinfo {author} {\bibfnamefont {G.}~\bibnamefont {Khaliullin}},\
  }\bibfield  {title} {\bibinfo {title} {Zigzag magnetic order in the iridium
  oxide ${\mathrm{na}}_{2}{\mathrm{iro}}_{3}$},\ }\href
  {https://doi.org/10.1103/PhysRevLett.110.097204} {\bibfield  {journal}
  {\bibinfo  {journal} {Phys. Rev. Lett.}\ }\textbf {\bibinfo {volume} {110}},\
  \bibinfo {pages} {097204} (\bibinfo {year} {2013})}\BibitemShut {NoStop}%
\bibitem [{\citenamefont {Rau}\ \emph {et~al.}(2014)\citenamefont {Rau},
  \citenamefont {Lee},\ and\ \citenamefont {Kee}}]{Rau2014}%
  \BibitemOpen
  \bibfield  {author} {\bibinfo {author} {\bibfnamefont {J.~G.}\ \bibnamefont
  {Rau}}, \bibinfo {author} {\bibfnamefont {E.~K.-H.}\ \bibnamefont {Lee}},\
  and\ \bibinfo {author} {\bibfnamefont {H.-Y.}\ \bibnamefont {Kee}},\
  }\bibfield  {title} {\bibinfo {title} {Generic spin model for the honeycomb
  iridates beyond the kitaev limit},\ }\href
  {https://doi.org/10.1103/PhysRevLett.112.077204} {\bibfield  {journal}
  {\bibinfo  {journal} {Phys. Rev. Lett.}\ }\textbf {\bibinfo {volume} {112}},\
  \bibinfo {pages} {077204} (\bibinfo {year} {2014})}\BibitemShut {NoStop}%
\bibitem [{\citenamefont {Rau}\ and\ \citenamefont
  {Kee}(2014)}]{rau2014trigonal}%
  \BibitemOpen
  \bibfield  {author} {\bibinfo {author} {\bibfnamefont {J.~G.}\ \bibnamefont
  {Rau}}\ and\ \bibinfo {author} {\bibfnamefont {H.-Y.}\ \bibnamefont {Kee}},\
  }\href@noop {} {\bibinfo {title} {Trigonal distortion in the honeycomb
  iridates: Proximity of zigzag and spiral phases in na2iro3}} (\bibinfo {year}
  {2014}),\ \Eprint {https://arxiv.org/abs/1408.4811} {arXiv:1408.4811
  [cond-mat.str-el]} \BibitemShut {NoStop}%
\bibitem [{\citenamefont {Sizyuk}\ \emph {et~al.}(2014)\citenamefont {Sizyuk},
  \citenamefont {Price}, \citenamefont {W\"olfle},\ and\ \citenamefont
  {Perkins}}]{Sizyuk2014}%
  \BibitemOpen
  \bibfield  {author} {\bibinfo {author} {\bibfnamefont {Y.}~\bibnamefont
  {Sizyuk}}, \bibinfo {author} {\bibfnamefont {C.}~\bibnamefont {Price}},
  \bibinfo {author} {\bibfnamefont {P.}~\bibnamefont {W\"olfle}},\ and\
  \bibinfo {author} {\bibfnamefont {N.~B.}\ \bibnamefont {Perkins}},\
  }\bibfield  {title} {\bibinfo {title} {{Importance of anisotropic exchange
  interactions in honeycomb iridates: Minimal model for zigzag
  antiferromagnetic order in ${\mathrm{Na}}_{2}{\mathrm{IrO}}_{3}$}},\ }\href
  {https://doi.org/10.1103/PhysRevB.90.155126} {\bibfield  {journal} {\bibinfo
  {journal} {Phys. Rev. B}\ }\textbf {\bibinfo {volume} {90}},\ \bibinfo
  {pages} {155126} (\bibinfo {year} {2014})}\BibitemShut {NoStop}%
\bibitem [{\citenamefont {Rousochatzakis}\ \emph {et~al.}(2015)\citenamefont
  {Rousochatzakis}, \citenamefont {Reuther}, \citenamefont {Thomale},
  \citenamefont {Rachel},\ and\ \citenamefont
  {Perkins}}]{Rousochatzakis2015PRX}%
  \BibitemOpen
  \bibfield  {author} {\bibinfo {author} {\bibfnamefont {I.}~\bibnamefont
  {Rousochatzakis}}, \bibinfo {author} {\bibfnamefont {J.}~\bibnamefont
  {Reuther}}, \bibinfo {author} {\bibfnamefont {R.}~\bibnamefont {Thomale}},
  \bibinfo {author} {\bibfnamefont {S.}~\bibnamefont {Rachel}},\ and\ \bibinfo
  {author} {\bibfnamefont {N.~B.}\ \bibnamefont {Perkins}},\ }\bibfield
  {title} {\bibinfo {title} {{Phase Diagram and Quantum Order by Disorder in
  the Kitaev ${K}_{1}\ensuremath{-}{K}_{2}$ Honeycomb Magnet}},\ }\href
  {https://doi.org/10.1103/PhysRevX.5.041035} {\bibfield  {journal} {\bibinfo
  {journal} {Phys. Rev. X}\ }\textbf {\bibinfo {volume} {5}},\ \bibinfo {pages}
  {041035} (\bibinfo {year} {2015})}\BibitemShut {NoStop}%
\bibitem [{\citenamefont {Rousochatzakis}\ and\ \citenamefont
  {Perkins}(2017)}]{Rousochatzakis2017PRL}%
  \BibitemOpen
  \bibfield  {author} {\bibinfo {author} {\bibfnamefont {I.}~\bibnamefont
  {Rousochatzakis}}\ and\ \bibinfo {author} {\bibfnamefont {N.~B.}\
  \bibnamefont {Perkins}},\ }\bibfield  {title} {\bibinfo {title} {{Classical
  Spin Liquid Instability Driven By Off-Diagonal Exchange in Strong Spin-Orbit
  Magnets}},\ }\href {https://doi.org/10.1103/PhysRevLett.118.147204}
  {\bibfield  {journal} {\bibinfo  {journal} {Phys. Rev. Lett.}\ }\textbf
  {\bibinfo {volume} {118}},\ \bibinfo {pages} {147204} (\bibinfo {year}
  {2017})}\BibitemShut {NoStop}%
\bibitem [{\citenamefont {Fujimoto}(2009)}]{Fujimoto2009}%
  \BibitemOpen
  \bibfield  {author} {\bibinfo {author} {\bibfnamefont {S.}~\bibnamefont
  {Fujimoto}},\ }\bibfield  {title} {\bibinfo {title} {Hall effect of spin
  waves in frustrated magnets},\ }\href
  {https://doi.org/10.1103/PhysRevLett.103.047203} {\bibfield  {journal}
  {\bibinfo  {journal} {Phys. Rev. Lett.}\ }\textbf {\bibinfo {volume} {103}},\
  \bibinfo {pages} {047203} (\bibinfo {year} {2009})}\BibitemShut {NoStop}%
\bibitem [{\citenamefont {Chern}\ \emph {et~al.}(2021)\citenamefont {Chern},
  \citenamefont {Zhang},\ and\ \citenamefont {Kim}}]{LiChern2021}%
  \BibitemOpen
  \bibfield  {author} {\bibinfo {author} {\bibfnamefont {L.~E.}\ \bibnamefont
  {Chern}}, \bibinfo {author} {\bibfnamefont {E.~Z.}\ \bibnamefont {Zhang}},\
  and\ \bibinfo {author} {\bibfnamefont {Y.~B.}\ \bibnamefont {Kim}},\
  }\bibfield  {title} {\bibinfo {title} {Sign structure of thermal hall
  conductivity and topological magnons for in-plane field polarized kitaev
  magnets},\ }\href {https://doi.org/10.1103/PhysRevLett.126.147201} {\bibfield
   {journal} {\bibinfo  {journal} {Phys. Rev. Lett.}\ }\textbf {\bibinfo
  {volume} {126}},\ \bibinfo {pages} {147201} (\bibinfo {year}
  {2021})}\BibitemShut {NoStop}%
\bibitem [{\citenamefont {Zhang}\ \emph {et~al.}(2021)\citenamefont {Zhang},
  \citenamefont {Chern},\ and\ \citenamefont {Kim}}]{Zhang2021}%
  \BibitemOpen
  \bibfield  {author} {\bibinfo {author} {\bibfnamefont {E.~Z.}\ \bibnamefont
  {Zhang}}, \bibinfo {author} {\bibfnamefont {L.~E.}\ \bibnamefont {Chern}},\
  and\ \bibinfo {author} {\bibfnamefont {Y.~B.}\ \bibnamefont {Kim}},\
  }\bibfield  {title} {\bibinfo {title} {Topological magnons for thermal hall
  transport in frustrated magnets with bond-dependent interactions},\ }\href
  {https://doi.org/10.1103/PhysRevB.103.174402} {\bibfield  {journal} {\bibinfo
   {journal} {Phys. Rev. B}\ }\textbf {\bibinfo {volume} {103}},\ \bibinfo
  {pages} {174402} (\bibinfo {year} {2021})}\BibitemShut {NoStop}%
\bibitem [{\citenamefont {Zhang}\ \emph {et~al.}(2023)\citenamefont {Zhang},
  \citenamefont {Wilke},\ and\ \citenamefont {Kim}}]{Zhang2023}%
  \BibitemOpen
  \bibfield  {author} {\bibinfo {author} {\bibfnamefont {E.~Z.}\ \bibnamefont
  {Zhang}}, \bibinfo {author} {\bibfnamefont {R.~H.}\ \bibnamefont {Wilke}},\
  and\ \bibinfo {author} {\bibfnamefont {Y.~B.}\ \bibnamefont {Kim}},\
  }\bibfield  {title} {\bibinfo {title} {Spin excitation continuum to
  topological magnon crossover and thermal hall conductivity in kitaev
  magnets},\ }\href {https://doi.org/10.1103/PhysRevB.107.184418} {\bibfield
  {journal} {\bibinfo  {journal} {Phys. Rev. B}\ }\textbf {\bibinfo {volume}
  {107}},\ \bibinfo {pages} {184418} (\bibinfo {year} {2023})}\BibitemShut
  {NoStop}%
\bibitem [{\citenamefont {Baskaran}\ \emph {et~al.}(2008)\citenamefont
  {Baskaran}, \citenamefont {Sen},\ and\ \citenamefont
  {Shankar}}]{BaskaranPRB2008}%
  \BibitemOpen
  \bibfield  {author} {\bibinfo {author} {\bibfnamefont {G.}~\bibnamefont
  {Baskaran}}, \bibinfo {author} {\bibfnamefont {D.}~\bibnamefont {Sen}},\ and\
  \bibinfo {author} {\bibfnamefont {R.}~\bibnamefont {Shankar}},\ }\bibfield
  {title} {\bibinfo {title} {Spin-$s$ kitaev model: Classical ground states,
  order from disorder, and exact correlation functions},\ }\href
  {https://doi.org/10.1103/PhysRevB.78.115116} {\bibfield  {journal} {\bibinfo
  {journal} {Phys. Rev. B}\ }\textbf {\bibinfo {volume} {78}},\ \bibinfo
  {pages} {115116} (\bibinfo {year} {2008})}\BibitemShut {NoStop}%
\bibitem [{\citenamefont {Chandra}\ \emph {et~al.}(2010)\citenamefont
  {Chandra}, \citenamefont {Ramola},\ and\ \citenamefont {Dhar}}]{Chandra2010}%
  \BibitemOpen
  \bibfield  {author} {\bibinfo {author} {\bibfnamefont {S.}~\bibnamefont
  {Chandra}}, \bibinfo {author} {\bibfnamefont {K.}~\bibnamefont {Ramola}},\
  and\ \bibinfo {author} {\bibfnamefont {D.}~\bibnamefont {Dhar}},\ }\bibfield
  {title} {\bibinfo {title} {Classical heisenberg spins on a hexagonal lattice
  with kitaev couplings},\ }\href {https://doi.org/10.1103/PhysRevE.82.031113}
  {\bibfield  {journal} {\bibinfo  {journal} {Phys. Rev. E}\ }\textbf {\bibinfo
  {volume} {82}},\ \bibinfo {pages} {031113} (\bibinfo {year}
  {2010})}\BibitemShut {NoStop}%
\bibitem [{\citenamefont {Samarakoon}\ \emph {et~al.}(2017)\citenamefont
  {Samarakoon}, \citenamefont {Banerjee}, \citenamefont {Zhang}, \citenamefont
  {Kamiya}, \citenamefont {Nagler}, \citenamefont {Tennant}, \citenamefont
  {Lee},\ and\ \citenamefont {Batista}}]{Samarakoon2017PRB}%
  \BibitemOpen
  \bibfield  {author} {\bibinfo {author} {\bibfnamefont {A.~M.}\ \bibnamefont
  {Samarakoon}}, \bibinfo {author} {\bibfnamefont {A.}~\bibnamefont
  {Banerjee}}, \bibinfo {author} {\bibfnamefont {S.-S.}\ \bibnamefont {Zhang}},
  \bibinfo {author} {\bibfnamefont {Y.}~\bibnamefont {Kamiya}}, \bibinfo
  {author} {\bibfnamefont {S.~E.}\ \bibnamefont {Nagler}}, \bibinfo {author}
  {\bibfnamefont {D.~A.}\ \bibnamefont {Tennant}}, \bibinfo {author}
  {\bibfnamefont {S.-H.}\ \bibnamefont {Lee}},\ and\ \bibinfo {author}
  {\bibfnamefont {C.~D.}\ \bibnamefont {Batista}},\ }\bibfield  {title}
  {\bibinfo {title} {{Comprehensive study of the dynamics of a classical Kitaev
  spin liquid}},\ }\href {https://doi.org/10.1103/PhysRevB.96.134408}
  {\bibfield  {journal} {\bibinfo  {journal} {Phys. Rev. B}\ }\textbf {\bibinfo
  {volume} {96}},\ \bibinfo {pages} {134408} (\bibinfo {year}
  {2017})}\BibitemShut {NoStop}%
\bibitem [{\citenamefont {Rousochatzakis}\ \emph {et~al.}(2018)\citenamefont
  {Rousochatzakis}, \citenamefont {Sizyuk},\ and\ \citenamefont
  {Perkins}}]{RousochatzakisNC2018}%
  \BibitemOpen
  \bibfield  {author} {\bibinfo {author} {\bibfnamefont {I.}~\bibnamefont
  {Rousochatzakis}}, \bibinfo {author} {\bibfnamefont {Y.}~\bibnamefont
  {Sizyuk}},\ and\ \bibinfo {author} {\bibfnamefont {N.~B.}\ \bibnamefont
  {Perkins}},\ }\bibfield  {title} {\bibinfo {title} {Quantum spin liquid in
  the semiclassical regime},\ }\href
  {https://doi.org/10.1038/s41467-018-03934-1} {\bibfield  {journal} {\bibinfo
  {journal} {Nat. Commun.}\ }\textbf {\bibinfo {volume} {9}},\ \bibinfo {pages}
  {1575} (\bibinfo {year} {2018})}\BibitemShut {NoStop}%
\bibitem [{\citenamefont {Samarakoon}\ \emph {et~al.}(2018)\citenamefont
  {Samarakoon}, \citenamefont {Wachtel}, \citenamefont {Yamaji}, \citenamefont
  {Tennant}, \citenamefont {Batista},\ and\ \citenamefont
  {Kim}}]{Samarakoon2018PRB}%
  \BibitemOpen
  \bibfield  {author} {\bibinfo {author} {\bibfnamefont {A.~M.}\ \bibnamefont
  {Samarakoon}}, \bibinfo {author} {\bibfnamefont {G.}~\bibnamefont {Wachtel}},
  \bibinfo {author} {\bibfnamefont {Y.}~\bibnamefont {Yamaji}}, \bibinfo
  {author} {\bibfnamefont {D.~A.}\ \bibnamefont {Tennant}}, \bibinfo {author}
  {\bibfnamefont {C.~D.}\ \bibnamefont {Batista}},\ and\ \bibinfo {author}
  {\bibfnamefont {Y.~B.}\ \bibnamefont {Kim}},\ }\bibfield  {title} {\bibinfo
  {title} {{Classical and quantum spin dynamics of the honeycomb
  $\mathrm{\ensuremath{\Gamma}}$ model}},\ }\href
  {https://doi.org/10.1103/PhysRevB.98.045121} {\bibfield  {journal} {\bibinfo
  {journal} {Phys. Rev. B}\ }\textbf {\bibinfo {volume} {98}},\ \bibinfo
  {pages} {045121} (\bibinfo {year} {2018})}\BibitemShut {NoStop}%
\bibitem [{\citenamefont {Saha}\ \emph {et~al.}(2019)\citenamefont {Saha},
  \citenamefont {Fan}, \citenamefont {Zhang},\ and\ \citenamefont
  {Chern}}]{Chern2019PRL}%
  \BibitemOpen
  \bibfield  {author} {\bibinfo {author} {\bibfnamefont {P.}~\bibnamefont
  {Saha}}, \bibinfo {author} {\bibfnamefont {Z.}~\bibnamefont {Fan}}, \bibinfo
  {author} {\bibfnamefont {D.}~\bibnamefont {Zhang}},\ and\ \bibinfo {author}
  {\bibfnamefont {G.-W.}\ \bibnamefont {Chern}},\ }\bibfield  {title} {\bibinfo
  {title} {{Hidden Plaquette Order in a Classical Spin Liquid Stabilized by
  Strong Off-Diagonal Exchange}},\ }\href
  {https://doi.org/10.1103/PhysRevLett.122.257204} {\bibfield  {journal}
  {\bibinfo  {journal} {Phys. Rev. Lett.}\ }\textbf {\bibinfo {volume} {122}},\
  \bibinfo {pages} {257204} (\bibinfo {year} {2019})}\BibitemShut {NoStop}%
\bibitem [{\citenamefont {Elitzur}(1975)}]{Elitzur1975}%
  \BibitemOpen
  \bibfield  {author} {\bibinfo {author} {\bibfnamefont {S.}~\bibnamefont
  {Elitzur}},\ }\bibfield  {title} {\bibinfo {title} {Impossibility of
  spontaneously breaking local symmetries},\ }\href
  {https://doi.org/10.1103/PhysRevD.12.3978} {\bibfield  {journal} {\bibinfo
  {journal} {Phys. Rev. D}\ }\textbf {\bibinfo {volume} {12}},\ \bibinfo
  {pages} {3978} (\bibinfo {year} {1975})}\BibitemShut {NoStop}%
\bibitem [{\citenamefont {Gohlke}\ \emph {et~al.}(2018)\citenamefont {Gohlke},
  \citenamefont {Wachtel}, \citenamefont {Yamaji}, \citenamefont {Pollmann},\
  and\ \citenamefont {Kim}}]{Gohlke2018}%
  \BibitemOpen
  \bibfield  {author} {\bibinfo {author} {\bibfnamefont {M.}~\bibnamefont
  {Gohlke}}, \bibinfo {author} {\bibfnamefont {G.}~\bibnamefont {Wachtel}},
  \bibinfo {author} {\bibfnamefont {Y.}~\bibnamefont {Yamaji}}, \bibinfo
  {author} {\bibfnamefont {F.}~\bibnamefont {Pollmann}},\ and\ \bibinfo
  {author} {\bibfnamefont {Y.~B.}\ \bibnamefont {Kim}},\ }\bibfield  {title}
  {\bibinfo {title} {Quantum spin liquid signatures in kitaev-like frustrated
  magnets},\ }\href {https://doi.org/10.1103/PhysRevB.97.075126} {\bibfield
  {journal} {\bibinfo  {journal} {Phys. Rev. B}\ }\textbf {\bibinfo {volume}
  {97}},\ \bibinfo {pages} {075126} (\bibinfo {year} {2018})}\BibitemShut
  {NoStop}%
\bibitem [{\citenamefont {Gordon}\ \emph {et~al.}(2019)\citenamefont {Gordon},
  \citenamefont {Catuneanu}, \citenamefont {S{\o}rensen},\ and\ \citenamefont
  {Kee}}]{Gordon2019NC}%
  \BibitemOpen
  \bibfield  {author} {\bibinfo {author} {\bibfnamefont {J.~S.}\ \bibnamefont
  {Gordon}}, \bibinfo {author} {\bibfnamefont {A.}~\bibnamefont {Catuneanu}},
  \bibinfo {author} {\bibfnamefont {E.~S.}\ \bibnamefont {S{\o}rensen}},\ and\
  \bibinfo {author} {\bibfnamefont {H.-Y.}\ \bibnamefont {Kee}},\ }\bibfield
  {title} {\bibinfo {title} {{Theory of the field-revealed Kitaev spin
  liquid}},\ }\href {https://doi.org/10.1038/s41467-019-10405-8} {\bibfield
  {journal} {\bibinfo  {journal} {Nat. Commun.}\ }\textbf {\bibinfo {volume}
  {10}},\ \bibinfo {pages} {2470} (\bibinfo {year} {2019})}\BibitemShut
  {NoStop}%
\bibitem [{\citenamefont {Gohlke}\ \emph {et~al.}(2020)\citenamefont {Gohlke},
  \citenamefont {Chern}, \citenamefont {Kee},\ and\ \citenamefont
  {Kim}}]{Gohlke2020PRR}%
  \BibitemOpen
  \bibfield  {author} {\bibinfo {author} {\bibfnamefont {M.}~\bibnamefont
  {Gohlke}}, \bibinfo {author} {\bibfnamefont {L.~E.}\ \bibnamefont {Chern}},
  \bibinfo {author} {\bibfnamefont {H.-Y.}\ \bibnamefont {Kee}},\ and\ \bibinfo
  {author} {\bibfnamefont {Y.~B.}\ \bibnamefont {Kim}},\ }\bibfield  {title}
  {\bibinfo {title} {{Emergence of nematic paramagnet via quantum
  order-by-disorder and pseudo-Goldstone modes in Kitaev magnets}},\ }\href
  {https://doi.org/10.1103/PhysRevResearch.2.043023} {\bibfield  {journal}
  {\bibinfo  {journal} {Phys. Rev. Res.}\ }\textbf {\bibinfo {volume} {2}},\
  \bibinfo {pages} {043023} (\bibinfo {year} {2020})}\BibitemShut {NoStop}%
\bibitem [{\citenamefont {Wachtel}\ and\ \citenamefont
  {Orgad}(2019)}]{Wachtel2019PRB}%
  \BibitemOpen
  \bibfield  {author} {\bibinfo {author} {\bibfnamefont {G.}~\bibnamefont
  {Wachtel}}\ and\ \bibinfo {author} {\bibfnamefont {D.}~\bibnamefont
  {Orgad}},\ }\bibfield  {title} {\bibinfo {title} {{Confinement transition in
  a Kitaev-like honeycomb model with bond anisotropy}},\ }\href
  {https://doi.org/10.1103/PhysRevB.99.115104} {\bibfield  {journal} {\bibinfo
  {journal} {Phys. Rev. B}\ }\textbf {\bibinfo {volume} {99}},\ \bibinfo
  {pages} {115104} (\bibinfo {year} {2019})}\BibitemShut {NoStop}%
\bibitem [{\citenamefont {Chern}\ \emph {et~al.}(2020)\citenamefont {Chern},
  \citenamefont {Kaneko}, \citenamefont {Lee},\ and\ \citenamefont
  {Kim}}]{Chern2020PRR}%
  \BibitemOpen
  \bibfield  {author} {\bibinfo {author} {\bibfnamefont {L.~E.}\ \bibnamefont
  {Chern}}, \bibinfo {author} {\bibfnamefont {R.}~\bibnamefont {Kaneko}},
  \bibinfo {author} {\bibfnamefont {H.-Y.}\ \bibnamefont {Lee}},\ and\ \bibinfo
  {author} {\bibfnamefont {Y.~B.}\ \bibnamefont {Kim}},\ }\bibfield  {title}
  {\bibinfo {title} {{Magnetic field induced competing phases in spin-orbital
  entangled Kitaev magnets}},\ }\href
  {https://doi.org/10.1103/PhysRevResearch.2.013014} {\bibfield  {journal}
  {\bibinfo  {journal} {Phys. Rev. Res.}\ }\textbf {\bibinfo {volume} {2}},\
  \bibinfo {pages} {013014} (\bibinfo {year} {2020})}\BibitemShut {NoStop}%
\bibitem [{\citenamefont {Yamada}\ \emph {et~al.}(2020)\citenamefont {Yamada},
  \citenamefont {Suzuki},\ and\ \citenamefont {Suga}}]{Yamada2020PRB}%
  \BibitemOpen
  \bibfield  {author} {\bibinfo {author} {\bibfnamefont {T.}~\bibnamefont
  {Yamada}}, \bibinfo {author} {\bibfnamefont {T.}~\bibnamefont {Suzuki}},\
  and\ \bibinfo {author} {\bibfnamefont {S.-i.}\ \bibnamefont {Suga}},\
  }\bibfield  {title} {\bibinfo {title} {{Ground-state properties of the
  $K\ensuremath{-}\mathrm{\ensuremath{\Gamma}}$ model on a honeycomb
  lattice}},\ }\href {https://doi.org/10.1103/PhysRevB.102.024415} {\bibfield
  {journal} {\bibinfo  {journal} {Phys. Rev. B}\ }\textbf {\bibinfo {volume}
  {102}},\ \bibinfo {pages} {024415} (\bibinfo {year} {2020})}\BibitemShut
  {NoStop}%
\bibitem [{\citenamefont {Liu}\ \emph {et~al.}(2021)\citenamefont {Liu},
  \citenamefont {Sadoune}, \citenamefont {Rao}, \citenamefont {Greitemann},\
  and\ \citenamefont {Pollet}}]{Liu2021PRR}%
  \BibitemOpen
  \bibfield  {author} {\bibinfo {author} {\bibfnamefont {K.}~\bibnamefont
  {Liu}}, \bibinfo {author} {\bibfnamefont {N.}~\bibnamefont {Sadoune}},
  \bibinfo {author} {\bibfnamefont {N.}~\bibnamefont {Rao}}, \bibinfo {author}
  {\bibfnamefont {J.}~\bibnamefont {Greitemann}},\ and\ \bibinfo {author}
  {\bibfnamefont {L.}~\bibnamefont {Pollet}},\ }\bibfield  {title} {\bibinfo
  {title} {{Revealing the phase diagram of Kitaev materials by machine
  learning: Cooperation and competition between spin liquids}},\ }\href
  {https://doi.org/10.1103/PhysRevResearch.3.023016} {\bibfield  {journal}
  {\bibinfo  {journal} {Phys. Rev. Res.}\ }\textbf {\bibinfo {volume} {3}},\
  \bibinfo {pages} {023016} (\bibinfo {year} {2021})}\BibitemShut {NoStop}%
\bibitem [{\citenamefont {Buessen}\ and\ \citenamefont
  {Kim}(2021)}]{Buessen2021PRB}%
  \BibitemOpen
  \bibfield  {author} {\bibinfo {author} {\bibfnamefont {F.~L.}\ \bibnamefont
  {Buessen}}\ and\ \bibinfo {author} {\bibfnamefont {Y.~B.}\ \bibnamefont
  {Kim}},\ }\bibfield  {title} {\bibinfo {title} {{Functional renormalization
  group study of the Kitaev-$\mathrm{\ensuremath{\Gamma}}$ model on the
  honeycomb lattice and emergent incommensurate magnetic correlations}},\
  }\href {https://doi.org/10.1103/PhysRevB.103.184407} {\bibfield  {journal}
  {\bibinfo  {journal} {Phys. Rev. B}\ }\textbf {\bibinfo {volume} {103}},\
  \bibinfo {pages} {184407} (\bibinfo {year} {2021})}\BibitemShut {NoStop}%
\bibitem [{\citenamefont {Chaloupka}\ and\ \citenamefont
  {Khaliullin}(2015)}]{ChaloupkaPRB2015}%
  \BibitemOpen
  \bibfield  {author} {\bibinfo {author} {\bibfnamefont {J.}~\bibnamefont
  {Chaloupka}}\ and\ \bibinfo {author} {\bibfnamefont {G.}~\bibnamefont
  {Khaliullin}},\ }\bibfield  {title} {\bibinfo {title} {Hidden symmetries of
  the extended kitaev-heisenberg model: Implications for the honeycomb-lattice
  iridates ${A}_{2}{\mathrm{iro}}_{3}$},\ }\href
  {https://doi.org/10.1103/PhysRevB.92.024413} {\bibfield  {journal} {\bibinfo
  {journal} {Phys. Rev. B}\ }\textbf {\bibinfo {volume} {92}},\ \bibinfo
  {pages} {024413} (\bibinfo {year} {2015})}\BibitemShut {NoStop}%
\bibitem [{\citenamefont {Rayyan}\ \emph {et~al.}(2021)\citenamefont {Rayyan},
  \citenamefont {Luo},\ and\ \citenamefont {Kee}}]{RayyanPRB2021}%
  \BibitemOpen
  \bibfield  {author} {\bibinfo {author} {\bibfnamefont {A.}~\bibnamefont
  {Rayyan}}, \bibinfo {author} {\bibfnamefont {Q.}~\bibnamefont {Luo}},\ and\
  \bibinfo {author} {\bibfnamefont {H.-Y.}\ \bibnamefont {Kee}},\ }\bibfield
  {title} {\bibinfo {title} {Extent of frustration in the classical
  kitaev-$\mathrm{\ensuremath{\Gamma}}$ model via bond anisotropy},\ }\href
  {https://doi.org/10.1103/PhysRevB.104.094431} {\bibfield  {journal} {\bibinfo
   {journal} {Phys. Rev. B}\ }\textbf {\bibinfo {volume} {104}},\ \bibinfo
  {pages} {094431} (\bibinfo {year} {2021})}\BibitemShut {NoStop}%
\bibitem [{\citenamefont {Chen}\ \emph {et~al.}(2023)\citenamefont {Chen},
  \citenamefont {Luo}, \citenamefont {Zhou}, \citenamefont {He}, \citenamefont
  {Xi}, \citenamefont {Jia}, \citenamefont {Luo},\ and\ \citenamefont
  {Zhao}}]{Chen2023NJP}%
  \BibitemOpen
  \bibfield  {author} {\bibinfo {author} {\bibfnamefont {K.}~\bibnamefont
  {Chen}}, \bibinfo {author} {\bibfnamefont {Q.}~\bibnamefont {Luo}}, \bibinfo
  {author} {\bibfnamefont {Z.}~\bibnamefont {Zhou}}, \bibinfo {author}
  {\bibfnamefont {S.}~\bibnamefont {He}}, \bibinfo {author} {\bibfnamefont
  {B.}~\bibnamefont {Xi}}, \bibinfo {author} {\bibfnamefont {C.}~\bibnamefont
  {Jia}}, \bibinfo {author} {\bibfnamefont {H.-G.}\ \bibnamefont {Luo}},\ and\
  \bibinfo {author} {\bibfnamefont {J.}~\bibnamefont {Zhao}},\ }\bibfield
  {title} {\bibinfo {title} {Triple-meron crystal in high-spin {Kitaev}
  magnets},\ }\href {https://doi.org/10.1088/1367-2630/acb5bb} {\bibfield
  {journal} {\bibinfo  {journal} {New J. Phys.}\ }\textbf {\bibinfo {volume}
  {25}},\ \bibinfo {pages} {023006} (\bibinfo {year} {2023})}\BibitemShut
  {NoStop}%
\bibitem [{\citenamefont {Rao}\ \emph {et~al.}(2021)\citenamefont {Rao},
  \citenamefont {Liu}, \citenamefont {Machaczek},\ and\ \citenamefont
  {Pollet}}]{Rau2021PRR}%
  \BibitemOpen
  \bibfield  {author} {\bibinfo {author} {\bibfnamefont {N.}~\bibnamefont
  {Rao}}, \bibinfo {author} {\bibfnamefont {K.}~\bibnamefont {Liu}}, \bibinfo
  {author} {\bibfnamefont {M.}~\bibnamefont {Machaczek}},\ and\ \bibinfo
  {author} {\bibfnamefont {L.}~\bibnamefont {Pollet}},\ }\bibfield  {title}
  {\bibinfo {title} {Machine-learned phase diagrams of generalized kitaev
  honeycomb magnets},\ }\href
  {https://doi.org/10.1103/PhysRevResearch.3.033223} {\bibfield  {journal}
  {\bibinfo  {journal} {Phys. Rev. Res.}\ }\textbf {\bibinfo {volume} {3}},\
  \bibinfo {pages} {033223} (\bibinfo {year} {2021})}\BibitemShut {NoStop}%
\bibitem [{\citenamefont {Agarwal}\ \emph {et~al.}(2022)\citenamefont
  {Agarwal}, \citenamefont {Mierle},\ and\ \citenamefont
  {Team}}]{Agarwal_Ceres_Solver_2022}%
  \BibitemOpen
  \bibfield  {author} {\bibinfo {author} {\bibfnamefont {S.}~\bibnamefont
  {Agarwal}}, \bibinfo {author} {\bibfnamefont {K.}~\bibnamefont {Mierle}},\
  and\ \bibinfo {author} {\bibfnamefont {T.~C.~S.}\ \bibnamefont {Team}},\
  }\href {https://github.com/ceres-solver/ceres-solver} {\bibinfo {title}
  {{Ceres Solver}}} (\bibinfo {year} {2022})\BibitemShut {NoStop}%
\bibitem [{\citenamefont {Janssen}\ \emph {et~al.}(2017)\citenamefont
  {Janssen}, \citenamefont {Andrade},\ and\ \citenamefont
  {Vojta}}]{JanssenPRB2017}%
  \BibitemOpen
  \bibfield  {author} {\bibinfo {author} {\bibfnamefont {L.}~\bibnamefont
  {Janssen}}, \bibinfo {author} {\bibfnamefont {E.~C.}\ \bibnamefont
  {Andrade}},\ and\ \bibinfo {author} {\bibfnamefont {M.}~\bibnamefont
  {Vojta}},\ }\bibfield  {title} {\bibinfo {title} {Magnetization processes of
  zigzag states on the honeycomb lattice: Identifying spin models for
  $\ensuremath{\alpha}\text{\ensuremath{-}}{\mathrm{rucl}}_{3}$ and
  ${\mathrm{na}}_{2}{\mathrm{iro}}_{3}$},\ }\href
  {https://doi.org/10.1103/PhysRevB.96.064430} {\bibfield  {journal} {\bibinfo
  {journal} {Phys. Rev. B}\ }\textbf {\bibinfo {volume} {96}},\ \bibinfo
  {pages} {064430} (\bibinfo {year} {2017})}\BibitemShut {NoStop}%
\bibitem [{\citenamefont {Ducatman}\ \emph {et~al.}(2018)\citenamefont
  {Ducatman}, \citenamefont {Rousochatzakis},\ and\ \citenamefont
  {Perkins}}]{Ducatman2018}%
  \BibitemOpen
  \bibfield  {author} {\bibinfo {author} {\bibfnamefont {S.}~\bibnamefont
  {Ducatman}}, \bibinfo {author} {\bibfnamefont {I.}~\bibnamefont
  {Rousochatzakis}},\ and\ \bibinfo {author} {\bibfnamefont {N.~B.}\
  \bibnamefont {Perkins}},\ }\bibfield  {title} {\bibinfo {title} {{Magnetic
  structure and excitation spectrum of the hyperhoneycomb Kitaev magnet
  $\ensuremath{\beta}\text{\ensuremath{-}}{\mathrm{Li}}_{2}{\mathrm{IrO}}_{3}$}},\
  }\href {https://doi.org/10.1103/PhysRevB.97.125125} {\bibfield  {journal}
  {\bibinfo  {journal} {Phys. Rev. B}\ }\textbf {\bibinfo {volume} {97}},\
  \bibinfo {pages} {125125} (\bibinfo {year} {2018})}\BibitemShut {NoStop}%
\bibitem [{\citenamefont {Rousochatzakis}()}]{Rousochatzakis2020KITP}%
  \BibitemOpen
  \bibfield  {author} {\bibinfo {author} {\bibfnamefont {I.}~\bibnamefont
  {Rousochatzakis}},\ }\href {https://doi.org/https://doi.org/10.26081/K62328}
  {\bibinfo {title} {{The spin-1/2 $K$-$\Gamma$ Honeycomb Model: Semiclassical,
  strong coupling and exact diagonalization results}}},\ \bibinfo
  {howpublished} {Invited talk, KITP Program: Correlated Systems with
  Multicomponent Local Hilbert Spaces (Sep 28 - Dec 18, 2020),
  https://doi.org/10.26081/K62328}\BibitemShut {NoStop}%
\bibitem [{\citenamefont {Li}\ \emph {et~al.}(2023)\citenamefont {Li},
  \citenamefont {Rao}, \citenamefont {von Delft}, \citenamefont {Pollet},\ and\
  \citenamefont {Liu}}]{JhengArxiv2023}%
  \BibitemOpen
  \bibfield  {author} {\bibinfo {author} {\bibfnamefont {J.-W.}\ \bibnamefont
  {Li}}, \bibinfo {author} {\bibfnamefont {N.}~\bibnamefont {Rao}}, \bibinfo
  {author} {\bibfnamefont {J.}~\bibnamefont {von Delft}}, \bibinfo {author}
  {\bibfnamefont {L.}~\bibnamefont {Pollet}},\ and\ \bibinfo {author}
  {\bibfnamefont {K.}~\bibnamefont {Liu}},\ }\href@noop {} {} (\bibinfo {year}
  {2023}),\ \Eprint {https://arxiv.org/abs/2206.08946} {arXiv:2206.08946
  [cond-mat.str-el]} \BibitemShut {NoStop}%
\bibitem [{\citenamefont {de~Gennes}(1975)}]{DesGennes1975}%
  \BibitemOpen
  \bibfield  {author} {\bibinfo {author} {\bibfnamefont {P.~G.}\ \bibnamefont
  {de~Gennes}},\ }\href@noop {} {\emph {\bibinfo {title} {Fluctuations,
  Instabilities, and Phase transitions}}}\ (\bibinfo  {publisher} {Plenum, New
  York},\ \bibinfo {year} {1975})\BibitemShut {NoStop}%
\bibitem [{\citenamefont {Bak}(1982)}]{Bak1982}%
  \BibitemOpen
  \bibfield  {author} {\bibinfo {author} {\bibfnamefont {P.}~\bibnamefont
  {Bak}},\ }\bibfield  {title} {\bibinfo {title} {Commensurate phases,
  incommensurate phases and the devil's staircase},\ }\href
  {http://stacks.iop.org/0034-4885/45/i=6/a=001} {\bibfield  {journal}
  {\bibinfo  {journal} {Rep. Prog. Phys.}\ }\textbf {\bibinfo {volume} {45}},\
  \bibinfo {pages} {587} (\bibinfo {year} {1982})}\BibitemShut {NoStop}%
\bibitem [{\citenamefont {McMillan}(1976)}]{McMillan1976}%
  \BibitemOpen
  \bibfield  {author} {\bibinfo {author} {\bibfnamefont {W.~L.}\ \bibnamefont
  {McMillan}},\ }\bibfield  {title} {\bibinfo {title} {Theory of
  discommensurations and the commensurate-incommensurate charge-density-wave
  phase transition},\ }\href {https://doi.org/10.1103/PhysRevB.14.1496}
  {\bibfield  {journal} {\bibinfo  {journal} {Phys. Rev. B}\ }\textbf {\bibinfo
  {volume} {14}},\ \bibinfo {pages} {1496} (\bibinfo {year}
  {1976})}\BibitemShut {NoStop}%
\bibitem [{\citenamefont {Schaub}\ and\ \citenamefont
  {Mukamel}(1985)}]{Schaub1985}%
  \BibitemOpen
  \bibfield  {author} {\bibinfo {author} {\bibfnamefont {B.}~\bibnamefont
  {Schaub}}\ and\ \bibinfo {author} {\bibfnamefont {D.}~\bibnamefont
  {Mukamel}},\ }\bibfield  {title} {\bibinfo {title} {Phase diagrams of systems
  exhibiting incommensurate structures},\ }\href
  {https://doi.org/10.1103/PhysRevB.32.6385} {\bibfield  {journal} {\bibinfo
  {journal} {Phys. Rev. B}\ }\textbf {\bibinfo {volume} {32}},\ \bibinfo
  {pages} {6385} (\bibinfo {year} {1985})}\BibitemShut {NoStop}%
\bibitem [{\citenamefont {Henley}(1984)}]{Henley1984}%
  \BibitemOpen
  \bibfield  {author} {\bibinfo {author} {\bibfnamefont {C.~L.}\ \bibnamefont
  {Henley}},\ }\bibfield  {title} {\bibinfo {title} {Defect concepts for vector
  spin glasses},\ }\href
  {https://doi.org/https://doi.org/10.1016/0003-4916(84)90038-1} {\bibfield
  {journal} {\bibinfo  {journal} {Ann. Phys. (N. Y.)}\ }\textbf {\bibinfo
  {volume} {156}},\ \bibinfo {pages} {368} (\bibinfo {year}
  {1984})}\BibitemShut {NoStop}%
\end{thebibliography}%

\appendix

\section{Commensuration of $\bm{A_{3\times n}}$ armchair clusters with $\bm{P_{l\times l}}$ primitive clusters }\label{appx:commensuration_of_clu}
The $A_{3\times n}$ cluster needs to be padded $v$ times along  $3\mathbf{a}_1^A$ and $u$ times along  $\mathbf{a}_2^A$  so that it commensurates with the $P_{l\times l}$ primitive cluster.  Given fixed integer $n$, to find the commensuration one needs to solve the equations  
\begin{equation}
        l \mathbf{a}_1^P + l \mathbf{a}_2^P = u n \mathbf{a}_2^{A}, \ l \mathbf{a}_1^P = v 3 \mathbf{a}_1^A + \dfrac{u}{2} n \mathbf{a}_2^A
\end{equation}
with $u$, $v$, $l$ in the positive integers. Given a solution $u$, $v$, $l$, any integer multiplication of all three is also a solution, and we are interested in the minimum $l$ that admits a solution.  In Tab.~\ref{tab:commensuration_of_clusters} the minimum $l$ that solves the above equation for $n$ up to 60 are shown.

\begin{table}\renewcommand{\arraystretch}{0.8}
\begin{center}
\begin{tabular}{cc|cc|cc||cc|cc|cc}
\hline
\multicolumn{2}{c|}{$(n-1)|3$} & 
\multicolumn{2}{c|}{$(n-2)|3$} &
\multicolumn{2}{c||}{$n|3$} &
\multicolumn{2}{c|}{$(n-1)|3$} & 
\multicolumn{2}{c|}{$(n-2)|3$} &
\multicolumn{2}{c}{$n|3$}\\
$n$ & $l$ & $n$ & $l$ & $n$ & $l$ & $n$ & $l$ & $n$ & $l$ & $n$ & $l$ \\
\hline
 1 & 6 & 2 & 12 & 3 & 6 & 31 & 186 & 32 & 192 & 33 & 66 \\
 4 & 24 & 5 & 30 & 6 & 12 & 34 & 204 & 35 & 210 & 36 & 72 \\
 7 & 42 & 8 & 48 & 9 & 18 & 37 & 222 & 38 & 228 & 39 & 78 \\
 10 & 60 & 11 & 66 & 12 & 24 & 40 & 240 & 41 & 246 & 42 & 84 \\
 13 & 78 & 14 & 84 & 15 & 30 & 43 & 258 & 44 & 264 & 45 & 90 \\
 16 & 96 & 17 & 102 & 18 & 36 & 46 & 276 & 47 & 282 & 48 & 96 \\
 19 & 114 & 20 & 120 & 21 & 42 & 49 & 294 & 50 & 300 & 51 & 102 \\
 22 & 132 & 23 & 138 & 24 & 48 & 52 & 312 & 53 & 318 & 54 & 108 \\
 25 & 150 & 26 & 156 & 27 & 54 & 55 & 330 & 56 & 336 & 57 & 114 \\
 28 & 168 & 29 & 174 & 30 & 60 & 58 & 348 & 59 & 354 & 60 & 120 \\
\hline  
\end{tabular}
\caption{Commensuration of armchair $A_{3\times n}\equiv(3\mathbf{a}_1^A, n\mathbf{a}_2^A)$ clusters with primitive $P_{l\times l}\equiv(l\mathbf{a}_1^P, l\mathbf{a}_2^P)$. Armchairs are split into three columns depending on if $n$ is divisible by 3, corresponding to the three families described in Sec.~\ref{sec:all_about_the_IS_state}.}
\label{tab:commensuration_of_clusters}
\end{center}
\end{table}

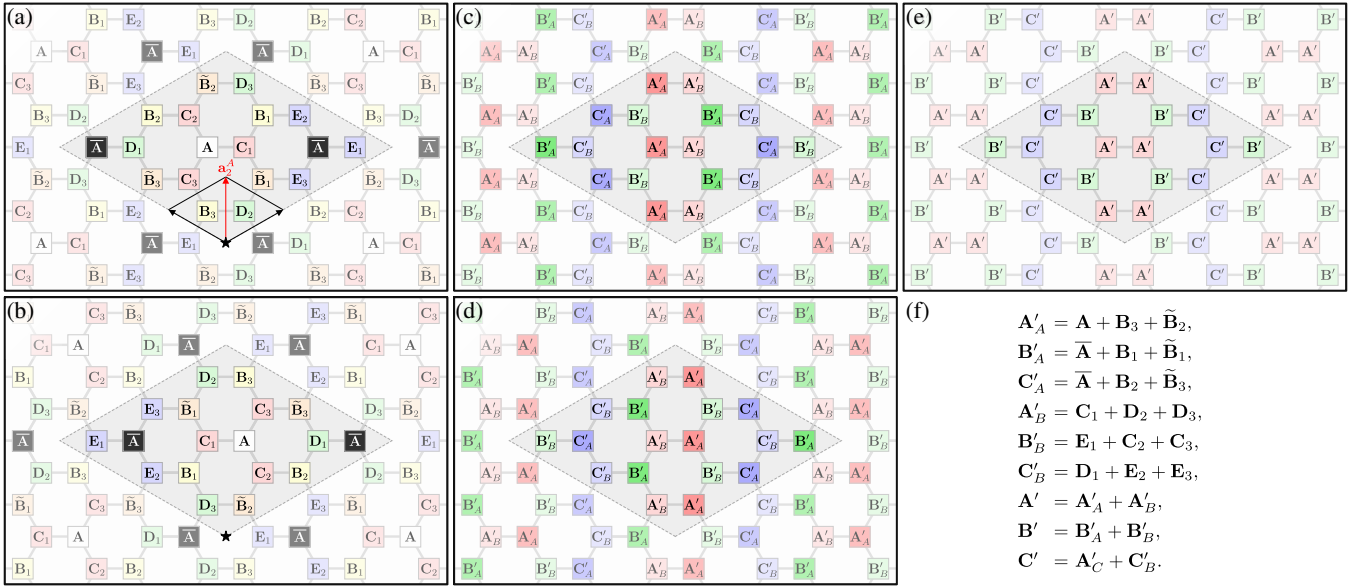
\begin{figure*}
	\centering
    \begin{overpic}[width=1.0\textwidth,percent,angle=0]{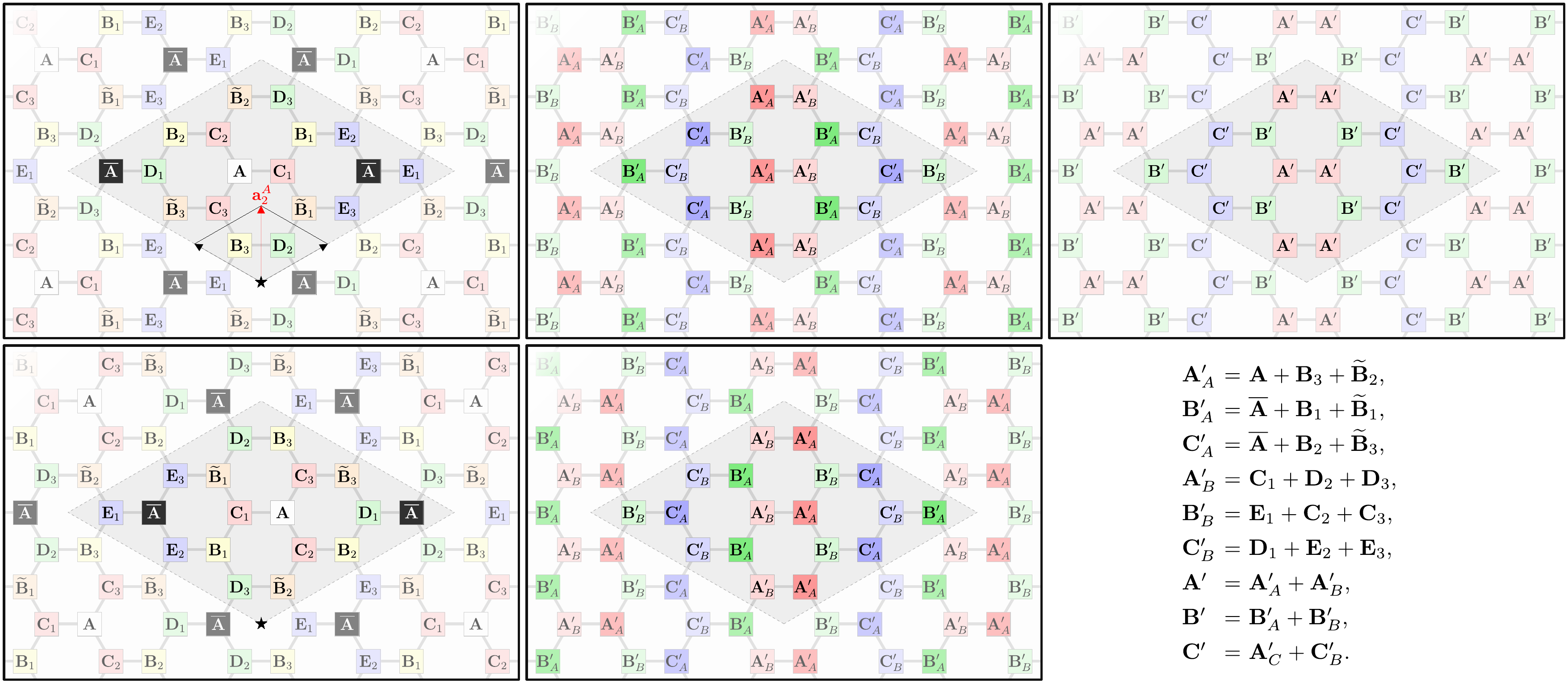}
        \put(0.4,41.8){(a)}
        \put(0.4,20){(b)}
        \put(33.7,41.8){(c)}
        \put(33.7,20){(d)}
        \put(67.1,41.8){(e)}
        \put(67.1,20){(f)}
    \end{overpic} 
    \caption{ Construction of the $6'$ ansatz from the 18C$_3$ ansatz. (a) The $\text{18C}_3^A$ ansatz, with the center on sublattice $A$. (b) The  $\text{18C}_3^B$ ansatz, with the center on sublattice $B$, constructed by crystallographic inversion, with the inversion center indicated by a star. The summation of three translations $(1+\mc{T}_{{\bf a}^A_2}+\mc{T}_{2{\bf a}^A_2})\!\cdot\!\text{18C}_3^A$ is shown in (c), and correspondingly the summation of $\text{18C}_3^B$ translations in (d). (e) Consists of the addition of (c) and (d), which is the complete construction $6' \propto (1+\mc{T}_{{\bf a}^A_2}+\mc{T}_{2{\bf a}^A_2})\!\cdot\!\left(\text{18C}_3^A\!+\!\text{18C}_3^B\right)$. The relations of the magnetic sublattices are shown in (f). }\label{fig:6pfrom18C3}
\end{figure*}

\section{Constructing $\bm{6'}$ from 18C$_{\bm{3}}$ }\label{appx:6pfrom18C3}

We find that the $6'$ state Eq.~(\ref{eq:Kanz}) can be constructed from the 18C$_3$ state Eq.~(\ref{eq:18c3anz}). To achieve this we need to remove two features of the 18C$_3$ state: 1) the $A$ vs $B$ sublattice imbalance, and 2) the inherent $\mathsf{C}_3$ character. 

The visual representation of the construction is shown in the various panels of Fig.~\ref{fig:6pfrom18C3}. The first goal of restoring sublattice symmetry will be achieved by the crystallographic inversion $I$, which naturally maps $A$ sublattice sites to $B$. So we begin with the 18C$_3^A$ state, with the center on sublattice $A$, in Fig.~\ref{fig:6pfrom18C3}(a), and construct the 18C$_3^B=I\cdot$18C$_3^A$ in Fig.~\ref{fig:6pfrom18C3}(b). 

For the second goal of lifting the inherent $\mathsf{C}_3$ character we aim to combining three sites at a time. The $\mathbf{a}_2^A\!=\!\mathbf{a}_1^P\!+\!\mathbf{a}_2^P$ 
lattice vector translations (parallel to the crystallographic $\mathbf{a}$ direction) is the necessary translation to achieve this goal
by forming the partial summation $(1+\mc{T}_{{\bf a}^A_2}+\mc{T}_{2{\bf a}^A_2})\!\cdot\!\text{18C}_3^A$ in panel (c), 
and the corresponding summation of $\text{18C}_3^B$ translations in Fig.~\ref{fig:6pfrom18C3}(d). Using Eq.~(\ref{eq:18c3anz}) to carry out the summation, six magnetic sublattices remain
\begin{equation}
\renewcommand*{\arraystretch}{1.3}
\begin{array}{lcrrrr}
\mathbf{A}'_A  & \!\!\!\!=\!\!\!\! &{[} & \!\!\!\!-1 + y + z ,&\!\! -1 + y + z ,&\!\! -1 + 2 x {]}, \\
\mathbf{B}'_A  & \!\!\!\!=\!\!\!\! &{[} & \!\!\!\!1 + x + y ,&\!\! 1 + x + y ,&\!\! 1 + 2 z {]},  \\
\mathbf{C}'_A  & \!\!\!\!=\!\!\!\! &{[} & \!\!\!\!1 + x + z ,&\!\! 1 + x + z ,&\!\! 1 + 2 y {]}, \\
 \mathbf{A}'_B & \!\!\!\!=\!\!\!\!&{[} & \!\!\!\!x_{\mathrm{r}} + x_{\mathrm{g}} + z_{\mathrm{g}} ,&\!\! x_{\mathrm{r}} + x_{\mathrm{g}} + z_{\mathrm{g}} ,&\!\! 2 x_{\mathrm{g}} + z_{\mathrm{r}}  {]}, \\
  \mathbf{B}'_B & \!\!\!\!=\!\!\!\! &{[} & \!\!\!\!x_{\mathrm{b}} + x_{\mathrm{r}} + z_{\mathrm{r}} ,&\!\! x_{\mathrm{b}} + x_{\mathrm{r}} + z_{\mathrm{r}} ,&\!\! 2 x_{\mathrm{r}} + z_{\mathrm{b}}  {]}, \\
  \mathbf{C}'_B & \!\!\!\!=\!\!\!\!&  {[} & \!\!\!\!x_{\mathrm{g}} + x_{\mathrm{b}} + z_{\mathrm{b}} ,&\!\! x_{\mathrm{g}} + x_{\mathrm{b}} + z_{\mathrm{b}} ,&\!\! 2 x_{\mathrm{b}} + z_{\mathrm{g}} {]},
\end{array}    
\end{equation}
with the spin directions fixed on the $\mathbf{ac}$ plane for all six magnetic sublattices. Finally, by combining these partial summations into the total operation
\be
\Big(1+\mc{T}_{{\bf a}^A_2}+\mc{T}_{2{\bf a}^A_2}\Big)\cdot
\Big(
\text{18C}_3^A\!+\!\text{18C}_3^B\Big)\,,
\ee
we arrive at a three magnetic sublattice structures $\mathbf{A}'\!=\!\mathbf{A}'_A \!+\!\mathbf{A}'_B$, $\mathbf{B}'\!=\!\mathbf{B}'_A \!+\!\mathbf{B}'_B$, and $\mathbf{C}'\!=\!\mathbf{C}'_A \!+\!\mathbf{C}'_B$, shown in Fig.~\ref{fig:6pfrom18C3}(e). By normalizing these three spin directions we arrived at the $6'$ state Eq.~(\ref{eq:Kanz}).

\section{Supplementary information on numerical results}\label{appx:additionla_numb_info}

\subsection{Evolution of the spin inertia eigenvalues}\label{appx:spin_inertia}
\begin{figure}
	\centering
    \begin{overpic}[width=1.0\columnwidth,percent]{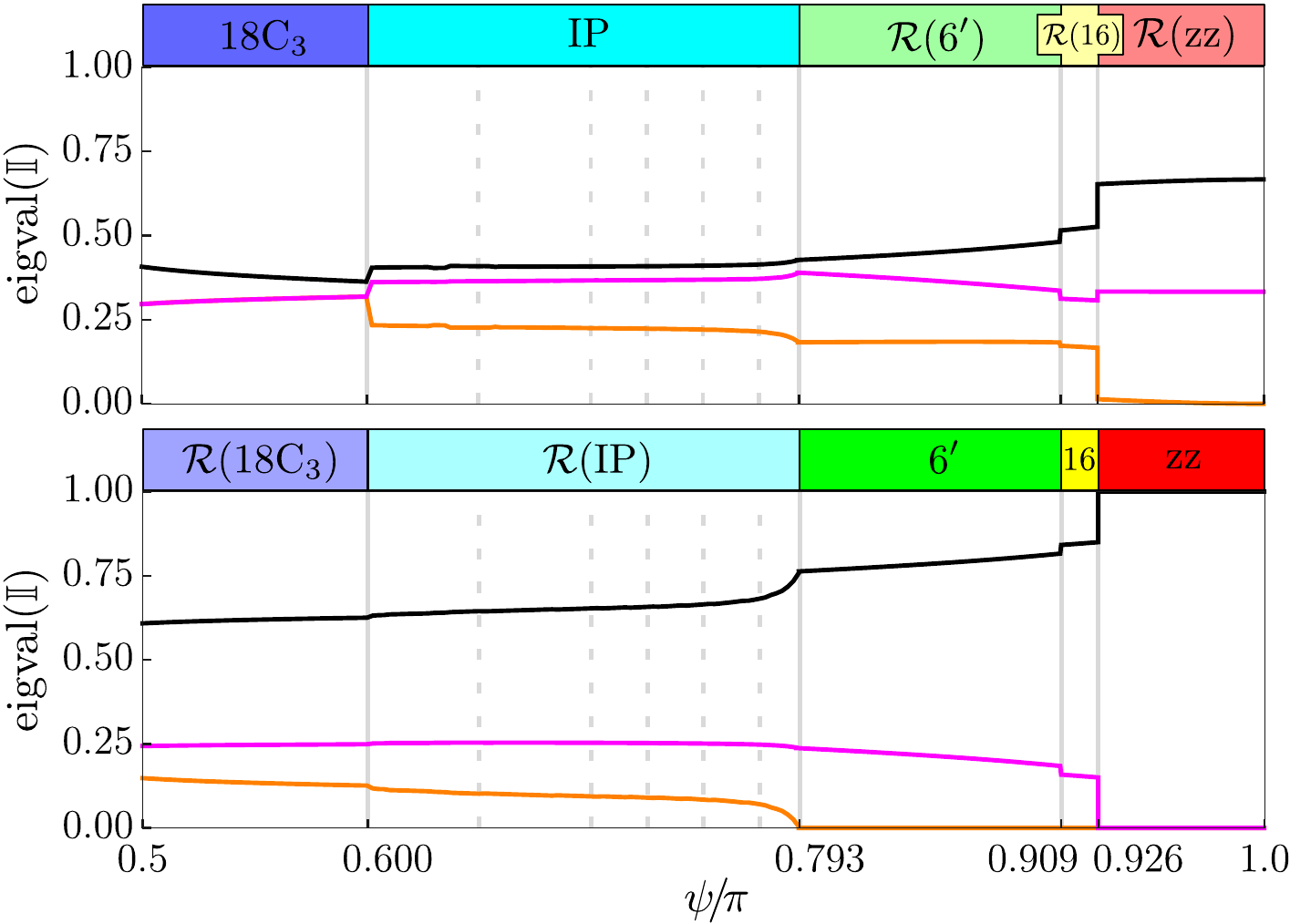}
        \put(0.5,69){(a)}
        \put(0.5,36){(b)}
    \end{overpic}  
\caption{ The eigenvalues of the spin inertia tensor $\mathbb{I}$ in the region $\psi/\pi \in [0.5,1.0]$, for states 18C$_3$, IP, $\mathcal{R}(6')$, $\mathcal{R}(16)$, and  $\mathcal{R}(\text{zz})$ in (a), and $\mathcal{R}(\text{18C}_3)$,  $\mathcal{R}(\text{IP})$, $6'$, $16$, and zz in (b). Dashed gray gridlines indicate the $\psi$ points for which $A_{3\times n}$ clusters where simulated in Monte Carlo, with other points in the IP are evaluated from Lagrange multiplier minimization. The commensurate 18C$_3$, $6'$, 16 and zz phases' inertia tensors evaluated from the respective ansatz. }\label{fig:spin_inertia}
\end{figure}

Many of the states discussed in this work have a non-coplanar nature, which can be also identified by examining the three eigenvalues of the so-called spin inertia tensor~\cite{Henley1984}
\be
    \mathbb{I}^{\alpha \beta}=\dfrac{1}{N}\displaystyle\sum\limits_{i}S_ {i}^{\alpha}S_ {i}^{\beta}.
\ee
This matrix is semi-definite positive and its trace equals 1, due to the spin length constrains. It follows that the three eigenvalues of $\mathbb{I}$ are non-negative and add to one. Moreover,  for collinear (respectively coplanar) configurations, two (respectively one) of the eigenvalues of $\mathbb{I}$  must vanish identically. Figure~\ref{fig:spin_inertia}\,(a) shows the evolution of the eigenvalues of $\mathbb{I}$ for the states 18C$_3$, IP, $\mc{R}(6')$, $\mc{R}(16)$, and $\mc{R}(zz)$, and Fig.~\ref{fig:spin_inertia}\,(b) shows the corresponding plot for the states $\mc{R}$(18C$_3$), $\mc{R}$(IP), $6'$, $16$ and $zz$. For the zz state, two of the eigenvalues of $\mathbb{I}$ vanish because this is a collinear state. Similarly, for the $6'$ and $16$ phases, one of the eigenvalues vanishes because these are coplanar states. Additionally, one of the two remaining eigenvalues is much larger than the other, indicating that these states are nearly collinear. For the remaining two phases, all three eigenvalues are nonzero, consistent with their general non-coplanar nature (although for  $\mc{R}$(18C$_3$) and $\mc{R}$(IP), one of the eigenvalues (in black) is evidently much larger than the other two). Finally, on the boundary of the phases we see a behavior, consistent with the one seen in the main text average chirality at the transition points: the smooth-ish transition at $\psi\!/\!\pi\!=\!0.6$ seems to be between IP to $\mc{R}(\text{18C}_3)$, while at $\psi\!/\!\pi\!=\!0.793$ the IP seems to transition into the $\mc{R}(6')$ state. This latter point is consistent with the findings in Sec.~\ref{sec:chir} as well.

\subsection{Decomposition of the SSF in cubic coordinates}\label{appx:SSF_decomposition}
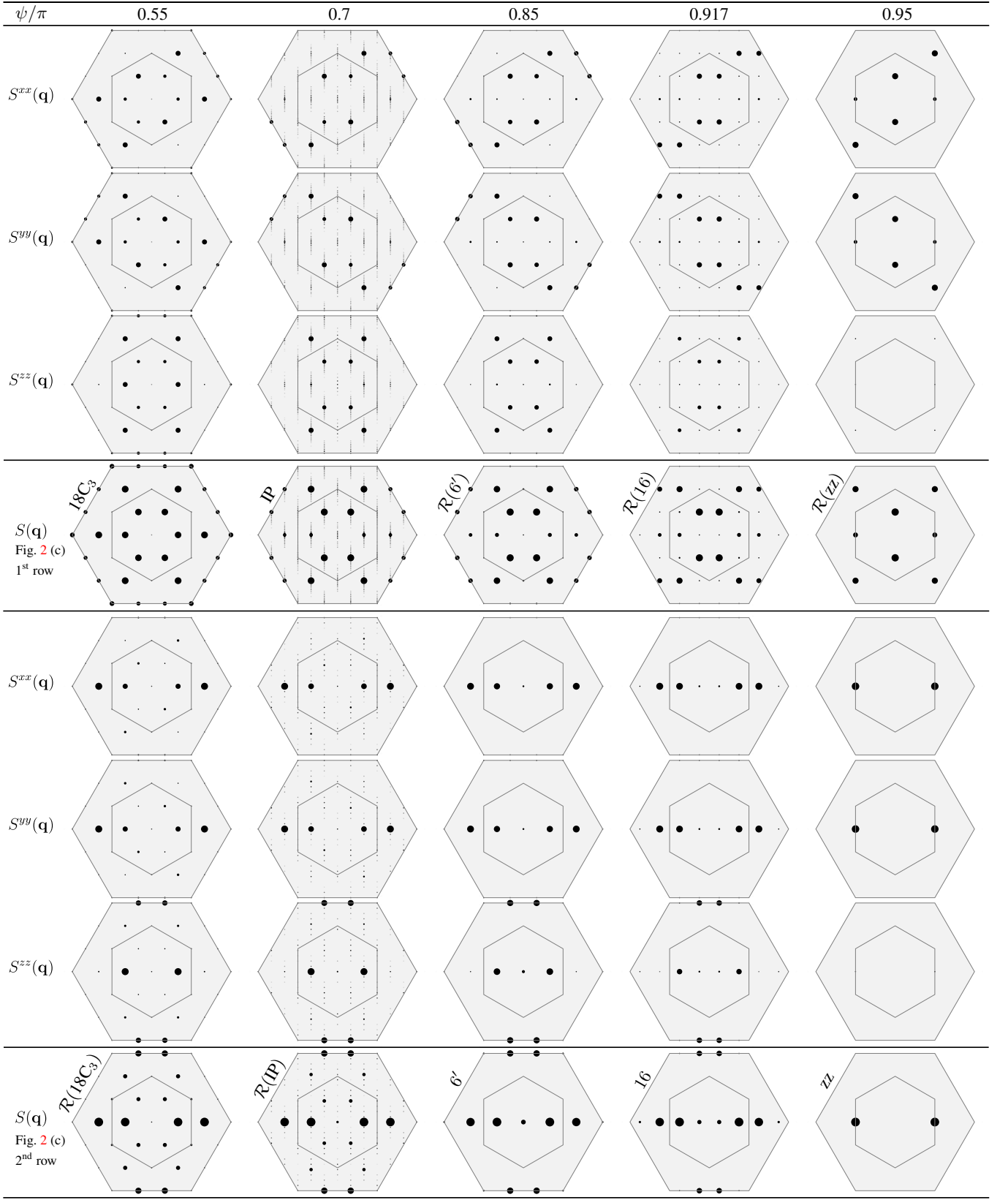
\begin{figure*}
    \centering
    \begin{overpic}[width=1.0\textwidth,percent]{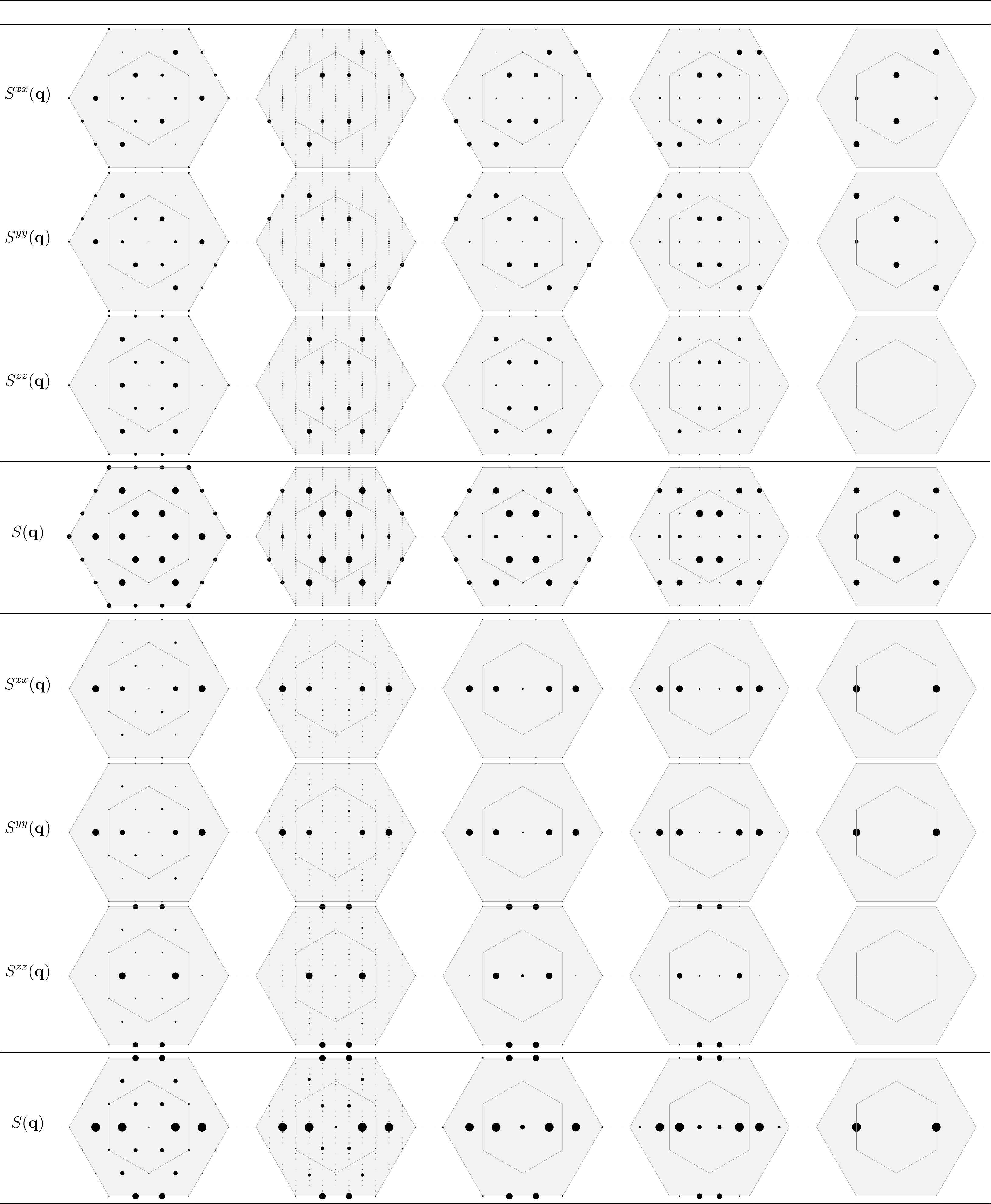}
        \put(1,98.6){$\psi/\pi$}
        \put(11.2,98.5){0.55}
        \put(27.15,98.5){0.7}
        \put(42.2,98.5){0.85}
        \put(57.3,98.5){0.917}
        \put(73.4,98.5){0.95}
             
        \put(1,53.8){\scriptsize Fig.~\ref{fig:all_the_energies} (c)}
        \put(1,52.1){\scriptsize 1\textsuperscript{st} row}
        \put(5.3,57.3){\rotatebox{60}{18C$_3$}}
        \put(21.4,58){\rotatebox{60}{IP}}
        \put(36,57.2){\rotatebox{60}{$\mc{R}(6')$}}
        \put(51.5,57){\rotatebox{60}{$\mc{R}(16)$}}
        \put(67.3,57.1){\rotatebox{60}{$\mc{R}(\text{zz})$}}
        
        \put(1,4.5){\scriptsize Fig.~\ref{fig:all_the_energies} (c)}
        \put(1,2.8){\scriptsize 2\textsuperscript{nd} row}        
        \put(4.4,7){\rotatebox{60}{$\mc{R}(\text{18C}_3)$}}
        \put(20.7,8.15){\rotatebox{60}{$\mc{R}(\text{IP})$}}
        \put(37,9.2){\rotatebox{60}{$6'$}}
        \put(52.6,9){\rotatebox{60}{$16$}}
        \put(68.2,9.1){\rotatebox{60}{zz}}
        
    \end{overpic}  
    \caption{Decomposition of the SSF in the cubic resolved  $S(\mathbf{q}) = S^{xx}(\mathbf{q})+S^{yy}(\mathbf{q})+S^{zz}(\mathbf{q})$, with one case from each state. The ordering (left to right) follows in tandem with Fig.~\ref{fig:all_the_energies} (c).}\label{fig:SSF_decomposition}
\end{figure*}

The SSF involves the correlator of the dot product of the spins on different sites. It can be decomposed into its three contributions
\be
\begin{array}{c}
     S^{\alpha\alpha}(\mathbf{q}) = \dfrac{1}{N}\sum\limits_{i,j}\mathrm{e}^{i \mathbf{q}\cdot(\mathbf{r}_i-\mathbf{r}_j)}S_i^{\alpha}S_j^{\alpha}, \\
     S(\mathbf{q}) = S^{xx}(\mathbf{q})+S^{yy}(\mathbf{q})+S^{zz}(\mathbf{q}),
\end{array}
\ee
working in the cubic axis decomposition. The results are shown in Fig.~\ref{fig:SSF_decomposition}. The collinear and coplanar orders zz,  $\mc{R}(6')$ and $16'$ show a characteristic $S^{xx}=S^{yy}\neq S^{zz}$, which reflects the fact that their spins lived purely on the $\mathbf{ac}$ plane leading to the form [x,x,z] in their respective ansatz. The inherent $C_3$ character of the 18C$_3$ state is now broken up into three parts, with the three contributions being $C_3$ related to each other, to add up to the star pattern with manifest $C_3$ character. Comparing the decomposed forms of 18C$_3$, IP, $6'$, and their $\mc{R}$ duals, we note that the decomposition of IP looks the most like the decomposition of $\mc{R}(6')$. This would be consistent with the findings in Sec.~\ref{sec:all_about_the_IS_state}.

\subsection{Numerical values of minimum energy in the IP phase from classical Monte Carlo simulation}\label{sec:appendixC3}

Here we present numerical values of minimum energy obtained from the classical Monte Carlo simulation for various $A_{3\times n}$ clusters in the IP phase. For comparison with the 18C$_3$ and $6'$ state, we also provide numerical values of their energy obtained from minimizing the ansatz Eq.~(\ref{eq:18c3anz}) and Eq.~(\ref{eq:Kanz}) in the main text. The results are tabulated in Table~\ref{tab:raw_num_mc}.

\begin{table*}
\begin{center}
\resizebox{\textwidth}{!}{%
\begin{tabular}{c|ccccc||c|ccccc}
\hline
State & \multicolumn{5}{|c||}{$\psi/\pi$} & State & \multicolumn{5}{|c}{$\psi/\pi$} \\
& 0.650 & 0.700 & 0.725 & 0.750 & 0.775 & & 0.650 & 0.700 & 0.725 & 0.750 & 0.775 \\
\hline
 $6'$    &         -1.004677747020  &         -0.974481042219  &         -0.951651791278  &         -0.923937337615  &         -0.891611575570  & 30      &         -1.008724471120  &         -0.975899130320  &         -0.952365304940  &         -0.924217705746  &         -0.891661503902  \\
 18C$_3$ &         -1.008606715530  &         -0.975743456564  &         -0.952243806538  &         -0.924135949894  &         -0.891594007435  & 31      &         -1.008674846450  &         -0.975851942481  &         -0.952358541099  &         -0.924243488658  &         -0.891686092798  \\
 1       &         -1.004677747020  &         -0.974481042219  &         -0.951651791278  &         -0.923937337615  &         -0.891611575570  & 32      &         -1.008711948450  &         -0.975916269549  &         -0.952405390316  &         -0.924268655787  &         -0.891698034786  \\
 2       &         -1.004677747020  &         -0.974481042219  &         -0.951651791278  &         -0.923937337615  &         -0.891611575570  & 33      &         -1.008726711580  &         -0.975912976207  &         -0.952383710900  &         -0.924236355913  &         -0.891673022421  \\
 3       &         -1.008606715530  &         -0.975743456564  &         -0.952243806538  &         -0.924135949894  &         -0.891611575570  & 34      &         -1.008699181830  &         -0.975855961990  &         -0.952350193735  &         -0.924237179678  &         -0.891681884891  \\
 4       &         -1.008070655770  &         -0.975076054471  &         -0.951723817053  &         -0.923937337615  &         -0.891611575570  & 35      &         -1.008704601590  &         -0.975909812771  &         -0.952402668617  &         -0.924269291293  &         -0.891699599287  \\
 5       &         -1.006287942300  &         -0.974481042219  &         -0.951651791278  &         -0.923937337615  &         -0.891611575570  & 36      &         -1.008726658320  &         -0.975919908239  &         -0.952395054405  &         -0.924249168580  &         -0.891681753091  \\
 6       &         -1.008606715530  &         -0.975743456564  &         -0.952243806538  &         -0.924135949894  &         -0.891611575570  & 37      &         -1.008713903120  &         -0.975881872769  &         -0.952344814276  &         -0.924231322545  &         -0.891677865423  \\
 7       &         -1.008640755980  &         -0.975760836137  &         -0.952221128554  &         -0.924099154506  &         -0.891614608897  & 38      &         -1.008698234140  &         -0.975902627788  &         -0.952398556287  &         -0.924268249761  & \textbf{-0.891699851092} \\
 8       &         -1.008070655770  &         -0.975076054471  &         -0.951723817053  &         -0.923937337615  &         -0.891611575570  & 39      &         -1.008725207970  & \textbf{-0.975922389050} &         -0.952401615402  &         -0.924257798349  &         -0.891688183734  \\
 9       &         -1.008606715530  &         -0.975743456564  &         -0.952243806538  &         -0.924135949894  &         -0.891611575570  & 40      &         -1.008724471120  &         -0.975899130320  &         -0.952365304940  &         -0.924225931250  &         -0.891674081616  \\
 10      &         -1.008724471120  &         -0.975899130320  &         -0.952365304940  &         -0.924217705746  &         -0.891661503902  & 41      &         -1.008692058450  &         -0.975895253532  &         -0.952393699332  &         -0.924266158745  &         -0.891699217219  \\
 11      &         -1.008470774700  &         -0.975550505303  &         -0.952037227478  &         -0.923988192315  &         -0.891611575570  & 42      &         -1.008720571970  &         -0.975921984956  &         -0.952404901333  &         -0.924263411159  &         -0.891692787658  \\
 12      &         -1.008606715530  &         -0.975743456564  &         -0.952243806538  &         -0.924135949894  &         -0.891611575570  & 43      &         -1.008726268970  &         -0.975910284970  &         -0.952379883149  &         -0.924232318654  &         -0.891670547347  \\
 13      &         -1.008725207970  & \textbf{-0.975922389050} &         -0.952401615402  &         -0.924257798349  &         -0.891688183734  & 44      &         -1.008705741650  &         -0.975887993143  &         -0.952388492174  &         -0.924263423433  &         -0.891697990433  \\
 14      &         -1.008640755980  &         -0.975760836137  &         -0.952221128554  &         -0.924099154506  &         -0.891614608897  & 45      &         -1.008716137980  &         -0.975919720696  & \textbf{-0.952405927775} &         -0.924266833321  &         -0.891695964541  \\
 15      &         -1.008606715530  &         -0.975743456564  &         -0.952243806538  &         -0.924135949894  &         -0.891611575570  & 46      & \textbf{-1.008727126580} &         -0.975917126621  &         -0.952390100555  &         -0.924243374972  &         -0.891677706779  \\
 16      &         -1.008711948450  &         -0.975916269549  &         -0.952405390316  &         -0.924268655787  &         -0.891698034786  & 47      &         -1.008716195840  &         -0.975885918232  &         -0.952383177369  &         -0.924260308962  &         -0.891696372712  \\
 17      &         -1.008699181830  &         -0.975855961990  &         -0.952316232291  &         -0.924173430925  &         -0.891638604955  & 48      &         -1.008711948450  &         -0.975916269549  &         -0.952405390316  &         -0.924268655787  &         -0.891698034786  \\
 18      &         -1.008606715530  &         -0.975743456564  &         -0.952243806538  &         -0.924135949894  &         -0.891611575570  & 49      &         -1.008726304280  &         -0.975920872147  &         -0.952397073578  &         -0.924251665597  &         -0.891683556879  \\
 19      &         -1.008698234140  &         -0.975902627788  &         -0.952398556287  &         -0.924268249761  & \textbf{-0.891699851092} & 50      &         -1.008724471120  &         -0.975899130320  &         -0.952377903819  &         -0.924256990474  &         -0.891694504306  \\
 20      &         -1.008724471120  &         -0.975899130320  &         -0.952365304940  &         -0.924217705746  &         -0.891661503902  & 51      &         -1.008706913350  &         -0.975912082226  &         -0.952403772315  & \textbf{-0.924269305439} &         -0.891699249692  \\
 21      &         -1.008640755980  &         -0.975760836137  &         -0.952243806538  &         -0.924135949894  &         -0.891614608897  & 52      &         -1.008725207970  & \textbf{-0.975922389050} &         -0.952401615400  &         -0.924257798349  &         -0.891688183734  \\
 22      &         -1.008686696370  &         -0.975887993143  &         -0.952388492174  &         -0.924263423433  &         -0.891697990433  & 53      &         -1.008725940130  &         -0.975908445263  &         -0.952377358721  &         -0.924253584258  &         -0.891692483136  \\
 23      & \textbf{-1.008727126580} &         -0.975917126621  &         -0.952390100555  &         -0.924243374972  &         -0.891677706779  & 54      &         -1.008709693630  &         -0.975907465642  &         -0.952401413780  &         -0.924269093680  &         -0.891699804628  \\
 24      &         -1.008684116450  &         -0.975832049169  &         -0.952291204777  &         -0.924152664386  &         -0.891629936789  & 55      &         -1.008721676770  &         -0.975922292872  &         -0.952404324835  &         -0.924262240919  &         -0.891691784674  \\
 25      &         -1.008677403820  &         -0.975874394728  &         -0.952377903819  &         -0.924256990474  &         -0.891694504306  & 56      &         -1.008726892220  &         -0.975914820552  &         -0.952386457476  &         -0.924250167541  &         -0.891690377814  \\
 26      &         -1.008725207970  & \textbf{-0.975922389050} &         -0.952401615402  &         -0.924257798349  &         -0.891688183734  & 57      &         -1.008717656950  &         -0.975902627788  &         -0.952398556287  &         -0.924268249761  & \textbf{-0.891699851092} \\
 27      &         -1.008709693630  &         -0.975874356536  &         -0.952336316587  &         -0.924190878585  &         -0.891646919299  & 58      &         -1.008718213000  &         -0.975921037012  &         -0.952405647518  &         -0.924265354588  &         -0.891694532683  \\
 28      &         -1.008669928890  &         -0.975862371610  &         -0.952367803793  &         -0.924250167541  &         -0.891690377814  & 59      &         -1.008726794920  &         -0.975918941255  &         -0.952393227872  &         -0.924246791292  &         -0.891688236470  \\
 29      &         -1.008718401720  &         -0.975921037012  &         -0.952405647518  &         -0.924265354588  &         -0.891694532683  & 60      &         -1.008724471120  &         -0.975899130320  &         -0.952395372851  &         -0.924266943849  &         -0.891699506635  \\
\hline  
\end{tabular}}
\caption{ Classical energies per site resulting from minimizing the 18C$_3$ ansatz of Eq.~(\ref{eq:18c3anz}), the $6'$ ansatz of Eq.~(\ref{eq:Kanz}), as well as from Monte Carlo simulations on the $A_{3\times n}$ clusters, for the representative set of parameter points $\psi/\pi=$ 0.65, 0.7, 0.725, 0.75, 0.775 parameter points. The lowest energies among all clusters are given in bold.}
\label{tab:raw_num_mc}
\end{center}
\end{table*}

\end{document}